\documentclass[a4paper,11pt]{article}
\usepackage{jheppub}

\usepackage{amsthm,amsmath,amssymb,amsfonts,tikz}
\usepackage{color}
\usepackage{braket}

\def\be{\begin{equation}\begin{gathered}}
\def\ee{\end{gathered}\end{equation}}
\newcommand{\rf}[1]{(\ref{#1})}
\def\stackreb#1#2{\mathrel{\mathop{#2}\limits_{#1}}}

\newcommand{\tr}{\operatorname{tr}}

\numberwithin{equation}{section}

\theoremstyle{definition}

\title{Irregular conformal blocks, Painlev\'e III and the blow-up equations}
\author[a,b]{Pavlo Gavrylenko}
\author[a,b,c]{Andrei Marshakov}
\author[a,b]{Artem Stoyan}

\affiliation[a]{Center for Advanced Studies, Skoltech, Moscow, Russia}
\affiliation[b]{Department of Mathematics, HSE University and Joint HSE/Skoltech Mathematical Physics Laboratory, Moscow, Russia}
\affiliation[c]{ITEP and Theory Department of Lebedev Institute, Moscow, Russia}

\emailAdd{pasha.145@gmail.com}
\emailAdd{andrei.marshakov@gmail.com}
\emailAdd{artem.stoyan1@gmail.com}

\abstract{
We study the relation of irregular conformal blocks with the Painlev\'e III\(_3\) equation.
The functional representation for the quasiclassical irregular block is shown to be consistent with the BPZ equations of conformal field theory and the Hamilton--Jacobi approach to Painlev\'e III\(_3\).
It leads immediately to a limiting case of the blow-up equations for dual Nekrasov partition function of 4d pure supersymmetric gauge theory, which can be even treated as a defining system of equations for both \(c=1\) and \(c\to\infty\) conformal blocks.
We extend this analysis to the domain of strong-coupling regime where original definition of conformal blocks and Nekrasov functions is not known and apply the results to spectral problem of the Matheiu equations.
Finally, we propose a construction of irregular conformal blocks in the strong coupling region by quantization of Painlev\'e III\(_3\) equation, and obtain in this way a general expression, reproducing \(c=1\) and quasiclassical \(c\to\infty\) results as its particular cases.
We have also found explicit integral representations for \(c=1\) and \(c=-2\) irregular blocks at infinity for some special points.
}

\date{}

\begin{document}

\maketitle

\section{Introduction}

Isomonodromy/CFT correspondence is now among the main modern puzzles and tools of mathematical physics. 
One of its main explicit formulations follows
original proposal of \cite{GIL1}, where the tau function of Painlev\'e VI has been expressed through a series of \(c=1\) conformal blocks~\footnote{
To be precise, the correspondence between isomonodromic deformations and quantum field theory was found in \cite{SJM3} even before the invention of conformal field theory (CFT) \cite{BPZ}.
Later it was noticed in \cite{Knizhnik1989} that isomonodromic tau functions should be related to the correlators in CFT.
Similar ideas in different context were also present in \cite{moore1990}.
}.
Independently in \cite{LLNZ} the regularized action on the Painlev\'e VI solution 
was identified with \(c\to\infty\), or quasi-classical conformal block.
It was proposed later \cite{NekrasovMovie,Jeong:2018qpc,Grassi:2016nnt} that these two isomonodromy/CFT connections can be related themselves by the Nakajima--Yoshioka blow-up relations \cite{NY1,Bershtein:2013oka},
involving Nekrasov partition functions at different values of $\Omega$-background parameters, or conformal blocks with different central charges.

In this paper we investigate these relations for the simplest case of Painlev\'e III\(_3\) (or just PIII\(_3\))
equation in the way, which initially does not refer to any constructions from CFT. In particular,
we exploit the definition of quasiclassical conformal blocks,
proposed in \cite{LLNZ} for the Painlev\'e VI case, as regularized action functional on the Painlev\'e solution, and extend it to our degenerate Painlev\'e III\(_3\) case.
This functional representation turns to be very useful for studying the properties of the solution in the vicinity of the Malgrange divisor,
though actually in this part we only reproduce the formulas, obtained already in \cite{L} without any references to conformal blocks~\footnote{
Namely, \cite[formula 4.36]{L} leads to (\ref{eq:blowup1}) below, whereas \cite[formula 4.51]{L} is our (\ref{eq:blowupComplicated}).
In this paper we actually present upgraded proofs of (\ref{eq:blowup1}) and (\ref{eq:blowupComplicated}),
replacing technically complicated re-summations of series expansions around zero and around movable pole by simple computations, involving ``Kiev formulas'' for the tau functions.
This upgrade becomes crucial when we move to expansion of PIII\(_3\) at \(t\to\infty\), since it looks that such re-summations do not work properly for the asymptotic series.}.
However, we demonstrate that in terms of isomonodromy tau functions description
of the Malgrange divisor becomes transparent, and automatically leads to the
blow-up equations, involving both \(c=1\) and \(c\to\infty\) conformal blocks, or even can be interpreted as system of equations for their definition.

We stress here that the approach, proposed below, actually derives rather complicated relations of 2d conformal field theory, or even 4d supersymmetric gauge theory in \(\Omega\)-background, by pure analytic methods of the theory of differential equations.
However, and this is one of the reasons to consider the Painlev\'e III\(_3\) case, these methods are extended below from \(t\to 0\) to \(t\to\infty\) domain of the tau function expansion, where most part of isomonodromy/CFT-correspondence ingredients, such as Nekrasov functions, are not known.
Nevertheless, following \cite{ILT} the analytic methods can be extended there, and we derive the analogs of the blow-up equations, hypothetically satisfied by partition functions for non-Lagrangian theories.
We overlap in this work with \cite{OlegMovie,OlegToAppear}, where the blow-up equations were proven independently, using different technique.

It has been also discovered that \(c\to\infty\) conformal blocks describe the spectra of 2-nd order differential equations \cite{Braverman:2004cr,NS1}, corresponding to quantum-mechanical version of the Seiberg--Witten integrable system \cite{GKMMM}, so that 
exact quantization conditions are written in terms of quasiclassical conformal blocks.
Together with relation of \(c=1\) conformal blocks with \(2\times 2\) first-order matrix differential equations, arising in the context of auxiliary linear problem by isomonodromy/CFT correspondence, this leads to idea that the blow-up relations for conformal blocks actually arise from the relation between 2-nd order differential equations and \(2\times 2\) systems, known already for a long time \cite{Novokshenov1986}.
Inspired by \cite[section 6]{Grassi:2019coc} we derive the quantization conditions for the quantum cosh-/cos-Mathieu systems~\footnote{
For more on this approach and extension of construction to the Lam\'e equation see \cite{BGG}.
} as some restrictions on monodromy data of the related \(2\times 2\) system, supplied with an extra relation on cancellation of apparent singularity, being actually
vanishing of the B\"acklund-transformed tau function.
It turns out, that in \(t\to\infty\) case the quasiclassical conformal blocks describe the exact perturbative spectrum of cosine potential.
To find this relation we use expressions for monodromy data in cluster coordinates, constructed by studying the WKB graphs~\footnote{
This procedure is just reverse engineering of the asymptotic analysis from \cite{FIKN}.}.

The paper is organized as follows. In sect.~\ref{sec:equationSystem} we start with \(2\times 2\) auxiliary linear problem for the Painlev\'e III\(_3\) equation and study its relation with the cosh-/cos-Mathieu systems.
The corresponding singularity of the Painlev\'e solution on Malgrange divisor is described as vanishing of a tau function, whose asymptotic properties are studied using explicit Kiev formulas from \cite{GIL2} (proven in \cite{GL,BShch1}).
The quasiclassical conformal block is introduced from the tau function vanishing condition, and following \cite{LLNZ} is written as regularized action functional.
In sect.~\ref{ss:blowup} we derive the blow-up equations, just by rewriting the basic formulas from sect.~\ref{sec:equationSystem}.

Sect.~\ref{sec:infinity} is devoted to the \(t\to\infty\) limit.
We start from the tau function expansion of \cite{ILT} and specify the poles of the solution, being governed by (derivatives of) a new function, to be identified further with the ``quasiclassical conformal block at infinity''.
To define the quasiclassical block at infinity explicitly we use the modified version of the action functional, and then, as in the \(t\to 0\) case, prove the derivative formulas (\ref{Finf}).
We also prove that quasiclassical blocks at zero and infinity are related via the generating function of canonical transformation.
Finally in this section we propose the analogs of the blow-up equations, relating ``\(c=1\)'' and ``\(c\to\infty\)'' blocks in the \(t\to\infty\) limit, see (\ref{blowup_infty}) and (\ref{tau1_zero}).

Sect.~\ref{sec:spectralTheory} is devoted to study of the spectral theory meaning of the quasiclassical conformal blocks at infinity. 
We find that, similarly to common description of spectra for \(\cos\) and \(\cosh\) potentials in terms of ``asymmetric single-$\epsilon$'' Nekrasov partition functions, they describe positions of bands in the \(\cos\) potential in the limit, when these bands become exponentially narrow.
The main tool in this section is the computation of monodromies and jumps using 
the WKB approximation, showing that the coordinates from \cite{ILT} are almost cluster coordinates, also noticed in \cite{Coman:2020qgf}.
In this section we also identify one of the tau functions with Zamolodchikov's polymer partition functions \cite{Zamolodchikov:1994uw}, or the spectral determinant from~\cite{AlbaFermiGas}.
This identification gives explicit integral representations for conformal blocks at infinity computed at some special charges.

In sect.~\ref{sec:CFT} we give an identification of our main ingredients with  actual (irregular) conformal blocks of 2d CFT, this is completely done in \(t\to 0\) limit.
We also perform some analysis for \(t\to\infty\), where conformal blocks are not algebraically defined on the CFT side, but nevertheless it is possible to find the behavior of matrix element with heavy degenerate field insertion, when position of this field moves to \(\infty\). It allows to identify our regularized action functional with the correlator without degenerate fields.

Finally, in sect.~\ref{sec:qpPainleve} we switch to quantum version of the Painlev\'e III\(_3\), which is solved by conformal blocks with \emph{arbitrary} central charge.
Conjecturing an expansion for the quantum tau function at infinity we get an analog of the \(\mathbb{C}^2/\mathbb{Z}_2\) blow-up relations for \(t\to\infty\) and solve them iteratively in order to find expansion of the conformal block.
We check that both its limits, \(c=1\) and \(c=\infty\), reproduce the known results.
We also check that the general conformal blocks also satisfy Nakajima--Yoshioka--type blow-up relations, supporting the idea that so defined objects are correct analogs of conformal blocks at \(t\to\infty\).

Some definitions, conventions and cumbersome results of the explicit computations are collected in Appendices.

\section{Isomonodromic deformations and Mathieu equations}

\label{sec:equationSystem}

\subsection{Scalar equation from \(2\times 2\) linear system}

Consider a linear system for the Painlev\'e III\(_3\) equation:
\begin{equation}
\label{eq:LinSystem1}
\frac{\partial}{\partial z}
\begin{pmatrix}
Y_1 \\
Y_2
\end{pmatrix}
=
A(z)
\begin{pmatrix}
Y_1 \\
Y_2
\end{pmatrix},
\end{equation}
with the connection matrix of the form~\footnote{
This matrix can be obtained from that one from \cite{GL} by transformation \(w(t)=-t/q(t)\) and \(z\mapsto-z\), i.e. by combination of a B\"acklund transformation and sign inversion.}
\begin{equation}
\label{eq:PainleveLax}
A(z)=
\begin{pmatrix}
0 & 1\\
0 & 0
\end{pmatrix}
+ \frac1{z}
\begin{pmatrix}
\frac{t w'(t)}{2w(t)} & -w(t)\\
-1 & -\frac{tw'(t)}{2w(t)}
\end{pmatrix}
+
\frac{t}{z^2w(t)}
\begin{pmatrix}
0 & 0\\
1 & 0
\end{pmatrix}.
\end{equation}
One can consider, first, \(w(t)\) and \(w'(t)\) as independent coordinates on the space of the matrices \eqref{eq:PainleveLax},
which will be then related by isomonodromic deformation equation \(\frac{d}{dt}w(t)=w'(t)\).
The isomonodromic deformations of \eqref{eq:LinSystem1} are given by
\begin{equation}
\label{eq:Deformation1}
\frac{\partial}{\partial t}
\begin{pmatrix}
Y_1(z) \\
Y_2(z)
\end{pmatrix}=
B(z)
\begin{pmatrix}
Y_1 \\
Y_2
\end{pmatrix},
\end{equation}
where
\begin{equation}
\label{eq:Bmatrix}
B(z)=
\begin{pmatrix}
0 & \frac{w(t)}{t}\\
-\frac1{zw(t)}& 0
\end{pmatrix}.
\end{equation}
Compatibility of (\ref{eq:LinSystem1}) and (\ref{eq:Deformation1}), i.e. the zero-curvature equation
\(\partial_tA(z)-\partial_zB(z)+[A(z),B(z)]=0\), gives rise to the Painlev\'e III\(_3\) equation~\footnote{
Notice that signs of the last two terms here are different from common conventions, but this choice of signs is more convenient from the CFT point of view in what follows.
}:
\begin{equation}
\label{eq:PIII}
w''(t)-\frac{w'(t)^2}{w(t)}+\frac{w'(t)}{t}+\frac{2w(t)^2}{t^2}-\frac{2}{t}=0.
\end{equation}
Let us now derive the second-order equation for the first component of the linear system \eqref{eq:LinSystem1}. For
\begin{equation}
\label{eq:tildeY}
\tilde{Y}(z)=\frac{Y_1(z)}{\sqrt{A_{12}(z)}}
\end{equation}
after direct computation, using \eqref{eq:LinSystem1} one gets
\begin{equation}\begin{gathered}
\label{eq:generalScalarEquation}
\frac{\partial^2\tilde{Y}(z)}{(\partial\log z)^2}=\\=\left(
\frac12\tr A(z)^2+\frac{\partial A_{11}}{\partial \log z} -
A_{11}\frac{\partial \log A_{12}}{\partial \log z}
+\frac{3}{4}\left( \frac{\partial \log A_{12}}{\partial \log z} \right)^2-\frac{1}{2A_{12}}\frac{\partial^2A_{12}}{(\partial\log z)^2}\right)\tilde{Y}(z).
\end{gathered}\end{equation}
For the matrix \(A(z)\) from (\ref{eq:PainleveLax}), equation (\ref{eq:generalScalarEquation}) acquires the form
\begin{equation}
\begin{gathered}
\label{eq:scalarEquation}
\left( z \frac{\partial}{\partial z} \right)^2 \tilde{Y}(z)=t\left(\frac{t w'(t)^2}{4w(t)^2}+\frac{w(t)}{t}+\frac{1}{w(t)}-\frac{1}{z}-\frac{z}{t}\right)\tilde{Y}(z)+\\+
\left( \frac{3 w(t)^2}{4(z-w(t))^2}+\frac{2 w(t)-tw'(t)}{2(z-w(t))}+\frac{w(t)-2tw'(t)}{4w(t)} \right) \tilde{Y}(z).
\end{gathered}
\end{equation}
In the first bracket in the r.h.s. one recognizes the PIII\(_3\) Hamiltonian
\begin{equation}
\label{eq:P3Hamiltonian1}
H=\frac{t w'(t)^2}{4w(t)^2}+\frac{w(t)}{t}+\frac{1}{w(t)} =\frac{p^2w^2}{t}+\frac{w(t)}{t}+\frac{1}{w(t)},
\end{equation}
when expressed as a function of \(w'\) and \(w\), which generates
the non-autonomous Hamiltonian equations of motion 
\begin{equation}\begin{gathered}
\label{eq:HamiltonianEquations}
w'=\frac{2w^2}{t}p,\\
p'=-\frac{2wp^2}{t}+\frac1{w^2}-\frac1{t},
\end{gathered}\end{equation}
equivalent to (\ref{eq:PIII}).

The second-order equation \eqref{eq:scalarEquation}, when compared to initial linear system, has an extra apparent singularity at the point \(z=w(t)\), coming from the fact that we divided in \eqref{eq:tildeY} by \(\sqrt{A_{12}(z)}\), vanishing at \(z=w(t)\).
Position of this singularity in \eqref{eq:scalarEquation} is exactly the Painlev\'e transcendent. To get rid of this singularity we have to move to the pole of solution at some point, say \(t=t_{\star}\).
Expansion of a solution to (\ref{eq:PIII}) around the pole has the form
\begin{equation}
\label{eq:poleExpansion}
w(t)=-\frac{t_{\star}^2}{(t-t_{\star})^2}-\frac{t_{\star}}{t-t_{\star}}+w_0+O\left((t-t_{\star})^2\right),
\end{equation}
and substituting it into the Hamiltonian, one gets
\begin{equation}
\label{eq:HailtonianW0}
\lim_{t\to t_{\star} }H(t)=\frac{1+12w_0}{4t_{\star}}.
\end{equation}
Now let us take the limit \(t\to t_{\star}\) in \eqref{eq:scalarEquation}. Expressing \(w'=2i \frac{w^{3/2}}{t}\sqrt{1-\frac{t}{w}H+\frac{t}{w^2}}\)
in the limit \(w(t)\stackreb{t\to t_{\star}}{\to}\infty\) we obtain
\begin{equation}
\label{eq:Mathieu1}
\left( z \frac{\partial}{\partial z} \right)^2 \tilde{Y}(z)=\left(t_{\star} H(t_{\star})-\frac{t_{\star}}{z}-z\right)\tilde{Y}(z),
\end{equation}
or the Mathieu equation.

One can easily transform equation (\ref{eq:Mathieu1}) to its usual form in two different ways, either by substitution \(z=\sqrt{t_{\star}} e^{ix}\):
\begin{equation}
\label{eq:sinMathieu}
\left(-\partial_x^2+2\sqrt{t_{\star}}\cos x\right)\tilde{Y}=t_{\star}H(t_{\star}) \tilde{Y},
\end{equation}
or by \(z=-\sqrt{t_{\star}}e^y\):
\begin{equation}
\label{eq:sinhMathieu}
\left(-\partial_y^2+2 \sqrt{t_{\star}}\cosh y \right)\tilde{Y}=-t_{\star}H_{\star}(t_{\star})\tilde{Y}.
\end{equation}
The quantum mechanical systems, described by these equations will be considered in sect.~\ref{sec:spectralTheory} below, where the quantization conditions are obtained from studying the monodromies of $2\times 2$ linear system, corresponding to the transitions along the unit circle in $z$-plane \rf{eq:sinMathieu} or from $z=0$ to $z=\infty$ \rf{eq:sinhMathieu}.
This perspective is developed in \cite{BGG}, where it is also further generalized to the \(2\times 2\) isomonodromic problem on torus with a single puncture.

\subsection{The Painlev\'e transcendent and tau functions}

It is well-known that the Painlev\'e III\(_3\) Hamiltonian \eqref{eq:P3Hamiltonian1} is given by the logarithmic derivative of  isomonodromic tau function:
\begin{equation}
\label{eq:HamiltonianGL}
H(w,w';t)=\frac{tw'^2}{4w^2}+\frac{w}{t}+\frac{1}{w}=\partial_t\log\tau(t),
\end{equation}
which, in its turn, gives \cite{GL} the PIII\(_3\) solution by
\begin{equation}
\label{eq:anotherW}
w(t)^{-1}=\partial_tt\partial_t\log\tau(t).
\end{equation}
For the B\"acklund--transformed solution \(w_1(t)=\frac{t}{w(t)}\) the analog of (\ref{eq:HamiltonianGL}) gives
\begin{equation}
\label{eq:HamiltonianGL1}
H(w_1,w_1';t)=\frac{(tw'-w)^2}{4tw^2}+\frac{w}{t}+\frac{1}{w}=\partial_t\log\tau_1(t),
\end{equation}
and subtracting it from (\ref{eq:HamiltonianGL}) we get
\begin{equation}
\frac12 \partial_t\log w(t)-\frac1{4t}=\partial_t\log \frac{\tau(t)}{\tau_1(t)},
\end{equation}
which is integrated to the formula
\begin{equation}
\label{eq:tauRatio}
w(t)=-t^{1/2}\frac{\tau(t)^2}{\tau_1(t)^2},
\end{equation}
where the constant is fixed from the asymptotics.

Below we shall intensively use the explicit ``Kiev formulas'' \cite{GIL2}
\begin{equation}
\label{eq:KyivFormula}
\tau(t)=\sum\limits_{n\in \mathbb{Z}} \frac{e^{4\pi in\eta}t^{(\sigma+n)^2} \mathcal{B}(\sigma+n,t)}{G(1+2(\sigma+n))G(1-2(\sigma+n))}
\end{equation}
for the isomonodromic tau function \(\tau(t)\)
and
\begin{equation}
\label{eq:BacklundKyivFormula}
\tau_1(t)=\sum\limits_{n\in \frac12+\mathbb{Z}} \frac{e^{4\pi in\eta}t^{(\sigma+n)^2} \mathcal{B}(\sigma+n,t)}{G(1+2(\sigma+n))G(1-2(\sigma+n))}
\end{equation}
for the B\"acklund--transformed \(\tau_1(t)\), which differs from \eqref{eq:KyivFormula} only by summing over the half-integers instead of integers,
see also \cite{BShch2} for bilinear relations between \(\tau\) and \(\tau_1\).
In \eqref{eq:KyivFormula} and \eqref{eq:BacklundKyivFormula} \(\mathcal{B}(\sigma,t)\) denote the irregular $c=1$ conformal blocks (normalized as \(\mathcal{B}(\sigma,t)=1+O(t)\)), or non-refined Nekrasov instanton partition functions in ``self-dual'' \(\Omega\)-background for pure $SU(2)$ supersymmetric 4d gauge theory, while $G(x)$ stays for the Barnes double $\Gamma$-function (see details in Appendix~\ref{ss:cft}).

In \eqref{eq:KyivFormula} and \eqref{eq:BacklundKyivFormula} parameters $\{\sigma,\eta\}$ are two integration constants of the second-order equation \eqref{eq:PIII}, or local coordinates on the monodromy space $\mathcal{M}$ for the linear system \eqref{eq:LinSystem1}, \eqref{eq:PainleveLax}, endowed with the symplectic form $\varpi=4\pi id\eta\wedge d\sigma$. For our purposes it is convenient to relate them to the asymptotics of the solution.
The asymptotics of the tau function (\ref{eq:KyivFormula}) for small positive \(0<\Re\sigma\ll 1\) is \begin{equation}
\label{tauass}
\tau(t)\ \stackreb{t\to 0}{\sim}\
t^{\sigma^2}
\left(
1-e^{-4\pi i\eta}\frac{\Gamma(2\sigma)^2}{\Gamma(2-2\sigma)^2}t^{-2\sigma+1}
+\frac{t}{2\sigma^2}
-e^{4\pi i\eta}\frac{\Gamma(-2\sigma)^2}{\Gamma(2+2\sigma)^2}t^{2\sigma+1}\right),
\end{equation}
which gives for the asymptotics of solution \eqref{eq:anotherW}
\begin{equation}
\label{eq:fastTimeAsymptotics}
w(t)\ \stackreb{t\to 0}{\sim}\
-\frac{\Gamma(1-2\sigma)^2}{\Gamma(2\sigma)^2}\frac{e^{4\pi i\eta}t^{2\sigma}}{\left(1-\frac{\Gamma(1-2\sigma)^2}{\Gamma(1+2\sigma)^2}e^{4\pi i\eta}t^{2\sigma}\right)^2}\
= -\kappa t^{2\sigma} + O(t^{2\sigma}),
\end{equation}
where \(\kappa=e^{4\pi i\eta}\frac{\Gamma(1-2\sigma)^2}{\Gamma(2\sigma)^2}\), and we have actually kept here \emph{all} orders in \(t^{2\sigma}\), but only the zeroth order in integer powers of \(t\).
In such limit our $w(t)$ satisfies the autonomous limiting 'Liouville' equation with the conserved energy \(\frac{(tw')^2}{4w^2}+w=
\sigma^2\).

One can easily find from \eqref{eq:fastTimeAsymptotics} that the B\"acklund transformation \(w(t)\mapsto w_1(t)=\frac{t}{w(t)}\) maps the parameters of solution as \(\eta\mapsto-\eta\), \(\sigma\mapsto \frac12-\sigma\).
Using obvious symmetry in the formula for the isomonodromic tau function (\ref{eq:KyivFormula}) we can rewrite this map as
\begin{equation}
\label{Bset}
\sigma_1=\sigma-\frac12,\quad \eta_1=\eta,
\end{equation}
mapping, in particular, tau function (\ref{eq:KyivFormula}) to the B\"acklund transformed (\ref{eq:BacklundKyivFormula}).

\subsection{Vanishing of the tau function}
\label{sec:tauzero}

Series \eqref{eq:KyivFormula} for \(\tau(t)\) is convergent in the whole $\mathbb{C}^{*}_t$, hence the isomonodromic tau function does not have poles as function of the variables \((t, \eta)\)~\footnote{
It has singularities as function of \(\sigma\) at points \(\sigma\in \frac12 \mathbb{Z}\).}.
Thus, the only poles of \(w(t)\) are zeros of \(\tau_1(\eta,\sigma,t)\), this locus is called as \emph{Malgrange divisor}.
It describes the situation when the B\"acklund--transformed Riemann--Hilbert problem does not have solution.

To denote specialization of some variables to Malgrange divisor we will use \(\star\)-sign, for example
\begin{equation}
\label{Malgrange}
\tau_1(\eta_\star(\sigma,t),\sigma,t)=0,
\end{equation}
or
\begin{equation}\begin{gathered}
\label{Malgrange1}
\tau_1(\eta,\sigma,t_{\star}(\eta,\sigma))=\tau_{1\star}=0,\quad \tau_{\star}(\sigma,t)=\tau(\eta_{\star},\sigma,t),\\
(\partial_t \tau_1)_{\star}(\sigma,t)=\left.\partial_t \tau_1\right|_{\star}(\sigma,t)=\left.\left(\partial_t\tau_{1}(\eta,\sigma,t)\right)\right|_{\eta=\eta_{\star}(\sigma,t)}.
\end{gathered}\end{equation}
Consider now the asymptotics of \(w(t)\) around the pole \eqref{eq:poleExpansion}.
Combining (\ref{eq:poleExpansion}) and (\ref{eq:HailtonianW0}) with (\ref{eq:tauRatio}) one gets~\footnote{
Plus-minus signs come from the fact the (\ref{eq:BacklundKyivFormula}) allows to change the sign of \(e^{2\pi i\eta}\) without changing the solution.
}:
\begin{equation}
\pm t^{1/4}\frac{\tau(\eta,\sigma,t)}{\tau_1(\eta,\sigma,t)}=\frac{t_{\star}}{t-t_\star}+\frac12-\frac1{12}\left(t_\star^{-1}+2H_{\star}\right)(t-t_{\star})+o(t-t_\star).
\end{equation}
Expanding the l.h.s. we obtain some relations between the tau functions and their derivatives, say,
in the leading order:
\begin{equation}
\label{eq:bup1}
\tau_{\star}=\pm t_{\star}^{3/4}(\partial_t\tau_1)_{\star}.
\end{equation}
Now let us look for the form of \(\eta_{\star}(\sigma,t)\).
In order to do this we substitute the ansatz
\begin{equation}
\begin{gathered}
\label{eq:etaStar}
e^{2\pi i\eta_{\star}}= \exp\left(
\frac12 \frac{\partial \mathcal{F}(\sigma,t)}{\partial \sigma}\right)
=-\frac{\Gamma\left(1+2\sigma\right)}{\Gamma\left(1-2\sigma\right)}
t^{-\sigma}
\exp\left(
\frac12 \frac{\partial f(\sigma,t)}{\partial \sigma}\right) =
\\
=-\frac{\Gamma\left(1+2\sigma\right)}{\Gamma\left(1-2\sigma\right)}
\exp\left({\frac12 \frac{\partial f_{cl}(\sigma,t)}{\partial \sigma}}+
{\frac12 \frac{\partial f(\sigma,t)}{\partial \sigma}}\right),
\end{gathered}
\end{equation}
into \eqref{Malgrange}, where~\footnote{
We also introduce here \(t^{-\sigma}=\exp\left(\frac12 \frac{\partial f_{cl}(\sigma,t)}{\partial \sigma}\right)\) to indicate that it is related to classical contribution to Nekrasov partition function, always appearing together with the ``instantonic'' part \(f(\sigma,t)\). We hope, it will not cause any confusion, when both $f(\sigma,t)$ and
$\mathcal{F}(\sigma,t)$, see also \eqref{F0} below, are referred to as quasiclassical conformal blocks, since the first one arises from quasiclassical limit of a conformal block in original normalization of \citep{BPZ}, while the second also absorbs the ``classical'' and ``perturbative'' parts, or the CFT structure constants.
},
\(f(\sigma,t)=\sum_{i=1}^{\infty}f_i(\sigma)t^i\), and get
\begin{equation}
f(\sigma,t)= - \frac{2t}{4\sigma^2-1}-\frac{(7+20\sigma^2)t^2}{4(\sigma^2-1)(4 \sigma^2-1)^3}-\frac{4(144\sigma^4+232\sigma^2+29)t^3}{3(4\sigma^2-1)^5(4\sigma^4-13\sigma^2+9)}+\ldots,
\end{equation}
which coincides with the expansion of quasiclassical conformal block. Other solutions,
due to obviously following from \eqref{eq:BacklundKyivFormula}
$\tau_1(t;\sigma+k,\eta)=e^{-4\pi ik\eta}\tau_1(t;\sigma,\eta)$, are
given by \(\eta=\eta_{\star}(\sigma+k,t)\) for $k\in \mathbb{Z}$~\footnote{
In the leading order at $t\to 0$ the value of \(\eta_{\star}(\sigma,t)\) is defined from cancellation between two neighboring terms in the tau function expansion, proportional to \(e^{4\pi i\eta}t^{(\sigma+1/2)^2}\) and \(e^{-4\pi i\eta}t^{(\sigma-1/2)^2} \), and
it occurs when \(e^{4\pi i\eta}\sim t^{-\sigma}\), as in \eqref{eq:etaStar}.
If one substitutes instead \(e^{4\pi i\eta}\sim t^{-\sigma-k}\), two other terms, namely --- proportional to  \(e^{4\pi i\eta}t^{(\sigma+k+1/2)^2}\) and \(e^{-4\pi i\eta}t^{(\sigma-k-1/2)^2}\), turn to be of the leading order.
Due to quasi-periodicity of the tau function under integer shift of \(\sigma\), it is clear that the whole solution for \(\eta_{\star}\) is then modified by \(\sigma\to\sigma+k\).
}.
Notice also that we have now fixed the sign ``\(+\)'' in formula (\ref{eq:bup1}).
To be precise, we were able to fix \(f(\sigma,t)\) up to the \(\sigma\)-independent part only, and
we are going to fill this gap in the next section.

\subsection{Conformal block as action functional
\label{ss:action}}

It is already known \cite{LLNZ,L} that the quasiclassical conformal block can be represented as action of the Painlev\'e equation on its solution.
Actually, let us define
\begin{equation}
\label{eq:action}
\tilde{f}(\sigma,t_{\star})=\int_0^{t_{\star}}dt\ \tilde{\mathcal{L}}(w,w',t),
\end{equation}
where
\begin{equation}
\label{eq:regLag}
\tilde{\mathcal{L}}(w,w',t)=\mathcal{L}(w,w',t)-\frac{2t_{\star}}{(t-t_{\star})^2}-\frac{\sigma^2}{t}
\end{equation}
is the regularized standard Lagrangian
\begin{equation}
\label{eq:Lagrangian}
\mathcal{L}(w,w',t)
=
\frac{t}{4}\left( \frac{w'}{w} \right)^2-\frac{w}{t}-\frac1{w},
\end{equation}
obtained by Legendre transformation of the Hamiltonian \eqref{eq:P3Hamiltonian1}.

The regularized action \eqref{eq:action} is well-defined on the solution
 \(w(t)=w(t;\sigma,t_{\star})\), with the integration constant \(\sigma\) fixed by the asymptotics \eqref{eq:fastTimeAsymptotics}, while the second one, \(\eta=\eta(\sigma,t_{\star})\), is fixed by \eqref{eq:poleExpansion} so that pole of \(w(t)\) is located at the point \(t=t_{\star}\).

Let us now compute the derivatives of the action \eqref{eq:action} w.r.t. \(t_{\star}\) and  \(\sigma\).
To do this on the solution to equations of motion one takes into account only the contributions of the boundary terms, therefore
\begin{equation}
\begin{gathered}
\label{eq:varprinc}
\frac{\partial\tilde{f}(\sigma,t_{\star})}{\partial t_{\star}}=\tilde{\mathcal{L}}(t_{\star})+\left.\left(\frac{t}{2}\frac{\partial \log w(t,\sigma,t_{\star})}{\partial t_{\star}}\frac{\partial\log w(t,\sigma,t_{\star})}{\partial t}+ \frac{2 t}{(t-t_{\star})^2} \right)\right|_0^{t_{\star}},
\\
\frac{\partial \tilde{f}(\sigma,t_{\star})}{\partial\sigma}=\left. \left(\frac{t}{2} \frac{\partial\log w(t,\sigma,t_{\star})}{\partial \sigma}\frac{\partial\log w(t,\sigma,t_{\star})}{\partial t}-2\sigma\log t \right) \right|_0^{t_{\star}}.
\end{gathered}
\end{equation}
Substituting explicit expansions \eqref{eq:fastTimeAsymptotics} and \eqref{eq:poleExpansion} of the solution around \(t=0\) and around \(t=t_{\star}\) we get
\begin{equation}
\begin{gathered}
\frac{\partial\tilde{f}(\sigma,t_{\star})}{\partial t_{\star}}=-H_{\star}-\frac{\sigma^2}{t_{\star}}-\sigma \frac{\partial }{\partial t_{\star}}\log \left( e^{4\pi i\eta_{\star}(\sigma,t_{\star})}\frac{\Gamma(1-2\sigma)^2}{\Gamma(2\sigma)^2} \right),\\
\frac{\partial \tilde{f}(\sigma,t_{\star})}{\partial \sigma}=-2\sigma\log t_{\star}-\sigma \frac{\partial }{\partial \sigma}\log \left( e^{4\pi i\eta_{\star}(\sigma,t_{\star})}\frac{\Gamma(1-2\sigma)^2}{\Gamma(2\sigma)^2}\right).
\end{gathered}
\end{equation}
Using expression (\ref{eq:etaStar}) for \(\eta_{\star}\) this can be rewritten as
\begin{equation}
\begin{gathered}
\label{eq:2}
\frac{\partial(f_{cl}+f)}{\partial\sigma}=\frac{\partial}{\partial\sigma}\left( \tilde{f}+\sigma\log \left( e^{4\pi i\eta_{\star}}\frac{\Gamma(1-2\sigma)^2}{\Gamma(1+2\sigma)^2} \right)+(1/4+\sigma^2)\log t_{\star}+2\sigma-1 \right),\\
-H_{\star}+\frac{1}{4t_{\star}}=\frac{\partial}{\partial t_{\star}}\left( \tilde{f}+\sigma\log \left( e^{4\pi i\eta_{\star}}\frac{\Gamma(1-2\sigma)^2}{\Gamma(1+2\sigma)^2} \right)+(1/4+\sigma^2)\log t_{\star}+2\sigma-1 \right).
\end{gathered}
\end{equation}
where the r.h. sides actually define the 
quasiclassical conformal block, if we know asymptotics of \(\tilde{f}\) when \(t_{\star}\to 0\).

To compute the integral (\ref{eq:action})
\begin{equation}\begin{gathered}
\tilde{f}(\sigma,t_{\star})\mathop{\sim}_{t_{\star}\to 0}\int_0^{t_{\star}}dt\left( -\frac{2 t_{\star}}{(t-t_{\star})^2} +t^{-1}\frac{8\sigma^2(t/t_{\star})^{2\sigma}}{(1-(t/t_{\star})^{2\sigma})^2}\right)=
\\
=\left.\frac{2t_{\star}}{t-t_{\star}}-\frac{4\sigma}{(t/t_{\star})^{2\sigma}-1}\right|_0^{t_{\star}}=1-2\sigma
\end{gathered}\end{equation}
in the limit \(t_{\star}\to 0\) we just use (\ref{eq:fastTimeAsymptotics}), when expressed in terms of \(t_{\star}\) and \(\sigma\):
\begin{equation}
w(t,\sigma,t_{\star})\sim \frac{-\sigma^2(t/t_{\star})^{2\sigma}}{(1-(t/t_{\star})^{2\sigma})^2}.
\end{equation}
This finally allows to define the quasiclassical conformal block as
\begin{equation}
\label{eq:confBlockViaIntegral}
f(\sigma,t)=\tilde{f}(\sigma,t)+\sigma\log \left( t^{2\sigma}e^{4\pi i\eta_{\star}}\frac{\Gamma(1-2\sigma)^2}{\Gamma(1+2\sigma)^2} \right) +2\sigma-1,
\end{equation}
with normalization condition \(\left.f(\sigma,t)\right|_{t=0}=0\).
There are also the following formulas for the first
derivatives:
\begin{equation}\begin{gathered}
\label{eq:confBlockDerivatives}
\frac{\partial f(\sigma,t)}{\partial t}=\frac{\sigma^2}{t}-H_{\star}(\sigma,t),\\
\frac{\partial f(\sigma,t)}{\partial \sigma}=4\pi i\eta_{\star}(\sigma,t)+2\sigma\log t+2\log \frac{\Gamma(1-2\sigma)}{\Gamma(1+2\sigma)},
\end{gathered}\end{equation}
which actually mean that the function
\begin{equation}
\label{F0}
\mathcal{F} = f(\sigma,t)-\sigma^2\log t + 2\int_0^\sigma
d\sigma'\log \frac{\Gamma(1+2\sigma')}{\Gamma(1-2\sigma')}
= f(\sigma,t) + f_{\rm cl}(\sigma,t) +f_{\rm pert}(\sigma,t)
\end{equation}
defines a Lagrangian submanifold (Malgrange divisor)
\begin{equation}
\label{eq:LagrangianSubmanifold}
4\pi i\eta_\star = \frac{\partial\mathcal{F}}{\partial\sigma},
\quad
- H_\star =  \frac{\partial\mathcal{F}}{\partial t}
\end{equation}
of the 2-form $4\pi id\eta\wedge d\sigma - dH\wedge dt$ on the extended 4-dimensional space \(\mathcal{M}\times \mathbb{C}_t^{*}\times \mathbb{C}_H\).
One can also compute the integral in \eqref{eq:confBlockDerivatives}
explicitly:
\begin{equation}
\label{eq:26}
f_{\rm pert}(\sigma,t)=-\log (G(1+2\sigma)G(1-2\sigma))+2\sigma\log \frac{\Gamma(1+2\sigma)}{\Gamma(1-2\sigma)}-4\sigma^2
\end{equation}
using formula \eqref{intlogamma} from Appendix~\ref{ss:cft}.

\section{Blow-up equations
\label{ss:blowup}}

Let us first recall the relations \eqref{Malgrange}, \eqref{eq:etaStar} we have already exploited above.
They follow just from the fact that solution $w(t)=w(t;\sigma,t_{\star})$ has a pole \eqref{eq:poleExpansion} at $t=t_{\star}$, or
the B\"acklund--transformed tau function $\tau_1(t)$ vanishes at $t=t_{\star}$, $\eta=\eta_{\star}$ or, more generally, on the Malgrange divisor. One can summarize this as
\begin{equation}\begin{gathered}
\label{eq:bu1}
\tau_1(t;\sigma,\eta_{\star})=0,
\\
e^{4\pi i\eta_{\star}}=\frac{\Gamma\left(1+2\sigma\right)^2}{\Gamma\left(1-2\sigma\right)^2}t^{-2\sigma}\exp\left({\frac{\partial f(\sigma,t)}{\partial \sigma}}\right),
\end{gathered}\end{equation}
or, explicitly
\begin{equation}\begin{gathered}
\label{eq:blowup1}
\sum\limits_{n\in \frac12+\mathbb{Z}}t^{n^2}
\frac{\Gamma\left(1+2\sigma\right)^{2n}}{\Gamma\left(1-2\sigma\right)^{2n}}
\prod_{\epsilon=\pm}\frac{1}{G(1+2\epsilon(\sigma+n))}
\exp\left(n{\frac{\partial f(\sigma,t)}{\partial \sigma}}\right)\mathcal{B}(\sigma+n,t) = 0.
\end{gathered}\end{equation}
This equation relates the $c=1$ conformal blocks, or non-refined (with opposite $\epsilon$-parameters) Nekrasov instanton partition functions $\mathcal{B}(\sigma,t)$ with the quasiclassical $c\to\infty$ conformal blocks $f(\sigma,t)$, or the same Nekrasov functions, but in asymmetric limit, when one of the $\epsilon$-parameters vanishes.
Such formulas are known as the blow-up relations \cite{NY1}, and what we found in \eqref{eq:blowup1} is just their very particular limiting case, which however has been derived without any effort --- almost only repeating the classical definitions. Below we are going to exploit the analogs of these blow-up equations at strong coupling domain, which can be used as definition of quasiclassical conformal block at \(t\to\infty\) in sect.~\ref{sec:infinity}, and serve as useful tool for testing formulas for generic irregular blocks at arbitrary values of central charge, see sect.~\ref{sec:qpPainleve}.

Let us now compute the integral (\ref{eq:action}) in terms of the tau function. Expressing the Lagrangian (see (\ref{eq:Lagrangian}), (\ref{eq:HamiltonianGL}), (\ref{eq:anotherW}) and (\ref{eq:PIII})) as
\begin{equation}
\label{eq:LagrangianTau}
\mathcal{L}=
H+\partial_{t}\left(\frac{tw'}{w}\right)-\frac{4}{w}=
\partial_t\left(\log\tau+\frac{tw'}{w}-4t \partial_t\log\tau\right),
\end{equation}
and substituting this into (\ref{eq:action}), one gets
\begin{equation}\begin{gathered}
\label{eq:confBlockViaTauFunction}
\tilde{f}(\sigma,t_{\star})=\left.\left(\log\tau(t)+\frac{tw'(t)}{w(t)}-4t\partial_t\log\tau(t)+\frac{2t_{\star}}{t-t_{\star}}-\sigma^2\log t\right)\right|_0^{t_{\star}}=
\\
=\log(t_{\star}^{-\sigma^2}\tau(t_{\star}))-
\lim_{t\to 0}\log(t^{-\sigma^2}\tau(t))-2\sigma+1-\left.\left(4t\partial_{t}\log\tau(t)\right)\right|_{t=t_{\star}}+4\sigma^2.
\end{gathered}\end{equation}
Now substituting here formulas (\ref{eq:HamiltonianGL}), (\ref{eq:confBlockDerivatives}), (\ref{eq:etaStar}), and (\ref{eq:KyivFormula}) we get
\begin{equation}\begin{gathered}
\label{eq:blowupTau0}
G(1+2\sigma)G(1-2\sigma) t_{\star}^{-\sigma^2}\tau(\eta_{\star}(\sigma,t_{\star}),\sigma,t_{\star})=
\\
=\exp \left( f(\sigma,t_{\star})-\sigma \frac{\partial f(\sigma,t_{\star})}{\partial\sigma}-4 t_{\star}\frac{\partial f(\sigma,t_{\star})}{\partial t_{\star}} \right),
\end{gathered}\end{equation}
or just
\begin{equation}
\label{eq:taustarF}
\log\tau(\eta_{\star}(\sigma,t),\sigma,t)= \mathcal{F} - \sigma\frac{\partial\mathcal{F}}{\partial\sigma} -4t\frac{\partial\mathcal{F}}{\partial t} = \mathcal{F} - \sigma\frac{\partial\mathcal{F}}{\partial\sigma} +4tH_\star.
\end{equation}
Notice, that this equation again relates the \(c=1\) and \(c\to\infty\) conformal blocks,
it can be rewritten more explicitly as
\begin{equation}\begin{gathered}
\label{eq:blowupComplicated}
\sum\limits_{n\in \mathbb{Z}}A_n(\sigma)t^{n^2}
\exp \left(n\frac{\partial f(\sigma,t)}{\partial \sigma} \right)\mathcal{B}(\sigma+n,t)=
\\
=\exp \left( f(\sigma,t)-\sigma \frac{\partial f(\sigma,t)}{\partial\sigma}-4 t\frac{\partial f(\sigma,t)}{\partial t} \right),
\end{gathered}\end{equation}
with
\begin{equation}
	A_n(\sigma) = \frac{\Gamma\left(1+2\sigma\right)^{2n}}{\Gamma\left(1-2\sigma\right)^{2n}}
\prod_{\epsilon=\pm}\frac{ G(1+2\epsilon\sigma)}{G(1+2\epsilon(\sigma+n))},
\end{equation}
and this is nothing but another particular case of the blow-up relations, derived here using almost only the methods of classical analysis.

\subsection*{Remark: conformal blocks from blow-up relations}

When the Hamiltonian (\ref{eq:HamiltonianGL}) is explicitly written as  logarithmic derivative (\ref{eq:KyivFormula}), the first equation in (\ref{eq:confBlockDerivatives}) takes the form
\begin{equation}
	\begin{gathered} \label{d_t_logtau}
	\sum_{n \in \mathbb{Z}} A_n(\sigma)\exp \Big(n\frac{\partial f(\sigma,t)}{\partial \sigma} \Big) t^{n^2} \Big(\frac{\partial}{\partial t} + \frac{n^2+2\sigma n}{t} \Big) \mathcal{B}(\sigma+n,t) = \\
	= -\frac{\partial f(\sigma,t)}{\partial t}\sum_{n \in \mathbb{Z}} A_n(\sigma) \exp \Big(n\frac{\partial f(\sigma,t)}{\partial \sigma} \Big) t^{n^2} \mathcal{B}(\sigma+n,t).
	\end{gathered}
\end{equation}
The $\tau_1$-vanishing condition (\ref{eq:BacklundKyivFormula}), (\ref{Malgrange}), (\ref{eq:etaStar}) in different normalization is written as
\begin{equation} \label{tau1_zero0}
	\sum_{n \in \mathbb{Z}+\frac{1}{2}} A_n(\sigma) \exp \Big(n\frac{\partial f(\sigma,t)}{\partial \sigma} \Big) t^{n^2} \mathcal{B}(\sigma+n,t) = 0.
\end{equation}
Equalities (\ref{eq:blowupComplicated}), (\ref{d_t_logtau}), (\ref{tau1_zero0}) constitute the system of equations on functions  $\mathcal{B}(\sigma,t)$ and $f(\sigma,t)$. When supplemented with normalization $f(\sigma,0) = 0$, this system has unique solution. Hence, one can consider this system as an alternative definition of both conformal blocks, and this will be important in the next section.

\section{Conformal blocks at infinity}

\label{sec:infinity}

\subsection{Solution and tau functions}

In \cite{ILT} an expansion of the Painlev\'e III\(_3\) tau function at \(t\to\infty\) has been proposed in the form
\begin{equation}
\begin{gathered}
\label{eq:tauILT}
\tau^{\infty}(\rho,\nu,r)=e^{\frac{r^2}{16}}r^{\frac1{4}}\sum\limits_{n\in \mathbb{Z}}
C(\nu+in) e^{4\pi i n\rho}e^{(\nu+in)r}r^{\frac12(\nu+i n)^2}\mathcal{B}^{\infty}(\nu+in,r),
\end{gathered}\end{equation}
where~\footnote{Below we hope to avoid confusion with using both variables $r$ and $t\sim r^4$ (up to numeric constant, imported for convenience from \cite{ILT}) at the strong-coupling domain  \(t\to\infty\) or  \(r\to\infty\). The terminology ``strong-coupling'' is taken from supersymmetric gauge theory, where power $4$ (for $SU(2)$ gauge group) distinguishes the expansion in non-Abelian theory at weak coupling, compare to expansion in the effective dual magnetic Abelian theory.}
\begin{equation}
\label{eq:Cdef}
C(\nu)=G(1+i\nu)2^{\nu^2}e^{\frac{i\pi\nu^2}4}(2\pi)^{-\frac{i\nu}{2}},\quad t=2^{-12}r^4,
\end{equation}
and
\begin{equation}\small
\begin{gathered}
\label{eq:infinityBlock}
\mathcal{B}^{\infty}(\nu,r)=
1+\frac{\nu(2\nu^2+1)}{8r}+\frac{\nu^2(4\nu^4-16\nu^2-11)}{128 r^2}+\frac{\nu(8\nu^8-108\nu^6+402\nu^4+269\nu^2-24)}{3\cdot 2^{10}\cdot r^3}+
\\
+
\frac{\nu^2(2\nu^{10}-56\nu^8+585\nu^6-2326\nu^4-\frac{7831}{8}\nu^2+612)}{3\cdot2^{12}\cdot r^4}
+
\\
+\frac{\nu(16\nu^{14}-760\nu^{12}+14920\nu^{10}-148220\nu^8+654377\nu^6-\frac{55975}{2}\nu^4-382488\nu^2+17280)}{15\cdot 2^{17}\cdot r^{5}}+O(r^{-6})
\end{gathered}\end{equation}
are the $c=1$ irregular ``blocks at infinity'' (here we presented one extra term of their expansion, see also Appendix~\ref{app:confBlock} for the general expression up to 7-th order).
Unlike \(t\to 0\) region, these ``conformal blocks'' \eqref{eq:infinityBlock} have no CFT definition yet, and this is not surprising for the such singular regions of the Painlev\'e solutions~\footnote{
For some less degenerate cases the CFT counterparts of tau functions were already defined in \cite{Nagoya:2018pgp,Nagoya:2016mlj,Nagoya:2015cja},
expressions, similar to \eqref{eq:infinityBlock} for different Painlev\'e equations, can be also found in \cite{Bonelli:2016qwg}.}.

The Poisson map from initial data \cite{ILT} is given by
\begin{equation}
\label{eq:monodromyMapping}
e^{4\pi i\rho}=\frac{\sin 2\pi\eta}{\sin 2\pi(\sigma+\eta)},\qquad e^{\pi\nu}=\frac{\sin 2\pi\eta}{\sin 2\pi\sigma},
\end{equation}
so we see that the B\"acklund transformation \eqref{Bset} maps \(\rho\mapsto\rho+\frac1{4}\), \(\nu\mapsto\nu+i\).
Actually, one can define two different tau functions ``at infinity'':
\begin{equation}\begin{gathered}
\label{eq:55}
\tau_+^{\infty}(\rho,\nu,r)=\tau^{\infty}(\rho,\nu,r),\\
\tau_-^{\infty}(\check{\rho},\nu,r)=e^{\frac{r^2}{16}}r^{\frac1{4}}\sum\limits_{n\in \mathbb{Z}}C_-(\nu+in)e^{4\pi in\check{\rho}}e^{(\nu+in)r}(-1)^{\frac12n(n-1)}r^{\frac12(\nu+in)^2}\mathcal{B}^{\infty}(\nu+in,r),
\end{gathered}\end{equation}
where
\begin{equation}
\label{eq:Cpm}
C_\pm(\nu)=G(1\pm i\nu)2^{\nu^2}e^{\frac{i\pi\nu^2}{4}}(2\pi)^{-\frac{i\nu}{2}}.
\end{equation}
Due to the identity \eqref{eq:BarnesDuality}
for the Barnes functions, there is a relation
\begin{equation}
\label{eq:53}
\tau_+^{\infty}(\rho,\nu,r)=\frac{G(1+i\nu)}{G(1-i\nu)}\tau_-^{\infty}(\check{\rho},\nu,r)
\end{equation}
after one substitutes~\footnote{
Such transformations of the tau functions were also considered in \cite{Coman:2020qgf} in the context of relation to the topological strings.}
\begin{equation}
\label{eq:rhoCheckRho}
e^{4\pi i\check{\rho}}=e^{4\pi i\rho}\frac{\sin i\pi\nu}{\pi} = \frac{1}{2\pi i}\frac{\sin 2\pi(\sigma-\eta)}{\sin 2\pi\sigma}
\end{equation}
for $\nu \notin i\mathbb{Z}$, see below.
It is useful to rewrite expansion \eqref{eq:tauILT} as
\begin{equation}\begin{gathered}
\label{eq:tauILT1}
\tau^{\infty}(\rho,\nu,r)=r^{\frac{\nu^2}{2}+\frac1{4}}e^{\frac{r^2}{16}+\nu r}\sum_{n\in \mathbb{Z}}C_n(\nu)e^{4\pi in(\rho+\rho_0)}e^{inr}r^{i\nu n-\frac{n^2}{2}}\mathcal{B}^{\infty}(\nu+in,r) =
\\
=r^{\frac{\nu^2}{2}+\frac1{4}}e^{\frac{r^2}{16}+\nu r} \sum_{n\in \mathbb{Z}}\left(e^{4\pi i(\rho+\rho_0)}e^{ir}r^{i\nu-\frac12}\right)^n r^{-n(n-1)/2}C_n(\nu)\mathcal{B}^{\infty}(\nu+in,r) =
\\
=r^{\frac{\nu^2}{2}+\frac1{4}}e^{\frac{r^2}{16}+\nu r} \sum_{n\in \mathbb{Z}}\mathcal{X}^n r^{-n(n-1)/2}C_n(\nu)\mathcal{B}^{\infty}(\nu+in,r)
\end{gathered}\end{equation}
with
\begin{equation}
\label{rho0tnu}
\mathcal{X}=e^{4\pi i(\rho+\rho_0)}e^{ir}r^{i\tilde{\nu}},\quad
 e^{4\pi i\rho_0}=\frac {\sqrt{2\pi}\,2^{2i\tilde{\nu}}}
 {\Gamma\left(\frac12+i\tilde{\nu}\right)e^{\frac{\pi\tilde{\nu}}{2}}},\quad
  \tilde{\nu}=\nu+i/2.
\end{equation}
The re-scaled structure constants are
\begin{equation}
\label{eq:32}
C_n(\nu)=\frac{C(\nu+in)}{C(\nu)}e^{-4\pi i\rho_0 n},
\end{equation}
e.g.
\begin{equation}\begin{gathered}
\label{eq:Crescaled}
\ldots, \quad C_{-2}(\nu)=\frac{\nu^2(\nu-i)}{64},\quad C_{-1}(\nu)=\frac{\nu}{4},\quad C_0(\nu)=C_1(\nu)=1,
\\
C_2(\nu)=\frac{\nu+i}{4}, \quad C_{3}(\nu)=\frac{(\nu+i)^2(\nu+2i)}{64}, \quad \ldots
\end{gathered}\end{equation}
In these terms solution for PIII\(_3\) has the form~\footnote{
Notice that overall sign here is opposite to \eqref{eq:tauRatio}.
This should follow from the connection constant for the tau functions computed in \cite{ILT}, but one can just check, that this expression satisfies the PIII\(_3\) equation.
}
\begin{equation}\begin{gathered}
\label{eq:infinitySolution}
w(r)=2^{-6}r^2\left(\frac{\tau^{\infty}(\rho,\nu,r)}
{\tau^{\infty}(\rho+\frac1{4},\nu,r)}\right)^2=
\\
=\frac{r^2}{64}
\left(\frac{\sum_{n\in \mathbb{Z}}\mathcal{X}^n r^{-n(n-1)/2}C_n(\tilde{\nu}-i/2)\mathcal{B}^{\infty}(\tilde{\nu}-i/2+in,r)}{\sum_{n\in \mathbb{Z}}(-1)^n\mathcal{X}^n r^{-n(n-1)/2}C_n(\tilde{\nu}-i/2)\mathcal{B}^{\infty}(\tilde{\nu}-i/2+in,r)}\right)^2.
\end{gathered}\end{equation}
We see that the denominator vanishes in the leading order in \(r\) if \(\mathcal{X}=e^{4\pi i(\rho+\rho_0)}e^{ir}r^{i\tilde{\nu}}=1\).
Expansion in \eqref{eq:tauILT1} and \eqref{eq:infinitySolution} effectively goes over the powers of \(\mathcal{X}=e^{4\pi i(\rho+\rho_0)}e^{ir}r^{i\tilde{\nu}}\) and \(r^{-1}\), and we would like, as in \eqref{tauass} and \eqref{eq:fastTimeAsymptotics} at $t\to 0$, to consider series in \(r^{-1}\), keeping exact dependence on \(\mathcal{X}\), e.g.
\begin{equation}\begin{gathered}
\label{eq:winfinityexpan}
w(r)=\frac{r^2}{64}\left(\frac{\tau^{\infty}(\rho,\nu,r)}
{\tau^{\infty}(\rho+\frac1{4},\nu,r)}\right)^2=
\\
=\frac{r^2}{64} \Big(\frac{1+\mathcal{X}}{1-\mathcal{X}} \Big)^2 - \frac{r}{128} \frac{(1+\mathcal{X})(2(\nu+i)\mathcal{X}^4 - (6i\nu^2 - 6\nu - i)\mathcal{X}^2 - 2\nu)}{\mathcal{X}(1-\mathcal{X})^3} + O(1).
\end{gathered}\end{equation}
Computing the Hamiltonian from \eqref{eq:tauILT1}, one gets
\begin{equation}
\label{eq:HAsymptotics}
tH= \frac{r}{4}\frac{\partial}{\partial r}\log\tau^{\infty}\ \stackreb{r\to\infty}{=}\
\frac{r^2}{32}+\frac{r\tilde{\nu}}{4}+\frac{ir(\mathcal{X}-1)}{8(\mathcal{X}+1)}+O(1).
\end{equation}
The leading term in \eqref{eq:winfinityexpan} corresponds to a solution of ``strong coupling'' autonomous
Toda equation (see e.g. \cite{BraMa}) with the critical $\nu$-independent Hamiltonian.

\subsection{Quasiclassical conformal blocks at infinity}

As in sect.~\ref{sec:tauzero}, let us now find some \(\rho=\rho_\star(\nu,r)\), so that solution \eqref{eq:infinitySolution} acquires pole at $r=r_\star$.
To do this we substitute into \(\tau^{\infty}\left(\rho_\star+\frac1{4},\nu,r\right)=0\) the following ansatz:
\begin{equation}
\label{eq:infinityRho}
e^{4i\pi\rho_{\star}}=
\exp\left(-i \frac{\partial \mathcal{F}^{\infty}\left(\tilde{\nu},r\right)}{\partial\tilde{\nu}}\right)
=
\frac{\Gamma\left(i\nu \right)e^{-ir}r^{-i\tilde{\nu}}}{\sqrt{2\pi}\,2^{2i\tilde{\nu}}
e^{-\frac{\pi\tilde{\nu}}{2}}}
\exp\left(-i \frac{\partial f^{\infty}\left(\tilde{\nu},r\right)}{\partial\tilde{\nu}}\right),
\end{equation}
where \(\tilde{\nu}=\nu+i/2\),  and
\begin{equation}
\begin{gathered}
\label{eq:infinityClassicalBock}
f^{\infty}(\tilde{\nu},r)=\frac{\tilde{\nu}(4\tilde{\nu}^2-3)}{16r}-\frac{10\tilde{\nu}^4-17\tilde{\nu}^2+\frac98}{64r^2}+
\frac{\tilde{\nu}(528\tilde{\nu}^4-1640\tilde{\nu}^2+405)}{3\cdot 2^{10}\cdot r^3}-
\\
-\frac{9\left(112\tilde{\nu}^6-560\tilde{\nu}^4+327\tilde{\nu}^2-\frac{27}{2}\right)}{2^{12}\cdot r^4}+
\frac{\tilde{\nu}\left(8432 \tilde{\nu}^6-62468\tilde{\nu}^4+69001\tilde{\nu}^2-\frac{41607}{4}\right)}{5\cdot 2^{12}\cdot r^5}+O(r^{-6}),
\end{gathered}
\end{equation}
where the $\nu$-independent part $f^{\infty}(0,r)=-\frac9{512\cdot r^2}+\ldots$ can be restored from the expansion
of \eqref{eq:HAsymptotics} under \eqref{eq:infinityRho}, see also (\ref{eq:Hpole}) and \eqref{fracfunc} below.

Following the logic as in \eqref{eq:etaStar},
we are going to call \eqref{eq:infinityClassicalBock} as ``quasiclassical conformal block at infinity'', though its CFT definition, as well as for \(\mathcal{B}^{\infty}(\nu,r)\), is yet unclear. Now the only thing to be checked immediately is that in the Seiberg--Witten limit \(\nu\mapsto\epsilon^{-1}\nu, r\mapsto\epsilon^{-1}r, \epsilon\to 0\):
\begin{equation}\begin{gathered}
\lim_{\epsilon\to 0}\epsilon^2\log \mathcal{B}\left(\frac{\nu}{\epsilon},\frac{r}{\epsilon}\right)
=
\lim_{\epsilon\to 0}\epsilon^2f^{\infty}\left(\frac{\nu}{\epsilon},\frac{r}{\epsilon}\right)=
\frac{\nu^3}{4r}-\frac{5\nu^4}{32r^2}+\frac{11\nu^5}{64r^3}-\frac{63\nu^6}{256r^4}+\frac{527 \nu^7}{1280r^5}+\ldots
\end{gathered}\end{equation}
expressions \eqref{eq:infinityClassicalBock} and \eqref{eq:infinityBlock} indeed coincide, and that
equation
\begin{equation}
\begin{gathered}
\label{blupinfty}
\tau^{\infty}\left(\rho_\star+\frac1{4},\nu,r\right)=
\\= r^{\frac{\nu^2}{2}+\frac1{4}}e^{\frac{r^2}{16}+\nu r} \sum_{n\in \mathbb{Z}}(-1)^ne^{-in \frac{\partial f^{\infty}\left(\nu+i/2,r\right)}{\partial\nu}} r^{-\frac12n(n-1)} e^{inr}C_n(\nu)\mathcal{B}^{\infty}(\nu+in,r) = 0
\end{gathered}
\end{equation}
relates them to each other exactly as an analog of the homogeneous blow-up equation \eqref{eq:blowup1} at infinity.
Relation (\ref{eq:infinityRho}) when written in terms of \(\check{\rho}\) 
\begin{equation}
\label{eq:56}
e^{4i\pi\check{\rho}_{\star}}=\frac{\sqrt{\pi/2}\, e^{-ir}r^{-i\tilde{\nu}}}{2^{2i\tilde{\nu}}e^{-\frac{\pi\tilde{\nu}}{2}}\Gamma(1-i\nu)}\exp \left( -i \frac{\partial f^{\infty}(\tilde{\nu},r)}{\partial \tilde{\nu}} \right)
\end{equation}
similarly leads to
\begin{equation}\begin{gathered}
\tau_-^{\infty}\left(\check{\rho}_{\star}+\frac1{4},\nu,r\right)=
\\
= e^{\frac{r^2}{16}}r^{\frac1{4}}\sum\limits_{n\in \mathbb{Z}}C_-(\nu+in)e^{4\pi in\check{\rho}_{\star}}e^{(\nu+in)r}(-1)^{\frac12n(n+1)}r^{\frac12(\nu+in)^2}\mathcal{B}^{\infty}(\nu+in,r) = 0.
\end{gathered}\end{equation}
To define the quasiclassical block at infinity completely, one has to compute the value of the Hamiltonian at the pole.
Using (\ref{eq:HAsymptotics}) and (\ref{eq:infinityRho}) we get
\begin{equation}
\begin{gathered}
\label{eq:Hpole}
t_{\star}H(t_{\star})=\frac{r^2}{32}+\frac{r\tilde{\nu}}{4}+
\frac{4\tilde{\nu}^2-1}{32}-\frac{4\tilde{\nu}^3-3\tilde{\nu}}{64 r}+
\frac{80\tilde{\nu}^4-136\tilde{\nu}^2+9}{1024r^2}-
\frac{526\tilde{\nu}^5-1640\tilde{\nu}^3+405\nu}{4096r^3}+\\+
\frac{9(224\tilde{\nu}^6-1120\tilde{\nu}^4+654\tilde{\nu}^2-27)}{r^4}+
\frac{\tilde{\nu}(33728\tilde{\nu}^6-249872\tilde{\nu}^4+276004\tilde{\nu}^2-41607)}{81920r^5}+\ldots,
\end{gathered}
\end{equation}
which fixes the \(\nu\)-independent part of the conformal block.
\subsection{Quasiclassical block at infinity as action functional}

Similarly to sect.~\ref{ss:action} we prove here that quasiclassical conformal block at infinity is given by
\begin{equation}
\begin{gathered}
\label{intinf}
f^\infty(\nu + i/2,r_{\star}) = -4\pi \nu \rho_\star - \frac{\pi i \nu}{2} \Big(\nu + \frac{i}{2} \Big) - i \nu \log \Big(2^{-5 i(\nu+i/2)} \frac{\Gamma(i \nu)}{\sqrt{2\pi}} \Big) - \\
		- 8 \Big(\nu + \frac{i}{2} \Big) t_{\star}^{1/4} - \frac{\nu}{4} \Big(\nu + \frac{i}{2} \Big)\log t_{\star} - \frac{i \nu}{2} - \frac{9}{8} + \log 2 
			+\int_{t_{\star}}^\infty dt\tilde{\mathcal{L}}^\infty\,,
\end{gathered}
\end{equation}
where the integral converges for $\Im \nu \in (-1/4, 1/4) $ after regularization of the Lagrangian
\eqref{eq:Lagrangian}:
\begin{equation}
	\begin{gathered}
		\tilde{\mathcal{L}}^\infty = \mathcal{L} - \frac{1}{8} \frac{d}{dt}\Big( \sqrt{t} \Big(1-\frac{t}{w^2} \Big) \frac{dw}{dt} - \frac{\sqrt{t}}{w} \Big) -
		\\
		- \frac{2 \nu^2+1}{16t}+\frac{2}{\sqrt{t}} - \frac{64 t_{\star} + \sqrt{t_{\star}}}{32(t-t_{\star})^2}-\frac{t_{\star}^{3/2}}{2(t-t_{\star})^3}-\frac{3 t_{\star}^{5/2}}{4(t-t_{\star})^4}.
	\end{gathered}
\end{equation}
Thus, computing derivatives of \eqref{intinf}, similarly to \eqref{eq:varprinc}, one gets~\footnote{
As in (\ref{eq:varprinc}), the derivatives of the on-shell action acquire only the boundary contributions. Notice also that derivatives of $\rho_\star$ in (\ref{intinf}) are canceled by derivatives of $\mathcal {X}$-variable, entering the solution, see Appendix~\ref{solution_infty}.
}
\begin{equation}\begin{gathered}
\label{fracfunc}
\frac{\partial f^\infty(\tilde{\nu},r)}{\partial t} =
H_\star - \frac{2}{t^{1/2}}  - \frac{2\tilde{\nu}}{t^{3/4}} - \frac{\tilde{\nu}^2-1/4}{8t},
\\
\frac{\partial f^\infty(\tilde{\nu},r)}{\partial \nu} = -4 \pi \rho_\star - \frac{i\pi \tilde{\nu}}{2} - \frac{\tilde{\nu}}{4} \log t - 8 t^{1/4} - i \log \Big(2^{-5 i\tilde{\nu}} \frac{\Gamma(\frac12 +i \tilde{\nu})}{\sqrt{2\pi}} \Big),
\end{gathered}
\end{equation}
where the last expression just coincides with \eqref{eq:infinityRho}, or
\begin{equation}\begin{gathered}
\label{Finf}
\frac{\partial \mathcal{F}^\infty(\tilde{\nu},r)}{\partial t} = H_\star,\quad
\frac{\partial \mathcal{F}^\infty(\tilde{\nu},r)}{\partial \nu} = -4 \pi \rho_\star,
\\
\mathcal{F}^\infty(\tilde{\nu},r) = f^\infty(\tilde{\nu},r) + 4t^{1/2} + 8\tilde{\nu}t^{1/4}
+ \frac{\tilde{\nu}^2-1/4}{8}\log t + \frac{\tilde{\nu}^2}{2}\left(1+\frac{i\pi}{2}+5\log 2\right)+
\\
+\left(i\tilde{\nu}-\frac12\right)\log\Gamma\left(\frac12+i\tilde{\nu}\right)-\log G\left(\frac12+i\tilde{\nu}\right) = f^\infty(\tilde{\nu},r) + f^\infty_{\rm cl}(\tilde{\nu},r)
+f^\infty_{\rm pert}(\tilde{\nu}),
\end{gathered}\end{equation}
where \(f_{cl}^{\infty}(\tilde{\nu},r)=\frac{r^2}{16}+\tilde{\nu} r +\frac{\tilde{\nu}^2-\frac1{4}}{2}\log r\), and we again used the integral formula \eqref{intlogamma}.

In Appendix~\ref{solution_infty} we explain how the integral in (\ref{intinf}) can be computed up to an arbitrary order in $t_{\star}$ and get explicitly 
in \eqref{eq:mainasympt} its principal asymptotics 
\begin{equation}
\int_{t_\star}^\infty \tilde{\mathcal{L}}^\infty dt = 4 i t_\star^{1/4} + \frac{9}{8} + \frac{i \nu}{2} - \log 2 + O(t_\star^{-1/4}),
\end{equation}
which
can be also extracted from (\ref{eq:tauILT}).
Using (\ref{eq:LagrangianTau}) we obtain
\begin{equation}\begin{gathered} \label{integraltau}
\int_{t_{\star}}^\infty \tilde{\mathcal{L}}^\infty dt = 4 i t_{\star}^{1/4} + \frac{9}{8} + \frac{i \nu}{2} + 4 t_{\star} \frac{\partial f^\infty(\tilde{\nu},t_{\star})}{\partial t_{\star}} - \log \tau(t_\star) +\\+ 4t_\star^{1/2} + 8\nu t_\star^{1/4} + \frac{1+2\nu^2}{16} \log(2^{12} t_\star) + \log C(\nu).
\end{gathered}
\end{equation}

\subsection{Blow-up equations at infinity}

Since the pairs $(\eta, \sigma)$ and $(\rho, \nu)$ (see \eqref{eq:monodromyMapping}) provide canonical coordinates for the same symplectic form on $\mathcal{M}$
\begin{equation}
\label{syformA}
    \varpi = 4 \pi i d\eta \wedge d\sigma = 4 \pi d\rho \wedge d\nu,
\end{equation}
they are related \cite{ILT} by canonical transformation
\begin{equation}
    4 \pi \rho d\nu = 4 \pi i \eta d\sigma  + d \mathcal{S}
\end{equation}
with the generating function
\begin{equation}
	\mathcal{S} = -\frac{i}{2\pi} \Big(\mathrm{Li}_2(-e^{2 \pi i \sigma + 2 \pi i \eta + \pi \nu}) + \mathrm{Li}_2(-e^{-2 \pi i \sigma - 2 \pi i \eta + \pi \nu}) + \pi^2 \nu^2 - 4 \pi^2 \eta^2 \Big).
\end{equation}
The function \eqref{Finf} defines the same Lagrangian submanifold (Malgrange divisor) in $\mathcal{M}\times \mathbb{C}^{*}_t\times \mathbb{C}_H$ \eqref{eq:LagrangianSubmanifold} as \eqref{F0} in coordinates $\{\nu,\rho\}$ ``at infinity'',
and it is related to \eqref{F0} by~\footnote{
Here we have parameterized the Malgrange divisor by \(\sigma\) and \(t\) and indicated all dependencies on these variables explicitly.
Below we always assume, that any two independent variables can be chosen as local coordinates, and all others can be expressed using monodromy map \eqref{eq:monodromyMapping} and tau function vanishing conditions \eqref{eq:etaStar}, \eqref{eq:infinityRho}.
}
\begin{equation}
\label{eq:classicalFusion}
\mathcal{F}(\sigma,t) + \mathcal{F}^\infty\Big(\tilde{\nu}\big(\sigma,\eta_{\star}(\sigma,t)\big),r\Big) + \mathcal{S}\Big(\sigma,\nu\big(\sigma,\eta_{\star}(\sigma,t)\big)\Big) = \mathcal{C},
\end{equation}
where the constant \(\mathcal{C}\) will be determined below in \eqref{eq:constantAnswer}.

Combining (\ref{intinf}), (\ref{integraltau}) and (\ref{fracfunc}) one gets, similarly to (\ref{eq:blowupTau0}), (\ref{eq:taustarF}) 
\begin{align}
\label{eq:blowupTauInf}
	\begin{split}
		\log \tau^{\infty}(\rho_\star(\nu, r_\star), \nu, r_\star) = &4 t_\star^{1/2} + 8 \nu t_\star^{1/4} + \frac{1+2\nu^2}{16} \log(2^{12} t_\star) + \log C(\nu) + \log 2 -\\
		&- f^\infty(\tilde{\nu}, t_\star) + \nu \frac{\partial f^\infty(\tilde{\nu},t_\star)}{\partial \nu} + 4 t_\star \frac{\partial f^\infty(\tilde{\nu},t_\star)}{\partial t_\star},
	\end{split}
\end{align}
or
\begin{equation}\begin{gathered}
\label{eq:tauFinf}
\log \tau^{\infty}(\rho_\star(\nu, r), \nu, r) = - \mathcal{F}^\infty + \nu\frac{\partial\mathcal{F}^\infty }{\partial\nu} + 4t \frac{\partial\mathcal{F}^\infty }{\partial t} + \mathcal{C}_0,
\\
\mathcal{C}_0 = \frac{1}{8} + \frac{9}{8}\log 2 + \frac{i\pi}{16},
\end{gathered}\end{equation}
which is actually an analog of the non-homogeneous blow-up equation \eqref{eq:blowupComplicated} 
\begin{align} \label{blowup_infty}
	\begin{split}
		\sum_{n \in \mathbb{Z}} \Big(2 e^{i\pi/4} \Big)^{n-n^2} \Gamma^n(i \nu) &\frac{G(1-n+i\nu)}{G(1+i\nu)} \exp \Big(-i n \frac{\partial f^\infty(\tilde{\nu},r)}{\partial \nu} \Big) r^{-n(n-1)/2} \mathcal{B}^\infty (\nu+in,r) = \\
		&=2 \exp \Big(- f^\infty(\tilde{\nu},r) + \nu \frac{\partial f^\infty(\tilde{\nu},r)}{\partial \nu} + r \frac{\partial f^\infty(\tilde{\nu},r)}{\partial r} \Big)
	\end{split}
\end{align}
at infinity.
The first equation in (\ref{fracfunc}), written in terms of the tau function, takes the form
\begin{align} \label{d_r_logtau}
	\begin{split}
		\sum_{n \in \mathbb{Z}} C_n \exp \Big(-i n \frac{\partial f^\infty(\tilde{\nu},r)}{\partial \nu} \Big) r^{-n(n-1)/2} \Big( \frac{\partial}{\partial r} + i n \Big(1 + \frac{\nu + i n/2}{r} \Big)\Big) \mathcal{B}^\infty (\nu+in,r) = \\=
		\Big(\frac{\partial f^\infty(\tilde{\nu},r)}{\partial r} + \frac{i}{2} \Big(1 + \frac{\nu+i}{r}\Big) \Big) \sum_{n \in \mathbb{Z}} C_n \exp \Big(-i n \frac{\partial f^\infty(\tilde{\nu},r)}{\partial \nu} \Big) r^{-n(n-1)/2} \mathcal{B}^\infty (\nu+in,r).
	\end{split}
\end{align}
Also, remember the $\tau_1$ vanishing condition (see (\ref{eq:infinitySolution})):
\begin{equation} \label{tau1_zero}
	\sum_{n \in \mathbb{Z}} (-1)^n C_n \exp \Big(-i n \frac{\partial f^\infty(\tilde{\nu},r)}{\partial \nu} \Big) r^{-n(n-1)/2} \mathcal{B}^\infty (\nu+in,r) = 0.
\end{equation}
Relations (\ref{blowup_infty}), (\ref{d_r_logtau}), (\ref{tau1_zero}) can be actually considered as system of equations for the functions $B^\infty(\nu,r)$ and $f^\infty(\nu,r)$, so that both ``conformal blocks'' $B^\infty(\nu,r)$ and $f^\infty(\nu,r)$, which do not have yet an algebraic formulation, can be defined as their solutions without any reference to original Painlev\'e equation.

In order to fix the constant in \eqref{eq:classicalFusion} let us subtract two blow-up relations, \eqref{eq:taustarF} and \eqref{eq:tauFinf}:
\begin{equation}\begin{gathered}
\label{eq:connection0}
\log\chi(\sigma,\nu;\eta)=\log \frac{\tau\left(\eta_{\star},\sigma,t\right)}{\tau^{\infty}\left(\rho_{\star},\nu,r\right)}=
\mathcal{F}+\mathcal{F}^{\infty} - \nu \frac{\partial \mathcal{F}^{\infty}}{\partial \nu}-\sigma \frac{\partial \mathcal{F}}{\partial \sigma}-\mathcal{C}_0
=\\=
\mathcal{C}-\mathcal{S}_{\star}+4\pi\nu\rho_\star-4\pi i\sigma\eta_\star-\frac{1}{8}-\frac{9}{8}\log 2-\frac{i\pi}{16}.
\end{gathered}
\end{equation}
It means that logarithm of the connection constant for the \(c=1\) tau functions (the l.h.s. of \eqref{eq:connection0}, see \cite{ILT}), when computed on Malgrange divisor\footnote{
Actually this constraint does not reduce the generality of \(\chi(\sigma,\nu;\eta)\), since it depends only on two variables, and Malgrange divisor itself is two-dimensional (parameterized locally, for example, by \(\sigma\) ant \(t\)).
} coincides, up to a numeric constant and the Legendre transform~\footnote{
In \cite{ILT} the term \(4\pi\nu\rho-4\pi i\sigma\) was crucial to solve the difference equations on \(\chi(\sigma,\nu;\eta)\).
}, with the ``connection constant'' for \(c\to\infty\) conformal blocks, being the generating function of canonical transformation between different variables.
Notice also that the above derivation, based on regularized action functionals, literally differs from the proof of \cite{ILP}, though they are quite similar ideologically.

To complete this computation we use the formula from \cite{ILT} and transform it to more convenient form using \eqref{eq:Li2zZinv} and \eqref{eq:Li2Barnes}:
\begin{equation}
\label{eq:connectionILT}
\log\chi(\sigma,\nu;\eta)=-\mathcal{S}-4\pi i\sigma\eta+4\pi\nu\rho+\frac{5i\pi}{24}-\frac{3}{4}\log 2-\frac12\log\pi-2\log G\left(\frac1{4}\right).
\end{equation}
Comparing \eqref{eq:connection0} and \eqref{eq:connectionILT} one finally concludes that
\begin{equation}
\label{eq:constantAnswer}
\mathcal{C}=\frac{1}{8}+\frac{13i\pi}{48}+\frac{3}{8}\log 2-\frac{1}{2}\log\pi-2\log G \left( \frac1{4} \right).
\end{equation}

\section{Spectral theory meaning of quasiclassical conformal blocks}

\label{sec:spectralTheory}

\subsection{Monodromies from exact WKB}

To understand the spectral theory meaning of quasiclassical conformal blocks 
one first needs to restore the monodromy data.
To do this it is convenient to use the WKB parameterization of monodromies.
All definitions and conventions are collected in the Appendix \ref{sec:WKB}, there is also an elementary overview of the construction~\footnote{
For the rigorous and detailed explanation of exact WKB analysis see \cite{iwaki2014exact1,iwaki2014exact2}.}, and here we proceed to direct computation of the transition matrices.

\subparagraph{The WKB graphs.}

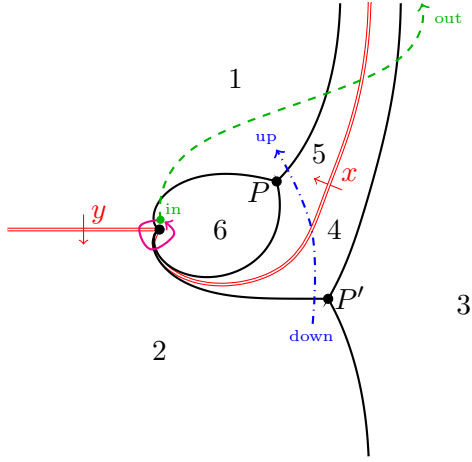
\begin{figure}[h!]
\begin{center}
\begin{tikzpicture}[scale=2]

\draw[double,red](0,0)--(-1,0);
\draw[red,->](-0.5,0.1)--(-0.5,-0.1);
\node at (-0.4,0.1) {\color{red} \(y\)};

\draw[red,->] (1.2,0.263)--+(158:0.2);
\node at (1.25,0.363){\color{red}\(x\)};

\draw[double,red]  plot[smooth]
coordinates {(1.38106, 1.50099) (1.38103, 1.49999) (1.3773, 1.40107) (1.37059, 1.3023) (1.36081, 1.20379) (1.34785, 1.10564) (1.33162, 1.00799) (1.31205, 0.910944) (1.28914, 0.814639) (1.26291, 0.71918) (1.23356, 0.624637) (1.20145, 0.530992) (1.16727, 0.438079) (1.13213, 0.345524) (1.09733, 0.252843) (1.06335, 0.159858) (1.02848, 0.0672064) (0.989348, -0.0226165) (0.940687, -0.108749) (0.879041, -0.186036) (0.803657, -0.249958) (0.717631, -0.298701) (0.624976, -0.333333) (0.52849, -0.355194) (0.430008, -0.364629) (0.331154, -0.360989) (0.23393, -0.342874) (0.141303, -0.308346) (0.0580565, -0.255194) (-0.00758836, -0.181643) (-0.0402029, -0.0891791) (-0.00862477, -0.00311165)};
\draw [thick] plot[smooth]
coordinates {(0.771152, 0.32014) (0.771862, 0.320845) (0.801732, 0.351771) (0.830305, 0.383901) (0.857572, 0.417147) (0.883535, 0.451421) (0.908206, 0.486636) (0.931605, 0.52271) (0.953756, 0.559564) (0.974687, 0.597123) (0.994429, 0.635321) (1.01302, 0.674095) (1.03048, 0.713388) (1.04685, 0.753148) (1.06217, 0.793327) (1.07646, 0.833882) (1.08975, 0.874773) (1.10208, 0.915966) (1.11348, 0.957427) (1.12397, 0.999127) (1.13358, 1.04104) (1.14234, 1.08314) (1.15027, 1.1254) (1.15739, 1.1678) (1.16374, 1.21033) (1.16931, 1.25297) (1.17415, 1.29569) (1.17827, 1.3385) (1.18169, 1.38136) (1.18443, 1.42427) (1.1865, 1.46722) (1.18765, 1.5002)};
\draw[thick]  plot[smooth]
coordinates {(0.771152, 0.32014) (0.770442, 0.319436) (0.780389, 0.272093) (0.78743, 0.223614) (0.790589, 0.17473) (0.789643, 0.125754) (0.784368, 0.0770561) (0.774546, 0.02907) (0.759998, -0.0176981) (0.740615, -0.0626753) (0.716408, -0.105249) (0.687551, -0.144818) (0.65439, -0.180856) (0.617411, -0.212968) (0.577176, -0.240895) (0.534259, -0.264496) (0.489201, -0.283701) (0.442497, -0.298463) (0.394603, -0.308727) (0.345952, -0.314413) (0.296981, -0.315402) (0.248155, -0.311534) (0.199998, -0.302615) (0.153128, -0.288419) (0.108304, -0.268704) (0.06649, -0.243235) (0.0289412, -0.211834) (-0.00265382, -0.174474) (-0.0259465, -0.131483) (-0.0375289, -0.0840511) (-0.0321676, -0.0357108) (-0.00935861, -0.00368052)};
\draw[thick]  plot[smooth]
coordinates {(0.768202, 0.319369) (0.767237, 0.319632) (0.734357, 0.328286) (0.701315, 0.336296) (0.668109, 0.343598) (0.634743, 0.350127) (0.601223, 0.355816) (0.567562, 0.360597) (0.533777, 0.364399) (0.499889, 0.367149) (0.46593, 0.368771) (0.431934, 0.369187) (0.397948, 0.368317) (0.364024, 0.366074) (0.330229, 0.362372) (0.296641, 0.357117) (0.263353, 0.350215) (0.230475, 0.341567) (0.198141, 0.331072) (0.166507, 0.318625) (0.135763, 0.304122) (0.106135, 0.287459) (0.0778961, 0.26854) (0.0513798, 0.247277) (0.0269943, 0.223605) (0.00524587, 0.197493) (-0.0132293, 0.168978) (-0.0276186, 0.138208) (-0.0368625, 0.105534) (-0.0395431, 0.0717085) (-0.0336945, 0.0383378) (-0.0164535, 0.00937995) (-0.00862278, 0.00319633)};
\draw[thick]  plot[smooth]
coordinates {(1.10991, -0.461423) (1.1104, -0.462293) (1.12778, -0.493821) (1.14443, -0.525737) (1.16038, -0.558012) (1.17563, -0.590619) (1.19022, -0.623532) (1.20415, -0.656728) (1.21743, -0.690185) (1.2301, -0.723884) (1.24215, -0.757807) (1.2536, -0.791936) (1.26447, -0.826256) (1.27476, -0.860752) (1.2845, -0.895411) (1.29368, -0.930219) (1.30233, -0.965164) (1.31045, -1.00024) (1.31805, -1.03542) (1.32515, -1.07072) (1.33175, -1.10611) (1.33787, -1.14158) (1.3435, -1.17714) (1.34867, -1.21277) (1.35338, -1.24846) (1.35764, -1.2842) (1.36145, -1.32) (1.36483, -1.35584) (1.36778, -1.39172) (1.37032, -1.42763) (1.37244, -1.46357) (1.37415, -1.49953) (1.37419, -1.50053)};
\draw[thick]  plot[smooth]
coordinates {(1.10991, -0.461423) (1.10942, -0.460553) (1.14093, -0.401532) (1.17183, -0.34096) (1.20053, -0.27932) (1.22722, -0.216779) (1.25211, -0.153498) (1.27542, -0.0896217) (1.29742, -0.0252792) (1.31836, 0.0394175) (1.33846, 0.104378) (1.35792, 0.169534) (1.37687, 0.234841) (1.39536, 0.300276) (1.41341, 0.365837) (1.43096, 0.431533) (1.44794, 0.49738) (1.46424, 0.563397) (1.47977, 0.629599) (1.49444, 0.695998) (1.50816, 0.762597) (1.52088, 0.829397) (1.53254, 0.896389) (1.5431, 0.963564) (1.55253, 1.03091) (1.56081, 1.0984) (1.56792, 1.16602) (1.57388, 1.23376) (1.57867, 1.30159) (1.5823, 1.3695) (1.58478, 1.43745) (1.58607, 1.50044)};
\draw[thick]  plot[smooth]
coordinates {(1.10612, -0.459205) (1.10512, -0.459199) (1.05812, -0.458981) (1.01112, -0.458851) (0.964121, -0.458768) (0.917121, -0.45868) (0.870122, -0.458521) (0.823123, -0.45821) (0.776126, -0.45765) (0.729135, -0.456728) (0.682157, -0.455315) (0.635202, -0.453267) (0.588289, -0.450425) (0.541444, -0.446615) (0.494709, -0.44165) (0.448137, -0.435331) (0.401806, -0.427446) (0.355816, -0.41777) (0.310301, -0.406068) (0.265433, -0.392092) (0.221435, -0.375584) (0.178592, -0.35628) (0.137272, -0.333908) (0.0979456, -0.308198) (0.0612236, -0.278896) (0.0279042, -0.245785) (-0.00095403, -0.208735) (-0.0239198, -0.167789) (-0.0390056, -0.123363) (-0.0433516, -0.076706) (-0.0326322, -0.0312569) (-0.0091632, -0.00331677)};

\draw[radius=0.3mm,fill](0,0)circle;
\draw[radius=0.3mm,fill](-22.5:1.2)circle;
\draw[radius=0.3mm,fill](22.5:1/1.2)circle;

\draw[dashed,thick,green!70!black,->]  plot[smooth, tension=.7] coordinates {(0.006,0.0648) (0.2391,0.5508) (1.5128,1.0868) (1.723,1.4999)};

\draw[thick,magenta,->]  plot[smooth, tension=.7] coordinates {(0.006,0.0648) (-0.1311,0.0176) (-0.0586,-0.1305) (0.094,-0.0355) (0.0323,0.0525)};

\draw[thick,green!70!black,fill,->]  (0.006,0.0648) circle[radius=0.2mm];

\draw[dashdotted,thick,blue,->]  plot[smooth, tension=.7] coordinates {(1.0065,-0.6255) (1,0) (0.7676,0.5303)};

\node at(0.5,1){\(1\)};
\node at(0,-0.8){\(2\)};
\node at(2,-0.5){\(3\)};
\node at(1.15,0){\(4\)};
\node at(1.05,0.5){\(5\)};
\node at(0.4,0){\(6\)};

\node at(0.65,0.25){\(P\)};
\node at(1.25,-0.45){\(P'\)};

\node [thick,green!70!black] at  (0.09,0.14) {\tiny in};
\node [thick,green!70!black] at  (1.9,1.4) {\tiny out};

\node at (0.7,0.6) {\color{blue}\tiny up};
\node at (1,-0.7) {\color{blue}\tiny down};

\end{tikzpicture}
\end{center}
\caption{WKB graph for real \(r\in \mathbb{R}_{>0}\).
The dashed arrow corresponds to the matrix $V$ relating solutions at zero and at infinity.
Solid arrow represents monodromy $M_0$ around $z=0$.
Dashed line goes to \(-\infty\), since all three anti-Stokes lines finally bend in that direction and go parallel to the real axis, see also Fig.~\ref{fig:WKBrealScaled} for the re-scaled picture. \label{fig:WKBreal} }
\end{figure}

\begin{figure}[h!]
\begin{center}
\begin{tabular}{ccc}

\begin{tikzpicture}[scale=1]
\draw[double,red]  plot[smooth]
coordinates {(1.99078, 1.50071) (1.99041, 1.49979) (1.95222, 1.40736) (1.9123, 1.31568) (1.87058, 1.2248) (1.82701, 1.1348) (1.78151, 1.04574) (1.73405, 0.957731) (1.68455, 0.870842) (1.63298, 0.78517) (1.57929, 0.70081) (1.52345, 0.61785) (1.46549, 0.536369) (1.40543, 0.456417) (1.34338, 0.378) (1.27951, 0.301051) (1.21408, 0.225434) (1.14727, 0.151031) (1.07891, 0.0780426) (1.00803, 0.00752013) (0.933567, -0.0576807) (0.852256, -0.115799) (0.764534, -0.163673) (0.671623, -0.200495) (0.57511, -0.226475) (0.476319, -0.241639) (0.376447, -0.245258) (0.276988, -0.235722) (0.180365, -0.210478) (0.0911482, -0.165853) (0.0194222, -0.09702) (-0.0043355, -0.0088384)};
\draw[double,red]  plot[smooth]
coordinates {(-1.50031, 0.756762) (-1.4996, 0.756059) (-1.45754, 0.714689) (-1.41521, 0.673581) (-1.37262, 0.632759) (-1.32972, 0.592251) (-1.2865, 0.55209) (-1.24293, 0.512309) (-1.19897, 0.472951) (-1.15461, 0.434058) (-1.1098, 0.395682) (-1.0645, 0.35788) (-1.01867, 0.320719) (-0.972275, 0.284273) (-0.925262, 0.248627) (-0.87758, 0.21388) (-0.829176, 0.180147) (-0.779994, 0.14756) (-0.729975, 0.116272) (-0.67906, 0.086466) (-0.627192, 0.0583539) (-0.574315, 0.0321887) (-0.520385, 0.00827172) (-0.466326, -0.0127269) (-0.410266, -0.0310904) (-0.353162, -0.045885) (-0.295133, -0.0564594) (-0.236417, -0.062009) (-0.177449, -0.0615309) (-0.11901, -0.0537611) (-0.0625012, -0.0370629) (-0.0107844, -0.00900259) (-0.00682405, -0.00595085)};
\draw [thick] plot[smooth]
coordinates {(0.771429, 0.319774) (0.772298, 0.32027) (0.816146, 0.34631) (0.859092, 0.373812) (0.901083, 0.402753) (0.942077, 0.433088) (0.98205, 0.464757) (1.02099, 0.497689) (1.05889, 0.53181) (1.09576, 0.567041) (1.13162, 0.603304) (1.16648, 0.640526) (1.20037, 0.678637) (1.23332, 0.717568) (1.26534, 0.757261) (1.29647, 0.797658) (1.32673, 0.838707) (1.35615, 0.880362) (1.38476, 0.92258) (1.41259, 0.96532) (1.43965, 1.00855) (1.46597, 1.05223) (1.49158, 1.09633) (1.51649, 1.14083) (1.54073, 1.1857) (1.56432, 1.23092) (1.58726, 1.27647) (1.60959, 1.32232) (1.63132, 1.36846) (1.65246, 1.41487) (1.67304, 1.46154) (1.68956, 1.50015)};
\draw[thick]  plot[smooth]
coordinates {(0.771429, 0.319774) (0.770561, 0.319278) (0.768611, 0.280952) (0.764886, 0.242137) (0.758681, 0.203641) (0.749909, 0.165648) (0.738499, 0.128363) (0.724396, 0.0920117) (0.707573, 0.0568375) (0.688037, 0.0230951) (0.66584, -0.00895858) (0.64108, -0.0390759) (0.613902, -0.0670312) (0.584494, -0.0926301) (0.553074, -0.115714) (0.519876, -0.136161) (0.485142, -0.153874) (0.449113, -0.168778) (0.412024, -0.180804) (0.374105, -0.18988) (0.335587, -0.195925) (0.296707, -0.198839) (0.25772, -0.198496) (0.218914, -0.194743) (0.180629, -0.187392) (0.143282, -0.176219) (0.107416, -0.16096) (0.07376, -0.141314) (0.0433531, -0.116959) (0.0177747, -0.0876045) (-0.00031777, -0.0531912) (-0.00541617, -0.0148784) (-0.00424549, -0.00899784)};
\draw[thick]  plot[smooth]
coordinates {(0.76838, 0.319792) (0.767516, 0.320296) (0.730838, 0.340753) (0.693395, 0.359774) (0.655207, 0.37725) (0.616304, 0.39307) (0.576727, 0.407118) (0.536529, 0.419275) (0.495778, 0.429419) (0.454552, 0.437424) (0.412951, 0.443158) (0.37109, 0.44649) (0.329104, 0.447284) (0.287154, 0.445402) (0.245425, 0.440707) (0.204136, 0.433063) (0.163539, 0.422339) (0.123929, 0.408411) (0.0856464, 0.391168) (0.0490911, 0.370518) (0.0147271, 0.346401) (-0.0169032, 0.318801) (-0.0451653, 0.287767) (-0.0693158, 0.253442) (-0.0884827, 0.216112) (-0.101645, 0.176278) (-0.107614, 0.134767) (-0.105032, 0.0929329) (-0.0924284, 0.0530017) (-0.0684625, 0.0187663) (-0.0329337, -0.00280151) (-0.00822669, -0.00385636)};
\draw[thick]  plot[smooth]
coordinates {(1.1093, -0.461673) (1.10955, -0.462641) (1.12187, -0.514186) (1.13228, -0.566149) (1.14085, -0.618449) (1.14763, -0.671011) (1.15267, -0.723769) (1.15601, -0.776661) (1.15772, -0.829632) (1.15784, -0.88263) (1.1564, -0.935608) (1.15346, -0.988525) (1.14907, -1.04134) (1.14325, -1.09402) (1.13606, -1.14653) (1.12753, -1.19884) (1.1177, -1.25091) (1.10661, -1.30274) (1.0943, -1.35429) (1.08079, -1.40554) (1.06613, -1.45647) (1.05035, -1.50706) (1.03347, -1.5573) (1.01554, -1.60718) (0.996582, -1.65667) (0.976625, -1.70577) (0.9557, -1.75446) (0.933833, -1.80274) (0.911053, -1.85059) (0.887386, -1.89801) (0.862858, -1.945) (0.837494, -1.99153) (0.832616, -2.00026)};
\draw[thick]  plot[smooth]
coordinates {(1.1093, -0.461673) (1.10905, -0.460705) (1.16117, -0.405522) (1.21319, -0.348755) (1.26358, -0.29054) (1.31257, -0.231131) (1.36035, -0.170752) (1.40712, -0.109586) (1.45304, -0.0477753) (1.49822, 0.0145759) (1.54274, 0.0774016) (1.58663, 0.140665) (1.62991, 0.204353) (1.67255, 0.268466) (1.71453, 0.333012) (1.75583, 0.398004) (1.79639, 0.463452) (1.83619, 0.529365) (1.87521, 0.595748) (1.91341, 0.662603) (1.95078, 0.729926) (1.98731, 0.797712) (2.02297, 0.865952) (2.05778, 0.934636) (2.09172, 1.00375) (2.1248, 1.07328) (2.15702, 1.14322) (2.18838, 1.21354) (2.21889, 1.28424) (2.24856, 1.35529) (2.27741, 1.42668) (2.30543, 1.4984) (2.30614, 1.50027)};
\draw[thick]  plot[smooth]
coordinates {(1.10621, -0.458549) (1.10525, -0.458286) (1.06468, -0.447414) (1.024, -0.436968) (0.983209, -0.426955) (0.942315, -0.41738) (0.901322, -0.408239) (0.860238, -0.399518) (0.819071, -0.391194) (0.777833, -0.383229) (0.736537, -0.375572) (0.695198, -0.368154) (0.653831, -0.360888) (0.612456, -0.353672) (0.571093, -0.346382) (0.529768, -0.338882) (0.488512, -0.331014) (0.447362, -0.32261) (0.406366, -0.313484) (0.365587, -0.303434) (0.325107, -0.292244) (0.285032, -0.279682) (0.245504, -0.265496) (0.20671, -0.249414) (0.168903, -0.231138) (0.132422, -0.210346) (0.0977392, -0.186681) (0.0655245, -0.159764) (0.0367642, -0.129199) (0.0130006, -0.0946332) (-0.00310673, -0.055961) (-0.00587013, -0.0143967) (-0.00439036, -0.00858567)};

\draw[radius=0.3mm,fill](0,0)circle;
\draw[radius=0.3mm,fill](-22.5:1.2)circle;
\draw[radius=0.3mm,fill](22.5:1/1.2)circle;
\end{tikzpicture}
&
\begin{tikzpicture}[scale=1]
\draw[double,red]  plot[smooth]
coordinates {(2.00007, 0.599617) (1.99928, 0.598992) (1.94119, 0.55315) (1.88252, 0.508063) (1.82323, 0.463776) (1.76332, 0.420338) (1.70277, 0.377807) (1.64154, 0.336244) (1.57963, 0.295722) (1.51699, 0.256323) (1.4536, 0.218144) (1.38943, 0.181293) (1.32445, 0.145897) (1.25862, 0.112095) (1.19193, 0.0800308) (1.12438, 0.049826) (1.056, 0.021543) (0.987809, -0.00451413) (0.918048, -0.0291951) (0.847725, -0.0522251) (0.776873, -0.0735721) (0.705466, -0.0929813) (0.633456, -0.11001) (0.560817, -0.124101) (0.487579, -0.134632) (0.413857, -0.140903) (0.339884, -0.142096) (0.266072, -0.137191) (0.193152, -0.124831) (0.122494, -0.103079) (0.057044, -0.0688801) (0.00573045, -0.0164593) (0.00234468, -0.00921611)};
\draw[double,red]  plot[smooth]
coordinates {(-0.723945, 1.50027) (-0.724097, 1.49928) (-0.734652, 1.42806) (-0.744412, 1.35672) (-0.753267, 1.28527) (-0.761095, 1.2137) (-0.767758, 1.14201) (-0.773107, 1.07021) (-0.776976, 0.998312) (-0.779183, 0.926347) (-0.779528, 0.85435) (-0.777792, 0.782374) (-0.773738, 0.710492) (-0.767107, 0.638802) (-0.75762, 0.567435) (-0.744979, 0.49656) (-0.728864, 0.426395) (-0.708939, 0.357217) (-0.684852, 0.289379) (-0.656243, 0.223323) (-0.622754, 0.159607) (-0.584043, 0.0989272) (-0.539807, 0.0421565) (-0.490573, -0.00895177) (-0.43492, -0.0545551) (-0.373586, -0.0921442) (-0.307134, -0.119633) (-0.236804, -0.134484) (-0.16498, -0.133724) (-0.095947, -0.114195) (-0.0371263, -0.0733988) (-0.0016176, -0.0117537) (-0.00120043, -0.00979777)};
\draw [thick] plot[smooth]
coordinates {(0.771603, 0.319348) (0.77257, 0.319602) (0.818914, 0.3321) (0.86506, 0.345309) (0.91097, 0.359317) (0.956607, 0.374189) (1.00194, 0.389975) (1.04693, 0.406702) (1.09155, 0.42438) (1.13579, 0.443007) (1.17962, 0.462565) (1.22304, 0.483029) (1.26604, 0.504366) (1.30861, 0.52654) (1.35075, 0.549511) (1.39248, 0.573238) (1.43379, 0.597683) (1.47469, 0.622804) (1.51519, 0.648566) (1.5553, 0.674932) (1.59503, 0.701868) (1.63438, 0.729344) (1.67338, 0.757329) (1.71203, 0.785797) (1.75033, 0.814723) (1.78831, 0.844082) (1.82596, 0.873853) (1.8633, 0.904015) (1.90033, 0.934552) (1.93707, 0.965444) (1.97352, 0.996675) (2.00067, 1.02031)};
\draw[thick]  plot[smooth]
coordinates {(0.771603, 0.319348) (0.770635, 0.319095) (0.760771, 0.288262) (0.749799, 0.257143) (0.737267, 0.226619) (0.723152, 0.196794) (0.707442, 0.167778) (0.690136, 0.139686) (0.671248, 0.112631) (0.650809, 0.0867289) (0.628863, 0.0620906) (0.605472, 0.0388197) (0.580713, 0.0170106) (0.554674, -0.00325353) (0.527453, -0.021901) (0.499157, -0.0388708) (0.469892, -0.0541111) (0.43977, -0.0675756) (0.408898, -0.0792204) (0.377385, -0.0890005) (0.345341, -0.0968651) (0.312876, -0.102754) (0.280106, -0.106593) (0.247156, -0.10829) (0.214168, -0.107727) (0.18131, -0.104756) (0.148793, -0.0991871) (0.116895, -0.0907742) (0.0860111, -0.0791945) (0.0567392, -0.0640101) (0.030104, -0.0445981) (0.00827029, -0.0199881) (0.00257876, -0.0094406)};
\draw[thick]  plot[smooth]
coordinates {(0.768662, 0.320155) (0.767958, 0.320865) (0.728705, 0.357936) (0.68715, 0.392405) (0.643414, 0.42406) (0.597636, 0.452682) (0.549978, 0.478051) (0.500628, 0.499942) (0.449796, 0.51813) (0.397728, 0.532392) (0.344698, 0.542506) (0.291021, 0.548259) (0.237051, 0.549447) (0.183187, 0.54588) (0.129878, 0.537392) (0.0776282, 0.523842) (0.0269994, 0.50513) (-0.0213804, 0.481201) (-0.066813, 0.452068) (-0.108525, 0.417825) (-0.145662, 0.378674) (-0.177289, 0.334957) (-0.202386, 0.287199) (-0.219858, 0.236165) (-0.228554, 0.18294) (-0.227311, 0.12904) (-0.215038, 0.0765642) (-0.190901, 0.0284182) (-0.154699, -0.0113814) (-0.107675, -0.0373169) (-0.0543925, -0.0420505) (-0.00785444, -0.0170569) (-0.00295139, -0.00952722)};
\draw[thick]  plot[smooth]
coordinates {(1.10864, -0.461755) (1.10863, -0.462755) (1.10655, -0.527716) (1.10143, -0.592508) (1.09335, -0.656998) (1.08238, -0.72106) (1.06861, -0.784579) (1.05211, -0.847447) (1.03299, -0.909565) (1.01132, -0.970841) (0.987188, -1.03119) (0.960692, -1.09054) (0.931917, -1.14882) (0.900952, -1.20597) (0.867884, -1.26192) (0.832801, -1.31664) (0.795788, -1.37007) (0.75693, -1.42217) (0.716311, -1.47291) (0.67401, -1.52226) (0.630107, -1.57019) (0.584681, -1.61667) (0.537807, -1.6617) (0.489558, -1.70525) (0.440007, -1.74732) (0.389223, -1.78788) (0.337273, -1.82695) (0.284222, -1.8645) (0.230135, -1.90055) (0.175073, -1.93509) (0.119094, -1.96812) (0.0622562, -1.99965) (0.0613754, -2.00012)};
\draw[thick]  plot[smooth]
coordinates {(1.10864, -0.461755) (1.10865, -0.460755) (1.13821, -0.442196) (1.16921, -0.423895) (1.20004, -0.405293) (1.23069, -0.386419) (1.26119, -0.367298) (1.29155, -0.347952) (1.32178, -0.328404) (1.35189, -0.308671) (1.38189, -0.288769) (1.41179, -0.268714) (1.44159, -0.248517) (1.4713, -0.22819) (1.50093, -0.207739) (1.53048, -0.187173) (1.55995, -0.166496) (1.58934, -0.145715) (1.61866, -0.12483) (1.64792, -0.103846) (1.6771, -0.0827636) (1.70621, -0.0615839) (1.73525, -0.0403075) (1.76422, -0.0189346) (1.79311, 0.00253504) (1.82194, 0.0241017) (1.85069, 0.0457658) (1.87937, 0.0675279) (1.90797, 0.0893887) (1.9365, 0.111349) (1.96495, 0.133409) (1.99332, 0.155569) (2.0004, 0.161125)};
\draw[thick]  plot[smooth]
coordinates {(1.10647, -0.45794) (1.1056, -0.457436) (1.07089, -0.437569) (1.03583, -0.418314) (1.00042, -0.399709) (0.964655, -0.381793) (0.928537, -0.364606) (0.892063, -0.348186) (0.855239, -0.332568) (0.818073, -0.317781) (0.780581, -0.303843) (0.742782, -0.290761) (0.704702, -0.278521) (0.66637, -0.26709) (0.627823, -0.256411) (0.589096, -0.246399) (0.55023, -0.236946) (0.511262, -0.227918) (0.472233, -0.219159) (0.433183, -0.210493) (0.394155, -0.201728) (0.355198, -0.192655) (0.316369, -0.183048) (0.277741, -0.172665) (0.23941, -0.161237) (0.201506, -0.148467) (0.164213, -0.134014) (0.127801, -0.117473) (0.0926816, -0.0983481) (0.0595344, -0.0759948) (0.0296112, -0.0495128) (0.00587905, -0.0174768) (0.00222265, -0.00925897)};

\draw[radius=0.3mm,fill](0,0)circle;
\draw[radius=0.3mm,fill](-22.5:1.2)circle;
\draw[radius=0.3mm,fill](22.5:1/1.2)circle;
\end{tikzpicture}
&
\begin{tikzpicture}[scale=1]
\draw[double,red]  plot[smooth]
coordinates {(2.00076, 0.142269) (1.99979, 0.141999) (1.9352, 0.124192) (1.87044, 0.107025) (1.80549, 0.0905671) (1.74035, 0.0749049) (1.675, 0.0601455) (1.60942, 0.0464216) (1.5436, 0.0338979) (1.47753, 0.0227758) (1.41121, 0.0132939) (1.34464, 0.00571579) (1.27786, 0.000292953) (1.21094, -0.00281175) (1.14395, -0.00364098) (1.07696, -0.00256719) (1.00999, -0.000360786) (0.94403, 0.00183286) (0.877039, 0.00279428) (0.810061, 0.00131398) (0.743218, -0.00318955) (0.676613, -0.0104208) (0.610218, -0.0194009) (0.5439, -0.0289353) (0.477509, -0.0379378) (0.41094, -0.0455108) (0.344161, -0.050894) (0.277213, -0.0533656) (0.210236, -0.0521174) (0.143529, -0.0460658) (0.0777709, -0.0334349) (0.0151213, -0.0101329) (0.00737948, -0.00554729)};
\draw[double,red]  plot[smooth]
coordinates {(2.00075, 1.30395) (1.99982, 1.30358) (1.87632, 1.25697) (1.75192, 1.21285) (1.62654, 1.17157) (1.50015, 1.13353) (1.3727, 1.09919) (1.24419, 1.06906) (1.11467, 1.04364) (0.984252, 1.0233) (0.853147, 1.00802) (0.721619, 0.996914) (0.589924, 0.987947) (0.458299, 0.978039) (0.327099, 0.96364) (0.19703, 0.941319) (0.0693214, 0.908131) (-0.0541914, 0.861767) (-0.17107, 0.800625) (-0.278316, 0.723873) (-0.372445, 0.631539) (-0.449585, 0.524638) (-0.505617, 0.405351) (-0.536381, 0.277242) (-0.537971, 0.145555) (-0.507215, 0.0175754) (-0.4433, -0.0962372) (-0.346359, -0.18486) (-0.223783, -0.231206) (-0.094073, -0.217172) (0.00322075, -0.132029) (0.00559946, -0.00746169)};
\draw [thick] plot[smooth]
coordinates {(0.77166, 0.318892) (0.77266, 0.318886) (0.814658, 0.318532) (0.856656, 0.318071) (0.898653, 0.317611) (0.940652, 0.317253) (0.982651, 0.317092) (1.02465, 0.317216) (1.06665, 0.317701) (1.10864, 0.318616) (1.15061, 0.320015) (1.19257, 0.321945) (1.2345, 0.324437) (1.27638, 0.327514) (1.31822, 0.331188) (1.36, 0.33546) (1.40172, 0.340325) (1.44336, 0.345771) (1.48493, 0.351784) (1.52642, 0.358341) (1.56781, 0.365423) (1.60912, 0.373004) (1.65034, 0.381061) (1.69147, 0.38957) (1.73251, 0.398507) (1.77346, 0.407851) (1.81432, 0.417578) (1.85509, 0.427669) (1.89577, 0.438104) (1.93637, 0.448865) (1.97688, 0.459934) (2.00096, 0.466663)};
\draw[thick]  plot[smooth]
coordinates {(0.76901, 0.317384) (0.768503, 0.316522) (0.752782, 0.290973) (0.736086, 0.26605) (0.718416, 0.241809) (0.699779, 0.218303) (0.680186, 0.195588) (0.659656, 0.173715) (0.638215, 0.152736) (0.615895, 0.132694) (0.592732, 0.113633) (0.568768, 0.0955884) (0.544051, 0.0785912) (0.518629, 0.0626671) (0.492553, 0.0478375) (0.465875, 0.0341197) (0.438648, 0.0215285) (0.410922, 0.0100774) (0.382746, -0.000219321) (0.35417, -0.00934564) (0.325241, -0.0172812) (0.296005, -0.0239994) (0.26651, -0.0294648) (0.236803, -0.0336298) (0.206937, -0.0364309) (0.17697, -0.037782) (0.146975, -0.0375661) (0.117043, -0.0356197) (0.0873053, -0.0317063) (0.0579728, -0.0254593) (0.0294461, -0.0162334) (0.00789527, -0.00573562)};
\draw[thick]  plot[smooth]
coordinates {(0.769029, 0.320433) (0.768531, 0.3213) (0.724405, 0.389176) (0.673098, 0.451803) (0.615273, 0.508466) (0.551615, 0.558488) (0.482859, 0.60123) (0.409795, 0.636099) (0.333282, 0.662553) (0.254254, 0.680114) (0.173724, 0.688377) (0.092784, 0.687031) (0.0126087, 0.675867) (-0.0655486, 0.654803) (-0.14036, 0.623901) (-0.21043, 0.583388) (-0.274308, 0.533689) (-0.3305, 0.475448) (-0.377492, 0.409568) (-0.413781, 0.337245) (-0.437908, 0.26002) (-0.448518, 0.179824) (-0.444439, 0.099046) (-0.42479, 0.0206076) (-0.389161, -0.0519532) (-0.337874, -0.114391) (-0.272426, -0.161687) (-0.196203, -0.1881) (-0.115663, -0.1876) (-0.0420608, -0.155424) (0.00695733, -0.0922533) (0.00809711, -0.0132281) (0.00545798, -0.00784192)};
\draw[thick]  plot[smooth]
coordinates {(1.10798, -0.461665) (1.10772, -0.462628) (1.03815, -0.657389) (0.94061, -0.839754) (0.817244, -1.00574) (0.671067, -1.15206) (0.505596, -1.27614) (0.324576, -1.37621) (0.13179, -1.45119) (-0.0690777, -1.50062) (-0.274551, -1.52461) (-0.481429, -1.52376) (-0.686834, -1.49902) (-0.888237, -1.45166) (-1.08347, -1.38314) (-1.27072, -1.2951) (-1.44851, -1.18924) (-1.6157, -1.06731) (-1.77143, -0.931036) (-1.91512, -0.782112) (-2.04642, -0.622148) (-2.16517, -0.45266) (-2.27142, -0.275057) (-2.36533, -0.0906288) (-2.4472, 0.0994567) (-2.51741, 0.294154) (-2.57642, 0.492538) (-2.62474, 0.693797) (-2.66289, 0.897229) (-2.69146, 1.10223) (-2.71099, 1.30829) (-2.72156, 1.50099)};
\draw[thick]  plot[smooth]
coordinates {(1.10798, -0.461665) (1.10825, -0.460701) (1.13681, -0.451812) (1.16675, -0.443797) (1.19667, -0.435684) (1.22657, -0.427485) (1.25644, -0.419209) (1.2863, -0.410866) (1.31614, -0.402464) (1.34596, -0.39401) (1.37578, -0.38551) (1.40558, -0.376968) (1.43537, -0.368389) (1.46514, -0.359774) (1.49491, -0.351128) (1.52467, -0.34245) (1.55443, -0.333743) (1.58417, -0.325007) (1.61391, -0.316242) (1.64363, -0.307449) (1.67335, -0.298628) (1.70306, -0.289778) (1.73276, -0.280898) (1.76245, -0.271988) (1.79214, -0.263047) (1.82181, -0.254075) (1.85147, -0.245071) (1.88113, -0.236033) (1.91077, -0.226962) (1.9404, -0.217856) (1.97002, -0.208714) (1.99963, -0.199537) (2.00059, -0.19924)};
\draw[thick]  plot[smooth]
coordinates {(1.10687, -0.457417) (1.10617, -0.456707) (1.07696, -0.427936) (1.04706, -0.399884) (1.01645, -0.372613) (0.985098, -0.346192) (0.952989, -0.320698) (0.920102, -0.296218) (0.886419, -0.272844) (0.851931, -0.250677) (0.816635, -0.229821) (0.780542, -0.210378) (0.743677, -0.19244) (0.706087, -0.176078) (0.667835, -0.16133) (0.629001, -0.148187) (0.589678, -0.136589) (0.549962, -0.126419) (0.509942, -0.117513) (0.469701, -0.109668) (0.429305, -0.102657) (0.388811, -0.0962384) (0.348264, -0.0901639) (0.307703, -0.0841805) (0.267167, -0.0780287) (0.226701, -0.071435) (0.186364, -0.0640983) (0.146242, -0.0556655) (0.106479, -0.0456864) (0.067335, -0.0335146) (0.0293981, -0.0180207) (0.00759621, -0.00582665)};

\draw[radius=0.3mm,fill](0,0)circle;
\draw[radius=0.3mm,fill](-22.5:1.2)circle;
\draw[radius=0.3mm,fill](22.5:1/1.2)circle;
\end{tikzpicture}

\end{tabular}
\end{center}
\caption{\label{fig:WKBdeformation} Intermediate steps of the rotation  \(r\to i\mathfrak{r}\). }
\end{figure}
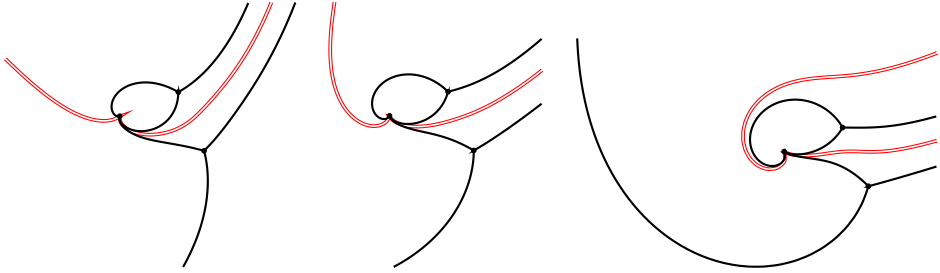

The WKB graph, corresponding to real values of \(r\), can be found in Fig.~\ref{fig:WKBreal}, and
corresponding WKB graph for imaginary \(r\in i\mathbb{R}_{>0}\) is shown in Fig.~\ref{fig:WKBimaginary}.
It can be obtained from Fig.~\ref{fig:WKBreal} by continuous rotation by \(\pi/2\) from \(r\in \mathbb{R}_{>0}\) to \(r\in i\mathbb{R}_{>0}\), see Fig. \ref{fig:WKBdeformation}. All these graphs have two triple points $P$ and ${P'}$, where derivative of the WKB phase vanishes, and two singularities at $z=0$ and $z=\infty$.
The anti-Stokes lines in general situation connect ``zero'' with ``infinity'', and in our case divide the $z$-plane into six domains.

\begin{figure}[h!]
\begin{center}
\begin{tikzpicture}[xscale=0.2, yscale=0.2]

\draw[red](0,0)--(-20,0);


\draw[red]  plot[smooth]
coordinates {(-20.0075, 17.6991) (-19.9982, 17.6954) (-19.5255, 17.5039) (-19.0536, 17.3105) (-18.5824, 17.1153) (-18.1121, 16.9182) (-17.6425, 16.7191) (-17.1739, 16.518) (-16.7061, 16.3148) (-16.2393, 16.1094) (-15.7734, 15.9019) (-15.3086, 15.692) (-14.8449, 15.4798) (-14.3822, 15.2651) (-13.9208, 15.048) (-13.4605, 14.8282) (-13.0016, 14.6058) (-12.544, 14.3806) (-12.0879, 14.1525) (-11.6332, 13.9215) (-11.1801, 13.6874) (-10.7287, 13.4501) (-10.279, 13.2094) (-9.83124, 12.9653) (-9.38542, 12.7176) (-8.9417, 12.4662) (-8.5002, 12.2109) (-8.06107, 11.9516) (-7.62447, 11.688) (-7.19056, 11.42) (-6.75955, 11.1474) (-6.33164, 10.8699) (-5.90706, 10.5874) (-5.48607, 10.2995) (-5.06896, 10.0061) (-4.65604, 9.70673) (-4.24768, 9.40122) (-3.84429, 9.08918) (-3.44631, 8.77026) (-3.05428, 8.44407) (-2.66877, 8.11019) (-2.29048, 7.76816) (-1.92018, 7.41749) (-1.55878, 7.05767) (-1.20734, 6.68811) (-0.867084, 6.30824) (-0.539487, 5.9174) (-0.226286, 5.51494) (0.0704308, 5.10019) (0.348133, 4.67248) (0.60373, 4.23122) (0.833404, 3.77596) (1.03238, 3.30649) (1.19465, 2.82315) (1.3125, 2.32717) (1.37607, 1.82144) (1.3728, 1.3119) (1.28857, 0.809553) (1.12533, 0.326704) (0.926612, -0.13054) (0.489808, -0.36268) (0.0187897, -0.220553) (-0.00517013, -0.0000719214)};
\draw [thick] plot[smooth]
coordinates {(0.771152, 0.32014) (0.77825, 0.327184) (1.16155, 1.17961) (1.14928, 2.12588) (0.909789, 3.04367) (0.523825, 3.91091) (0.0365526, 4.72592) (-0.524233, 5.49241) (-1.14033, 6.21531) (-1.79927, 6.89947) (-2.49219, 7.54924) (-3.21259, 8.16843) (-3.9556, 8.76033) (-4.71749, 9.32775) (-5.49533, 9.8731) (-6.28681, 10.3985) (-7.09005, 10.9057) (-7.90355, 11.3963) (-8.72604, 11.8717) (-9.55648, 12.333) (-10.394, 12.7814) (-11.2378, 13.2178) (-12.0873, 13.6431) (-12.942, 14.0579) (-13.8013, 14.463) (-14.6648, 14.8589) (-15.5322, 15.2463) (-16.4032, 15.6256) (-17.2775, 15.9973) (-18.1548, 16.3618) (-19.0349, 16.7194) (-19.9176, 17.0706) (-20.0014, 17.1035)};
\draw[thick]  plot[smooth]
coordinates {(0.771152, 0.32014) (0.764054, 0.313096) (0.780704, 0.267139) (0.787945, 0.21768) (0.790995, 0.167788) (0.789717, 0.11782) (0.783897, 0.0681773) (0.773318, 0.0193296) (0.757796, -0.0281767) (0.73723, -0.0737241) (0.711652, -0.116656) (0.681277, -0.156337) (0.646503, -0.192227) (0.607873, -0.223932) (0.565998, -0.251213) (0.521489, -0.273947) (0.474914, -0.29208) (0.426788, -0.305572) (0.377587, -0.314367) (0.327767, -0.318374) (0.277795, -0.31746) (0.22818, -0.311449) (0.179504, -0.300127) (0.132469, -0.283251) (0.0879498, -0.260569) (0.0470705, -0.231857) (0.0113224, -0.196984) (-0.0172548, -0.156058) (-0.0357651, -0.109752) (-0.0398906, -0.0601628) (-0.0227481, -0.0138632) (-0.00753372, -0.00107068)};
\draw[thick]  plot[smooth]
coordinates {(0.768202, 0.319369) (0.758555, 0.322003) (0.719822, 0.331986) (0.680862, 0.341046) (0.641678, 0.349076) (0.602278, 0.355968) (0.562679, 0.36161) (0.52291, 0.365886) (0.48301, 0.368677) (0.44303, 0.369859) (0.403038, 0.3693) (0.363115, 0.366865) (0.323369, 0.362411) (0.283926, 0.355786) (0.244946, 0.346836) (0.206624, 0.335398) (0.169198, 0.321304) (0.132964, 0.304385) (0.0982858, 0.284474) (0.0656199, 0.261414) (0.0355406, 0.235076) (0.00878065, 0.205379) (-0.0137091, 0.172341) (-0.0306699, 0.136166) (-0.0403894, 0.0974389) (-0.0404527, 0.0575613) (-0.027274, 0.0200802) (-0.00539282, 0.000176006)};
\draw[thick]  plot[smooth]
coordinates {(1.10991, -0.461423) (1.11484, -0.470123) (1.36973, -1.39828) (1.31171, -2.36399) (1.05473, -3.29808) (0.662201, -4.18439) (0.172772, -5.02141) (-0.388592, -5.81216) (-1.00502, -6.56089) (-1.66466, -7.2719) (-2.35895, -7.94917) (-3.08154, -8.5962) (-3.82757, -9.21607) (-4.59331, -9.81144) (-5.37581, -10.3846) (-6.1727, -10.9376) (-6.98207, -11.4722) (-7.80235, -11.9899) (-8.63223, -12.4921) (-9.47064, -12.9799) (-10.3166, -13.4544) (-11.1694, -13.9166) (-12.0284, -14.3672) (-12.8929, -14.8071) (-13.7625, -15.237) (-14.6366, -15.6573) (-15.515, -16.0688) (-16.3973, -16.4719) (-17.2832, -16.867) (-18.1724, -17.2546) (-19.0646, -17.6351) (-19.9597, -18.0088) (-20.006, -18.0279)};
\draw[thick]  plot[smooth]
coordinates {(1.10991, -0.461423) (1.10498, -0.452723) (1.44963, 0.50138) (1.58774, 1.50945) (1.48368, 2.52211) (1.19108, 3.49807) (0.763153, 4.42324) (0.236489, 5.29629) (-0.364162, 6.12036) (-1.02169, 6.89991) (-1.72394, 7.6395) (-2.46207, 8.34334) (-3.22947, 9.01517) (-4.02112, 9.65828) (-4.83312, 10.2755) (-5.66242, 10.8693) (-6.50654, 11.4419) (-7.3635, 11.995) (-8.23168, 12.5304) (-9.10972, 13.0494) (-9.99649, 13.5534) (-10.891, 14.0435) (-11.7926, 14.5206) (-12.7004, 14.9857) (-13.6138, 15.4395) (-14.5325, 15.8828) (-15.4558, 16.3162) (-16.3834, 16.7403) (-17.315, 17.1557) (-18.2503, 17.5628) (-19.1889, 17.9621) (-20.0012, 18.3006)};
\draw[thick]  plot[smooth]
coordinates {(1.10612, -0.459205) (1.09612, -0.459147) (1.04612, -0.458933) (0.996121, -0.45881) (0.946121, -0.458729) (0.896122, -0.458621) (0.846122, -0.4584) (0.796124, -0.457961) (0.74613, -0.457177) (0.696147, -0.455902) (0.646185, -0.453968) (0.596263, -0.45119) (0.546411, -0.447361) (0.496673, -0.442258) (0.447115, -0.435643) (0.397826, -0.427259) (0.348928, -0.416836) (0.300586, -0.404088) (0.253016, -0.388715) (0.206499, -0.370403) (0.161405, -0.34883) (0.118215, -0.323669) (0.0775564, -0.294601) (0.0402614, -0.261339) (0.00744091, -0.223668) (-0.0193899, -0.18154) (-0.0381111, -0.135267) (-0.0456483, -0.0859818) (-0.0373088, -0.0369859) (-0.00540901, -0.000154988)};

\draw[radius=0.3mm,fill](0,0)circle;
\draw[radius=0.3mm,fill](-22.5:1.2)circle;
\draw[radius=0.3mm,fill](22.5:1/1.2)circle;









\end{tikzpicture}
\end{center}
\caption{The same WKB graph, as in Fig.~\ref{fig:WKBreal}, at smaller scale, demonstrating behavior of the anti-Stokes lines at $z\to\infty$, where they bend finally to the negative real axis. \label{fig:WKBrealScaled} }
\end{figure}
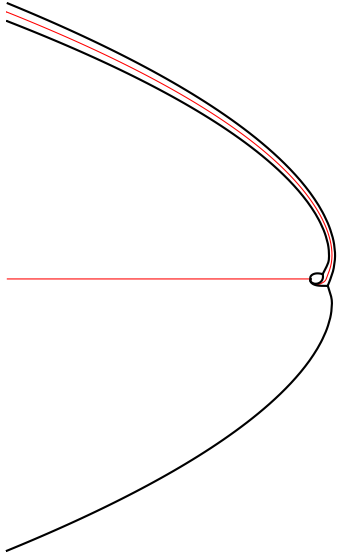

\begin{figure}[h!]
\begin{center}
\begin{tikzpicture}[scale=2]

\draw[red,->](2.0232,0.47)--+(84:0.2);
\node at (1.95,0.4411) {\color{red} \(y\)};

\draw[red,->] (1.9894,-0.3408)--+(84:0.2);
\node at (1.92,-0.3408){\color{red}\(x\)};

\draw[double,red]  plot[smooth]
coordinates {(3.00006, -0.309326) (2.99906, -0.30927) (2.89723, -0.303457) (2.7954, -0.297505) (2.69358, -0.291394) (2.59178, -0.285102) (2.48998, -0.278596) (2.38821, -0.271836) (2.28645, -0.26477) (2.18473, -0.257324) (2.08303, -0.249396) (1.98139, -0.240842) (1.87983, -0.231451) (1.77838, -0.220906) (1.67711, -0.208725) (1.57616, -0.194157) (1.47578, -0.176054) (1.37648, -0.152811) (1.27904, -0.122759) (1.18412, -0.0854722) (1.09137, -0.0430441) (1, 0) (0.90597, 0.0394662) (0.808258, 0.0684105) (0.707084, 0.0801203) (0.605272, 0.0752096) (0.504194, 0.0615998) (0.40347, 0.0455205) (0.302655, 0.0300182) (0.201555, 0.0165195) (0.100112, 0.00592387) (0.00930462, 0.000176523)};
\draw[double,red]  plot[smooth]
coordinates {(3.00093, 0.510536) (2.99993, 0.510563) (2.73208, 0.51948) (2.46441, 0.532684) (2.19711, 0.551943) (1.93057, 0.57974) (1.66553, 0.6193) (1.40331, 0.674399) (1.14609, 0.749356) (0.89614, 0.845874) (0.648651, 0.948661) (0.393222, 1.02907) (0.128099, 1.0652) (-0.138577, 1.04501) (-0.39336, 0.96397) (-0.620929, 0.82373) (-0.806376, 0.631255) (-0.936878, 0.39803) (-1.00296, 0.139094) (-0.999383, -0.128101) (-0.925723, -0.384951) (-0.786529, -0.613015) (-0.591206, -0.795292) (-0.353584, -0.917336) (-0.0913784, -0.968102) (0.174141, -0.940348) (0.417092, -0.830345) (0.599813, -0.637372) (0.650836, -0.380553) (0.521802, -0.151626) (0.28434, -0.0330429) (0.0189122, -0.000245042) (0.00991367, -0.000084537)};
\draw[thick]  plot[smooth]
coordinates {(0.771597, 0.318437) (0.772561, 0.318171) (0.84545, 0.296661) (0.917668, 0.272991) (0.989389, 0.247853) (1.06083, 0.221936) (1.13224, 0.195925) (1.20386, 0.170477) (1.27586, 0.146175) (1.3484, 0.123484) (1.4215, 0.102701) (1.49514, 0.0839477) (1.56927, 0.0671902) (1.64379, 0.0522869) (1.71863, 0.0390398) (1.79371, 0.0272344) (1.86897, 0.0166649) (1.94437, 0.0071467) (2.01988, -0.00147972) (2.09547, -0.00934909) (2.17112, -0.0165739) (2.24683, -0.0232474) (2.32258, -0.0294472) (2.39835, -0.0352376) (2.47416, -0.0406721) (2.54999, -0.0457955) (2.62583, -0.0506457) (2.70169, -0.0552543) (2.77757, -0.0596483) (2.85345, -0.0638509) (2.92934, -0.0678818) (3.00025, -0.0715075)};
\draw[thick]  plot[smooth]
coordinates {(0.768647, 0.317666) (0.767935, 0.316964) (0.747701, 0.29761) (0.726937, 0.278828) (0.705653, 0.260636) (0.683864, 0.243052) (0.661587, 0.226092) (0.63884, 0.209766) (0.615645, 0.194083) (0.592024, 0.17905) (0.568001, 0.164668) (0.5436, 0.150936) (0.518847, 0.13785) (0.493766, 0.125405) (0.468381, 0.11359) (0.442717, 0.102396) (0.416796, 0.0918104) (0.390639, 0.0818204) (0.364268, 0.0724128) (0.3377, 0.0635741) (0.310953, 0.0552917) (0.284044, 0.0475538) (0.256987, 0.0403501) (0.229795, 0.0336727) (0.202481, 0.0275162) (0.175055, 0.0218794) (0.147526, 0.0167661) (0.119903, 0.0121882) (0.0921937, 0.00816939) (0.0644034, 0.00475432) (0.0365369, 0.00203166) (0.00959615, 0.000270976)};
\draw[thick]  plot[smooth]
coordinates {(0.769454, 0.320606) (0.769198, 0.321573) (0.705004, 0.480686) (0.605672, 0.620659) (0.478241, 0.735691) (0.329558, 0.821548) (0.166495, 0.875334) (-0.00403986, 0.895423) (-0.175189, 0.881438) (-0.340293, 0.834216) (-0.493051, 0.755764) (-0.6277, 0.64918) (-0.739173, 0.518549) (-0.823258, 0.368815) (-0.876726, 0.205624) (-0.89744, 0.0351533) (-0.884438, -0.136075) (-0.837983, -0.301388) (-0.759585, -0.454155) (-0.652009, -0.587977) (-0.519263, -0.696857) (-0.366588, -0.775345) (-0.200504, -0.818655) (-0.0289609, -0.82274) (0.138199, -0.784361) (0.288104, -0.701357) (0.402452, -0.57418) (0.455689, -0.412225) (0.426265, -0.244649) (0.320516, -0.110997) (0.169339, -0.0308588) (0.00970707, -0.000291844)};
\draw[thick]  plot[smooth]
coordinates {(1.10737, -0.461408) (1.10687, -0.46227) (0.954166, -0.677201) (0.763836, -0.859519) (0.540241, -0.998945) (0.291936, -1.08703) (0.0303181, -1.11797) (-0.231524, -1.08914) (-0.479883, -1.00137) (-0.701447, -0.858912) (-0.884221, -0.66924) (-1.01834, -0.442536) (-1.0967, -0.191052) (-1.1154, 0.0717013) (-1.07394, 0.331847) (-0.975167, 0.576075) (-0.82497, 0.792539) (-0.631735, 0.971684) (-0.405554, 1.10695) (-0.157216, 1.19536) (0.103012, 1.23792) (0.366795, 1.23999) (0.62914, 1.21136) (0.889101, 1.16545) (1.14846, 1.11619) (1.40891, 1.07307) (1.67075, 1.03956) (1.93365, 1.01553) (2.19716, 0.999607) (2.46098, 0.990204) (2.72495, 0.985862) (2.98894, 0.98538) (3.00094, 0.985433)};
\draw[thick]  plot[smooth]
coordinates {(1.10737, -0.461408) (1.10788, -0.460546) (1.16976, -0.4597) (1.23276, -0.460434) (1.29575, -0.461445) (1.35874, -0.462737) (1.42172, -0.464304) (1.48469, -0.466132) (1.54766, -0.468203) (1.61061, -0.470498) (1.67356, -0.472996) (1.73651, -0.475676) (1.79944, -0.478521) (1.86237, -0.481511) (1.92529, -0.484631) (1.98821, -0.487865) (2.05112, -0.4912) (2.11403, -0.494624) (2.17693, -0.498125) (2.23983, -0.501694) (2.30273, -0.505322) (2.36562, -0.509002) (2.42851, -0.512725) (2.4914, -0.516487) (2.55428, -0.520282) (2.61717, -0.524104) (2.68005, -0.52795) (2.74293, -0.531815) (2.80581, -0.535696) (2.86869, -0.539589) (2.93157, -0.543493) (2.99445, -0.547404) (3.00044, -0.547777)};
\draw[thick]  plot[smooth]
coordinates {(1.1074, -0.457017) (1.1069, -0.456149) (1.08501, -0.419145) (1.062, -0.382818) (1.03785, -0.347243) (1.01251, -0.312505) (0.985937, -0.278703) (0.958076, -0.245954) (0.928876, -0.214393) (0.898284, -0.18418) (0.866247, -0.155506) (0.83272, -0.128591) (0.797674, -0.103687) (0.761114, -0.0810663) (0.723096, -0.0609944) (0.683743, -0.043687) (0.643247, -0.029254) (0.601849, -0.0176574) (0.559797, -0.00870643) (0.517313, -0.00209596) (0.474566, 0.00253407) (0.431674, 0.00554565) (0.38871, 0.00726945) (0.345717, 0.00799086) (0.302717, 0.00794865) (0.259722, 0.00734017) (0.216734, 0.00632911) (0.173752, 0.0050542) (0.130776, 0.00363878) (0.0877997, 0.00220371) (0.0448197, 0.000893682) (0.00982888, 0.000102367)};

\draw[radius=0.3mm,fill](0,0)circle;
\draw[radius=0.3mm,fill](-22.5:1.2)circle;
\draw[radius=0.3mm,fill](22.5:1/1.2)circle;

\draw[green!70!black,->,thick,dashed]  plot[smooth, tension=.7] coordinates {(0.0684,-0.087) (0.0059,-0.2225) (-0.2048,-0.1579) (-0.2329,0.1217) (0.045,0.1899) (0.3222,-0.2223) (1.25,-0.75) (2.933,-0.6923)};

\draw[thick,magenta,->]  plot[smooth, tension=.7] coordinates {(0.061,-0.0827) (0.0747,0.0673) (-0.0811,0.132) (-0.1385,-0.023) (0.0363,-0.0926)};

\draw[thick,green!70!black,fill]  (0.0684,-0.087) circle[radius=0.2mm];

\node [thick,green!70!black] at  (0.14,-0.17) {\tiny in};

\draw[dashdotted,thick,blue,->]  plot[smooth, tension=.7] coordinates {(0.8879,-0.8784) (1.2676,-0.3935) (0.7782,0.1241) (0.5308,0.5915)};

\node [thick,green!70!black] at  (2.8,-0.8) {\tiny out};

\draw[thick,blue,fill]  (0.8879-0.01,-0.8784-0.01) circle[radius=0.2mm];
\draw[thick,blue,fill]  (0.5308-0.01,0.5915+0.02) circle[radius=0.2mm];

\node at (0.8879-0.15,-0.8784-0.2) {\color{blue}\tiny\(\begin{pmatrix}+ \\ -\end{pmatrix}\)};

\node at (0.5308-0.2,0.5915+0.03) {\color{blue}\tiny\(\begin{pmatrix}+ \\ -\end{pmatrix}\)};

\node at (0.5308-0.05,0.5915-0.15) {\color{blue}\tiny up};
\node at (0.8879+0.15,-0.8784-0.1) {\color{blue}\tiny down};

\node at(1.2,0.5){\(1\)};
\node at(1.2,0.95){\(2\)};
\node at(1.5,-0.6){\(3\)};
\node at(1.4,-0.3){\(4\)};
\node at(1.3,0){\(5\)};
\node at(-0.5,0){\(6\)};

\node at(0.85,0.45){\(P\)};
\node at(0.95,-0.45){\(P'\)};

\end{tikzpicture}
\end{center}
\caption{WKB graph for imaginary 
\(r = i\mathfrak{r}\in i\mathbb{R}_{>0}\). The extra dash-dotted line corresponds to computation of the transition matrix $T$. \label{fig:WKBimaginary} }
\end{figure}
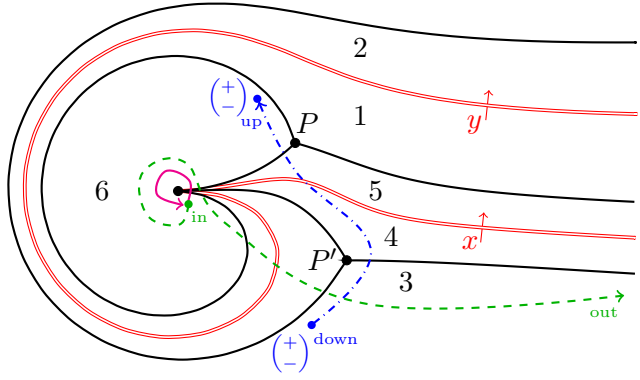

\begin{figure}[h!]
	\begin{center}
		\begin{tabular}{cc}
			\begin{tikzpicture}[scale=2]
			
			\draw[red,double](0,0)--(2,0);
			
			\draw[thick,double]  plot[smooth]
			coordinates {(0.968332, 0.258174) (0.969298, 0.257916) (1.00314, 0.248978) (1.03704, 0.240303) (1.07104, 0.231962) (1.10512, 0.224017) (1.13931, 0.216511) (1.17359, 0.209477) (1.20798, 0.202929) (1.24245, 0.19687) (1.277, 0.19129) (1.31162, 0.186173) (1.34631, 0.181494) (1.38105, 0.177225) (1.41583, 0.173336) (1.45065, 0.169797) (1.4855, 0.166578) (1.52038, 0.16365) (1.55528, 0.160987) (1.59019, 0.158564) (1.62512, 0.156359) (1.66007, 0.154351) (1.69502, 0.152521) (1.72998, 0.150854) (1.76495, 0.149334) (1.79992, 0.147947) (1.83489, 0.146681) (1.86988, 0.145527) (1.90486, 0.144474) (1.93985, 0.143513) (1.97484, 0.142636) (2.00083, 0.142036)};
			\draw[thick]  plot[smooth]
			coordinates {(0.964165, 0.257058) (0.963455, 0.256353) (0.938852, 0.232889) (0.913295, 0.21047) (0.886775, 0.189198) (0.8593, 0.169175) (0.830898, 0.150491) (0.801616, 0.133218) (0.771523, 0.1174) (0.740706, 0.103045) (0.709261, 0.0901223) (0.677286, 0.078569) (0.644878, 0.0682946) (0.612121, 0.0591926) (0.579087, 0.0511492) (0.545837, 0.0440515) (0.512419, 0.0377925) (0.478871, 0.0322741) (0.445221, 0.0274088) (0.411493, 0.0231196) (0.377704, 0.0193393) (0.343868, 0.0160105) (0.309994, 0.0130836) (0.276091, 0.0105169) (0.242165, 0.00827506) (0.208221, 0.00632865) (0.174263, 0.00465392) (0.140292, 0.00323256) (0.106313, 0.00205225) (0.0723261, 0.00110855) (0.0383333, 0.00041084) (0.00933569, 0.0000463757)};
			\draw[thick,double]  plot[smooth]
			coordinates {(0.965281, 0.261225) (0.965019, 0.26219) (0.89731, 0.44154) (0.796622, 0.604674) (0.666655, 0.745599) (0.512189, 0.859138) (0.338899, 0.941123) (0.153152, 0.988545) (-0.0382308, 0.999665) (-0.228221, 0.974077) (-0.409843, 0.912722) (-0.576431, 0.817857) (-0.721872, 0.692965) (-0.84083, 0.542632) (-0.928944, 0.372376) (-0.982984, 0.188444) (-1.00097, -0.000584363) (-0.982414, -0.19139) (-0.927822, -0.375158) (-0.839199, -0.545149) (-0.719791, -0.695125) (-0.573976, -0.819581) (-0.407104, -0.913946) (-0.225299, -0.974756) (-0.0352327, -0.999774) (0.156116, -0.988081) (0.34172, -0.940102) (0.514763, -0.857597) (0.668888, -0.743595) (0.798431, -0.602281) (0.89863, -0.438846) (0.965281, -0.261225)};
			\draw[thick]  plot[smooth]
			coordinates {(0.965281, -0.261225) (0.965543, -0.26026) (0.99796, -0.250337) (1.03186, -0.241614) (1.06583, -0.233218) (1.0999, -0.225208) (1.13407, -0.217633) (1.16834, -0.210524) (1.20271, -0.203901) (1.23717, -0.197768) (1.27171, -0.192116) (1.30632, -0.18693) (1.341, -0.182185) (1.37573, -0.177855) (1.41051, -0.17391) (1.44532, -0.17032) (1.48017, -0.167053) (1.51504, -0.164083) (1.54994, -0.161381) (1.58485, -0.158922) (1.61978, -0.156685) (1.65472, -0.154648) (1.68967, -0.152793) (1.72463, -0.151101) (1.7596, -0.149559) (1.79457, -0.148153) (1.82955, -0.14687) (1.86453, -0.145699) (1.89951, -0.144631) (1.9345, -0.143656) (1.96949, -0.142767) (2.00048, -0.142046)};
			\draw[thick]  plot[smooth]
			coordinates {(0.964165, -0.257058) (0.963455, -0.256353) (0.938852, -0.232889) (0.913295, -0.21047) (0.886775, -0.189198) (0.8593, -0.169175) (0.830898, -0.150491) (0.801616, -0.133218) (0.771523, -0.1174) (0.740706, -0.103045) (0.709261, -0.0901223) (0.677286, -0.078569) (0.644878, -0.0682946) (0.612121, -0.0591926) (0.579087, -0.0511492) (0.545837, -0.0440515) (0.512419, -0.0377925) (0.478871, -0.0322741) (0.445221, -0.0274088) (0.411493, -0.0231196) (0.377704, -0.0193393) (0.343868, -0.0160105) (0.309994, -0.0130836) (0.276091, -0.0105169) (0.242165, -0.00827506) (0.208221, -0.00632865) (0.174263, -0.00465392) (0.140292, -0.00323256) (0.106313, -0.00205225) (0.0723261, -0.00110855) (0.0383333, -0.00041084) (0.00933569, -0.0000463757)};

			\draw[radius=0.3mm,fill](0,0)circle;
			\draw[radius=0.3mm,fill](15:1)circle;
			\draw[radius=0.3mm,fill](-15:1)circle;
			
			\draw[dashdotted,thick,blue,->]  plot[smooth, tension=.7] coordinates {(0.9437,-0.5275) (1.0673,-0.1028) (0.8792,0.2305) (0.7849,0.5054)};
			
			\draw[thick,blue,fill]  (0.9437,-0.5275) circle[radius=0.2mm];
			\draw[thick,blue,fill]  (0.7849,0.5054+0.02) circle[radius=0.2mm];
			
			\node at(1.4,-0.6){\(3\)};
			\node at(1.2,-0.1){\(4\)};
			\node at(1.2,0.1){\(5\)};
			\node at(0,0.5){\(6\)};
			
			\node at(1.05,0.4){\(P\)};
			\node at(0.8,-0.3){\(P'\)};

			\end{tikzpicture}
			\qquad
			&\qquad
			\begin{tikzpicture}[scale=2]
			\draw[thick,double](0,0)--(2,0);
			\draw[radius=0.3mm,fill](0,0) circle;
			\draw[thick,double,radius=1cm](0,0)circle;
			\draw[thick,fill,radius=0.3mm](1,0)circle;
			
			\draw[dashdotted,thick,blue,->]  plot[smooth, tension=.7] coordinates {(0.9437,-0.5275) (1.0673,-0.1028) (0.8792,0.2305) (0.7849,0.5054)};
			
			\draw[thick,blue,fill]  (0.9437,-0.5275) circle[radius=0.2mm];
			\draw[thick,blue,fill]  (0.7849,0.5054+0.02) circle[radius=0.2mm];
			
			\end{tikzpicture}
		\end{tabular}

	\end{center}
	\caption{\label{fig:WKBdegeneration} Formal degeneration of the WKB graph for real energy and for \(r\to+i\infty\).
		At the degenerate picture on the right $P={P'}$ and the domains $1\sqcup 2$ and $4\sqcup 5$ collapse.
		In this case $\int\lambda dz\sim \sqrt{\tilde{z}}+ \frac{1}{\sqrt{\tilde{z}}}$.}
\end{figure}
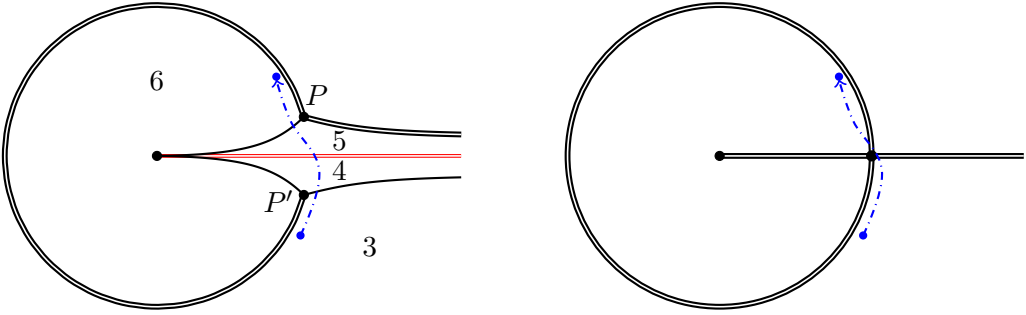

To find these WKB graphs one has to look at the expression for \(\lambda(z)dz\), where \(\pm\lambda\) are eigenvalues of the connection matrix $A(z)$ for the linear problem \eqref{eq:PainleveLax}, giving
\begin{equation}
\label{dS}
\lambda dz = \frac{dz}{z}\sqrt{tH-z-\frac{t}{z}}\approx \frac{rd \tilde{z}}{8 \tilde{z}}\sqrt{2+16 \frac{\tilde{\nu}}{r}-\tilde{z}-1/\tilde{z}}
\end{equation}
in the limit (\ref{eq:HAsymptotics}) with \(\tilde{z}=\frac{r^2}{64}z\), and where we put \(\mathcal{X}=1\). 
In Fig.~\ref{fig:WKBreal}, Fig.~\ref{fig:WKBimaginary} we vary \(\tilde{\nu}\) a little bit from \(\tilde{\nu}=-i \left( N+\frac12 \right)\), preserving topology of the graph and keeping positions of the saddle points.
At $r\to\infty$ the phase turns into \(\int \lambda dz \stackreb{r\to\infty}{\sim}  \frac{ir}{4}(\sqrt{\tilde{z}}+1/\sqrt{\tilde{z}})\), corresponding to degenerate picture on the right, Fig.~\ref{fig:WKBdegeneration}.


\subparagraph{Asymptotics of solutions.}

Let us now analyze the asymptotics of solutions, and introduce monodromy and transition matrices.
Bases of solutions in all six regions are defined by their WKB asymptotics, normalized to those coming from Airy-type asymptotics \eqref{eq:AiryNormalization} near the neighboring turning points:
\begin{equation}
\label{eq:30}
\Psi^{1,5,6}(z)\sim 
\begin{pmatrix}
\exp \left( \int_P^z \lambda dz \right) \ldots  \\ \exp \left( -\int_P^z \lambda dz \right) \ldots 
\end{pmatrix}, \quad
\Psi^{2,3,4}(z)\sim 
\begin{pmatrix}
\exp \left( \int_{P'}^z \lambda dz \right) \ldots  \\ \exp \left( -\int_{P'}^z \lambda dz \right) \ldots 
\end{pmatrix}.
\end{equation}
Branches \(\pm\int\lambda dz\) are chosen so that \(\int_P^z\lambda dz\) grows on the clockwise boundary of the anti-Stokes ray in the sector adjacent to the turning point \(P\) (the same condition for $P'$).
The signs in the exponentials \eqref{eq:30} are indicated by \(\left({}^+_-\right)\)
in Figs.~\ref{fig:WKBimaginary}, \ref{fig:dominantSubdominant}, where \(\pm\) signs indicate solutions, respectively, growing and decaying at the corresponding side of an anti-Stokes ray, when going \emph{out} of the turning point.

The bases of solutions around \(z\to0\) (denoted by ``in'' to specify precise direction) and \(z\to\infty\) (denoted by ``out'') are chosen as
\begin{equation}
\label{eq:28}
\Psi^{in}(z)=D_0\Psi^6(z),\quad \Psi^{out}=D_1\Psi^3(z),
\end{equation}
where $D_0$ and $D_1$ are certain diagonal ``normalization'' matrices to be specified below.
The corresponding monodromies \(\Psi^{in}(e^{2\pi i}z)=M_0\Psi^{in}(z)\) and \(\Psi^{out}(e^{2\pi i}z)=M_\infty\Psi^{out}(z)\) are given 
by monodromy matrices around $z=0$ and $z=\infty$, where $M_0$ is depicted by the solid line in Fig.~\ref{fig:WKBreal} and Fig.~\ref{fig:WKBimaginary}, 
and matrix \(\Psi^{out}(z)=V\Psi^{in}(z)\) corresponds to transition along the dashed line from ``in'' to ``out'' region at these pictures.

Fix now \(r=i \mathfrak{r}\in i \mathbb{R}_{>0}\) for definiteness,
the asymptotics of \(\Psi^{in,out}\) for \(\tilde{z}\in -i0+\mathbb{R}_{>0}\), corresponding to chosen in Fig.~\ref{fig:WKBimaginary} ``in'' and ``out'' directions, are given by
\begin{equation}
\label{eq:46}
\Psi^{in}(z)\sim 
\begin{pmatrix}
\exp\left(+\frac{\mathfrak{r}}{4\sqrt{\tilde{z}}}\right)\ldots  \\
\exp\left(-\frac{\mathfrak{r}}{4\sqrt{\tilde{z}}}\right)\ldots 
\end{pmatrix},\quad
\Psi^{out}(z)\sim 
\begin{pmatrix}
\exp\left(-\frac{\mathfrak{r}}{4}\sqrt{\tilde{z}}\right)\ldots  \\
\exp\left(+\frac{\mathfrak{r}}{4}\sqrt{\tilde{z}}\right)\ldots 
\end{pmatrix},
\end{equation}
and we shall see indeed, that monodromy matrices (formulas \eqref{eq:M01} and \eqref{eq:Minf1} below) actually add sub-dominant to the dominant solutions, also permuting them due to jumps in the square roots.

Turn now to solutions in the ``up'' and ``down'' regions,
used below to describe the spectral problem in degenerate limit, shown in Fig.~\ref{fig:WKBdegeneration}.
In this case classically allowed region is a short arc between the points \(P\) and \(P'\) (Stokes line, where both exponents oscillate), and classically forbidden region is the anti-Stokes line, also connecting these two points.
On the upper side we choose the region \(6\) from the two adjacent to \(P\), since solution in the region \(1\), collapsing with \(2\), does not survive in the degenerate limit.
Taking \(\Psi^3(z)\) for the ``down'' region, for the
asymptotics in degenerate limit one can write
\begin{equation}
\begin{gathered}
\label{eq:48}
\begin{pmatrix}
\psi_+^{up} \\ \psi_-^{up}
\end{pmatrix}=\Psi^{up}(z)=\Psi^6(z)\sim
\begin{pmatrix}
\exp \left(\frac{\mathfrak{r}}{8}\int\limits_{\phi_P}^{\phi} d\phi \sqrt{\cos\phi_P-\cos\phi}\right) \ldots  \\
\exp \left(-\frac{\mathfrak{r}}{8}\int\limits_{\phi_P}^{\phi} d\phi \sqrt{\cos\phi_P-\cos\phi}\right) \ldots 
\end{pmatrix},\\
\begin{pmatrix}
\psi_+^{down} \\ \psi_-^{down}
\end{pmatrix}=\Psi^{down}(z)=\Psi^3(z)\sim
\begin{pmatrix}
\exp \left(\frac{\mathfrak{r}}{8}\int\limits_{\phi_{P'}}^{\phi} d(-\phi) \sqrt{\cos\phi_{P'}-\cos\phi}\right) \ldots  \\
\exp \left(-\frac{\mathfrak{r}}{8}\int\limits_{\phi_{P'}}^{\phi} d(-\phi) \sqrt{\cos\phi_{P'}-\cos\phi}\right) \ldots
\end{pmatrix},
\end{gathered}
\end{equation}
where we have parameterized \(\tilde{z}=e^{i\phi}\), while \(\phi_P\) and \(\phi_{P'}\) are the angles of the points \(P\) and \({P'}\) themselves.
Solutions with these asymptotics are related by \(\Psi^{down}=T_{i\mathbb{R}}\Psi^{up}\), with the transition matrix along the dash-dotted line in Fig.~\ref{fig:WKBimaginary} (to be given by \eqref{eq:45} below).

Similar analysis can be performed for \(r\in \mathbb{R}\), then 
the allowed and forbidden regions will replace each other, see \cite{He:2010xa,DU}.

\subsection{Cluster parameterization of monodromies} 
\label{ss:cluster}

Let us now present the explicit expressions for the monodromy and transition matrices, using the definitions, collected in Appendix~\ref{sec:WKB}.
The monodromy matrix around zero, according to  Fig.~\ref{fig:WKBreal} and
Fig.~\ref{fig:matrices} together with \eqref{eq:WKBmatrices}, is given by
\begin{equation}
\label{eq:M0}
M_0 \sim LX(x)LX(y)L.
\end{equation}
Actually, in this way monodromy matrix is defined only up to conjugation by some diagonal matrix \eqref{Da}, so we choose
\begin{equation}
\label{eq:M01}
M_0 = D_0 LX(x)LX(y)LD_0^{-1} =
-\begin{pmatrix}
0 & i\\
i & \;\;\; \frac1{xy}+\frac{x}{y}+xy
\end{pmatrix},
\end{equation}
with $D_0 = D(e^{i\pi/4}\sqrt{x}\sqrt{y})$.

Transition matrix from zero to infinity, as shown by dashed arrow in Fig.~\ref{fig:WKBreal}, is
\begin{equation}
V\sim LX(-x)LL = LX(-x)R,
\end{equation}
and this matrix conjugates \(M_0\) to \(M_{\infty}\):
\begin{equation}
M_{\infty}\sim VM_0V^{-1} = RX(y)RX(x)R.
\end{equation}
We actually normalize \(M_{\infty}\) to be
\begin{equation}
\label{eq:Minf1}
M_{\infty}=D_1RX(y)RX(x)RD_1^{-1}=
-\begin{pmatrix}
\frac1{xy}+\frac{x}{y}+xy &\;\;\;  i\\
i & 0
\end{pmatrix},
\end{equation}
conjugated by $D_1 = D(e^{-i\pi/4}\sqrt{x}\sqrt{y})$, so that
\begin{equation}
\label{eq:Vcluster}
V=D_1LX(-x)RD_0^{-1}=
\begin{pmatrix}
\frac{i}{x} & y\\
-\frac{1+x^{-2}}{y} & \frac{i}{x}.
\end{pmatrix}.
\end{equation}
Finally, let us present the expression for transition matrix $T$ between two WKB regions with growing/decaying solutions.
This matrix describes the relation between solutions with given WKB asymptotics:
\begin{equation}
\label{eq:50}
\begin{pmatrix}
\psi_+^{up} \\ \psi_-^{up}
\end{pmatrix}=
T 
\begin{pmatrix}
\psi_+^{down} \\ \psi_-^{down}
\end{pmatrix},
\end{equation}
(as shown by dash--dotted line in Fig.~\ref{fig:WKBreal} and Fig.~\ref{fig:WKBimaginary}). For real $r\in \mathbb{R}$ from Fig.~\ref{fig:WKBreal} one
gets
\begin{equation}
\label{eq:T1}
T_{\mathbb{R}} \sim RX(x)L=
\begin{pmatrix}
x & \;\;\;  -\frac{1+x^2}x\\
x &  \;\;\; -x
\end{pmatrix},
\end{equation}
while for imaginary \(r\in i\mathbb{R}\) it follows from Fig.~\ref{fig:WKBimaginary}, that
\begin{equation}
\label{eq:45}
T_{i\mathbb{R}}\sim LX(x)R=
\begin{pmatrix}
\frac1{x} & \;\;\; -\frac1{x}\\
\frac{1+x^2}{x} & \;\;\; -\frac1{x}
\end{pmatrix}.
\end{equation}
Actually, this is the only transition matrix, which differs in real and imaginary cases, since it connects different regions in these two cases.
Other matrices are the same, but the path corresponding to \(V\) is different, see Fig.~\ref{fig:WKBreal},~\ref{fig:WKBimaginary}, and also discussion around~\eqref{eq:Dehn}.
As is seen from Fig.~\ref{fig:WKBdeformation}, the picture deforms continuously without flips, and therefore all other matrices \eqref{eq:M01}, \eqref{eq:Minf1} and \eqref{eq:Vcluster} remain the same.

Let us now identify the WKB parameters, or cluster coordinates, with common parameterization used before in the paper.
The simplest way is to extract from \cite{GL} the monodromy matrices around zero and infinity
\begin{equation}
\label{eq:47}
M_0=
\begin{pmatrix}
0 & -i\\
-i & 2\cos2\pi\sigma
\end{pmatrix},\quad
M_{\infty}=
\begin{pmatrix}
2\cos2\pi\sigma & -i\\
-i & 0
\end{pmatrix},
\end{equation}
which are slightly different from \cite{GL} by conventions: here we act by monodromy
matrices from the left, consider non--B\"acklund--transformed system, and added extra diagonal conjugation. For such conventions transition matrix from zero to infinity,
satisfying $VM_0V^{-1}=M_{\infty}$, acquires the form
\begin{equation}\begin{gathered}
\label{eq:Vnurho}
V=
\frac1{\sin2\pi\sigma}
\begin{pmatrix}
\sin2\pi\eta & i\sin2\pi(\eta-\sigma)\\
-i\sin2\pi(\eta+\sigma) & \sin2\pi\eta
\end{pmatrix} =
\\
=\begin{pmatrix}
e^{\pi\nu} & -ie^{4\pi i\rho-\pi\nu}\left( 1-e^{2\pi\nu} \right)\\
-i e^{\pi\nu-4\pi i\rho} & e^{\pi\nu}
\end{pmatrix},
\end{gathered}\end{equation}
where we have also used (\ref{eq:monodromyMapping}) to express it in terms of \(\nu\) and \(\rho\).

Now it is easy to compare these formulas with monodromies and transition matrices  from the previous section. 
Comparing (\ref{eq:Vnurho}) and (\ref{eq:Vcluster}), we immediately express the cluster variables in terms of \(\nu, \rho\):
\begin{equation}
\label{eq:nurhocluster}
x=ie^{-\pi\nu},\quad y=-ie^{4\pi i\rho-\pi\nu}\left(1-e^{2\pi\nu}\right)=2\pi e^{4\pi i\check{\rho}},
\end{equation}
(or \(\nu, \check{\rho}\) due to \eqref{eq:rhoCheckRho}),
so that $d\rho\wedge d\nu =d\check{\rho}\wedge d\nu \sim \frac{dx}{x}\wedge  \frac{dy}{y}$,
and unlike \(e^{4\pi i\rho}\), \(e^{4\pi i\check{\rho}}\) is a true cluster variable, related to corresponding WKB graph.
Notice also, that as follows from \eqref{eq:M01} and \eqref{eq:Minf1}, the invariant of the diagonal conjugation
\begin{equation}
-\mathrm{Tr} M_0 = -\mathrm{Tr} M_\infty = xy + \frac{x}{y} + \frac{1}{xy}
\end{equation}
coincides with the Hamiltonian of simplest relativistic Toda chain written in cluster variables.

\subsection{Quantization conditions}

\subparagraph{\(t\to 0\).}

The cos-Mathieu equation (\ref{eq:sinMathieu}) is solved by the quasi-periodic functions \(\tilde{Y}(x)=e^{i\sigma x}u(x)\), where \(u(x+2\pi)=u(x)\).
It is clear that quasi-periodicity factor is the same as monodromy around \(0\) or around \(\infty\), so to find the dependence of \(\mathcal{E}(\sigma,t)=tH(t,\sigma)\) one can just use \eqref{eq:confBlockDerivatives}.

The cosh-Mathieu equation (\ref{eq:sinhMathieu}) is different, since potential there is confining, and
one can look for the energy levels of this potential.
To do this it is necessary to find solutions of the linear equation, which decay at both infinities \(y\to\pm\infty\), or \(z\to 0\) and \(z\to\infty\) in the initial variable.
Starting from solution of the linear system, decaying near \(z\to 0\), it maps to basis of decaying and growing solutions at \(z\to\infty\) by the matrix \eqref{eq:Vnurho}.

Generally, decaying solution at the origin maps to the linear combination of growing and decaying solutions around infinity, but when the diagonal matrix elements vanish, i.e. \(2\eta\in \mathbb{Z}\), one gets only decaying solution at \(z\to\infty\).
Hence, to get normalizable solution of \eqref{eq:sinhMathieu} one has to impose the condition \(2\eta=k\in \mathbb{Z}\), see also \cite{Grassi:2019coc,BGG}.
Taking into account (\ref{eq:LagrangianSubmanifold}) we see that this condition is nothing but the quantization condition \cite{NS1}
\begin{equation}
\label{quant0}
\frac{\partial \mathcal{F}}{\partial \sigma} \in 2\pi i \mathbb{Z}.
\end{equation}

\subparagraph{\(t\to\infty\).}

Let us now rewrite the matrices \eqref{eq:T1} and \eqref{eq:45} using \eqref{eq:nurhocluster}, i.e.
\begin{equation}
\label{eq:T1nu}
T_{\mathbb{R}} =
\begin{pmatrix}
x & \;\;\; -\frac{1+x^2}x\\
x & \;\;\; -x
\end{pmatrix}=
\begin{pmatrix}
ie^{-\pi\nu} & ie^{\pi\nu}\left(1-e^{-2\pi\nu}\right)\\
ie^{\pi\nu} & -ie^{\pi\nu}
\end{pmatrix},
\end{equation}
and 
\begin{equation}
\label{T2nu}
T_{i\mathbb{R}}=
\begin{pmatrix}
\frac1{x} & \;\;\; -\frac1{x}\\
\frac{1+x^2}{x} & \;\;\; -\frac1{x}
\end{pmatrix}=
\begin{pmatrix}
-ie^{\pi\nu} & ie^{\pi\nu}\\
-ie^{\pi\nu}\left(1-e^{-2\pi\nu}\right) & ie^{\pi\nu}
\end{pmatrix}
\end{equation}
for imaginary \(r\in i\mathbb{R}_{>0}\). The latter one relates by \eqref{eq:50} the
growing and decaying solutions in two regions for $r\to +i\infty$, and
rewriting this in components one gets
\begin{equation}
\label{eq:52}
\psi_-^{up}=ie^{\pi\nu}\psi_-^{down}-ie^{\pi\nu}\left(1-e^{-2\pi\nu}\right)\psi_+^{down}.
\end{equation}
This means that decaying solution \(\psi_-^{up}\) continues to the decaying one \(\psi_-^{down}\) iff \(e^{-2\pi\nu}=1\), i.e.
\begin{equation}
\label{eq:WKBquantization}
\nu\in i\mathbb{Z}, 
\end{equation}
which is the analog of quantization \rf{quant0} at infinity.

\subsection{Vanishing of tau functions at infinity and spectral problems}

Let us now combine the WKB quantization condition (\ref{eq:WKBquantization}) with vanishing of the tau function, provided by 
expression (\ref{eq:infinityRho}) for \(\rho_{\star}\). Here is
a tricky point, related with possible singularities at \(\nu\in i \mathbb{Z}\). 

Starting to substitute \(\nu\in i \mathbb{Z}\),
we first notice that
\begin{equation}\begin{gathered}
\label{eq:58}
C_+(i\nu)=0 \text{ for } \nu\in i \mathbb{Z}_{>0},\\
C_-(i\nu)=0 \text{ for } \nu\in i \mathbb{Z}_{<0},
\end{gathered}\end{equation}
so that half of the structure constants vanish, and
both series for \(\tau^{\infty}_\pm\) terminate in one direction,
namely
\begin{equation}\begin{gathered}
\label{taucut}
\tau_+^{\infty}(\rho,-iN,r)=e^{-4\pi iN\rho}e^{\frac{r^2}{16}}r^{\frac1{4}}\sum\limits_{n\in \mathbb{Z}_{\geq 0}}C_+(-in)e^{-4\pi in\rho}e^{-inr}r^{-\frac12 n^2}\mathcal{B}^{\infty}(-in,r),\\
\tau_-^{\infty}(\check{\rho},iN,r)=(-1)^{\frac{N(N+1)}{2}}e^{-4\pi iN\check{\rho}}e^{\frac{r^2}{16}}r^{\frac1{4}}
\times \hspace{6cm}\\\hspace{3cm}\qquad\times
\sum\limits_{n\in \mathbb{Z}_{\geq 0}}C_-(in)e^{4\pi in(\check{\rho}+\frac{N}{4})}e^{inr}(-1)^{\frac12n(n-1)}r^{-\frac12 n^2}\mathcal{B}^{\infty}(in,r),
\end{gathered}\end{equation}
so that both \(\tau^{\infty}_\pm(\rho,\mp iN,r)\) are, up to a constant, actually given by their values at $\nu=iN=0$.
Initial tau functions (\ref{eq:tauILT}), \eqref{taucut}  are written for real \(r\in \mathbb{R}_{>0}\), but now we wish to continue them to the imaginary axis
\begin{equation}
\label{eq:iR}
r=e^{\frac{i\pi}{2}}\mathfrak{r}, \qquad \mathfrak{r}\in \mathbb{R}_{>0}.
\end{equation}
In order to write these tau functions we also introduce the new variables \(\rho^{i\mathbb{R}}\), \(\nu^{i\mathbb{R}}\), and \(\check{\rho}^{i\mathbb{R}}\), for the reason which is explained below.
Finally, the continued tau functions become
\begin{equation}\begin{gathered}
\label{eq:tauPM0}
\tau_+^{\infty}(\rho^{i\mathbb{R}} ,0,i\mathfrak{r})=e^{-\frac{\mathfrak{r}^2}{16}}(i\mathfrak{r})^{\frac1{4}}
\sum\limits_{n\in \mathbb{Z}_{\geq 0}}C_+(-in)e^{-4\pi in\rho^{i\mathbb{R}}}e^{n\mathfrak{r}}(i\mathfrak{r})^{-\frac12  n^2}\mathcal{B}^{\infty}(-in,i\mathfrak{r}),
\\
\tau_-^{\infty}(\check{\rho}^{i\mathbb{R}},0,i\mathfrak{r})=e^{-\frac{\mathfrak{r}^2}{16}}(i\mathfrak{r})^{\frac1{4}}
\sum\limits_{n\in \mathbb{Z}_{\geq 0}}
C_-(in)e^{4\pi i n\check{\rho}^{i\mathbb{R}}}e^{-n\mathfrak{r}}(-1)^{\frac12 n(n-1)}(i\mathfrak{r})^{-\frac12  n^2}\mathcal{B}^{\infty}(in,i\mathfrak{r}),
\end{gathered}\end{equation}
expansions over \(e^{\pm \mathfrak{r}}\). These formulas give us positions of the zeroes of the tau functions \(\tau_\pm^{\infty}\), so that
upper expression should be applied for \(\nu^{i\mathbb{R}} = -i-iN\in i\mathbb{Z}_{<0}\), whereas lower one works for \(\nu^{i\mathbb{R}}\in  i\mathbb{Z}_{\ge 0}\).

One should be careful at this point~\footnote{
We would like to thank A.~Grassi, whose questions and comments on the preliminary version of this paper allowed to clarify this point, see also sect.~\ref{sec:Alba} below.
} and check what happens with the monodromy data.
Since \(t=2^{-12}r^4\), multiplication of \(r\) by \(i\) leads to the same \(t\), but changes the monodromy:
\begin{equation}\begin{gathered}
\label{eq:Dehn}
\tau^{\infty}(\rho,\nu,e^{i\pi/2}\mathfrak{r})=
const\cdot\tau(\eta,\sigma, 2^{-12} e^{2\pi i}\mathfrak{r}^4)=
const\cdot\tau(\eta+\sigma,\sigma, 2^{-12}\mathfrak{r}^4).
\end{gathered}\end{equation}
To compensate this transformation~\footnote{
This is precisely an analog of the Dehn twist in degenerate situation, compare also Fig.~\ref{fig:WKBreal} and Fig.~\ref{fig:WKBimaginary}. In terms of the cluster variables have been used in sect.~\ref{ss:cluster} this is just a cluster mutation.
} we introduce \(\rho^{i\mathbb{R}}\), \(\nu^{i\mathbb{R}}\) and \(\check{\rho}^{i\mathbb{R}}\), defined so that 
\begin{equation}\begin{gathered}
\tau^{\infty}(\rho^{i\mathbb{R}},\nu^{i\mathbb{R}},e^{\frac{i\pi}{2}}\mathfrak{r})=
const\cdot\tau(\eta,\sigma, 2^{-12}\mathfrak{r}^4)\,,
\end{gathered}\end{equation}
these variables are just given by \(\eta\mapsto\eta-\sigma\) in \rf{eq:monodromyMapping} and \rf{eq:rhoCheckRho}, i.e.
\begin{equation}
\label{eq:nuRhoIR}
e^{4\pi i\rho^{i\mathbb{R}}}=\frac{\sin 2\pi(\eta-\sigma)}{\sin 2\pi\eta},\quad
e^{\pi\nu^{i\mathbb{R}}}=\frac{\sin 2\pi(\eta-\sigma)}{\sin 2\pi\sigma},\quad
e^{4\pi i\check{\rho}^{i\mathbb{R}}}=\frac{1}{2\pi i}\frac{\sin 2\pi(2\sigma-\eta)}{\sin 2\pi\sigma}.
\end{equation}
The spectral problem at the pole of solution \(w(r)\) acquires the form
\begin{equation}
\label{eq:60}
\left(-\partial_x^2+\frac{r^2}{32}\cos x\right)\tilde{Y}=\left(\frac{r^2}{32}+\frac{\tilde{\nu}r}{4}+\ldots \right)\tilde{Y}
\end{equation}
after one puts \(\mathcal{X}=1\) in (\ref{eq:HAsymptotics}) and substitutes it into (\ref{eq:sinMathieu}).
We find from r.h.s. of \eqref{eq:60} that real values of \(r\in\mathbb{R}\) correspond to a problem with energy near the top of cosine potential, sometimes called as ``magnetic'' region as follows from the picture of supersymmetric gauge theory, see also \cite{He:2010xa,BD}.
More interesting is the ``dyonic'' region, corresponding to the energies near the bottom of potential~\footnote{
It is actually hard to distinguish
``magnetic'' and ``dyonic'' here, moreover
usage of these notions is not consistent in the literature. Related problem is
that quantum energies  can have different sign from the classical one, see 
\rf{eq:sinhMathieu}.
} in \eqref{eq:60}, i.e. one should substitute \eqref{eq:iR} into \eqref{eq:60}.
There are two options to choose the sign in order to have positive energy shift \(\frac{\tilde{\nu}^{i\mathbb{R}}r}{4}+\ldots \) from the bottom of the potential in \eqref{eq:60}: for \(r\in i\mathbb{R}_{>0}\) one should take \(\tilde{\nu}^{i\mathbb{R}}\in i\mathbb{R}_{<0}\), whiles \(\tilde{\nu}^{i\mathbb{R}}\in i\mathbb{R}_{>0}\) for \(\mathfrak{r}\in i\mathbb{R}_{<0}\).

To clarify this point, let us compute the monodromy data for the solution in the limit \(\mathfrak{r}\to\infty\). From \eqref{eq:monodromyMapping} and \eqref{eq:infinityRho}:
\begin{equation}
\label{eq:nuirho}
\frac{\sin 2\pi(\eta-\sigma)}{\sin2\pi\eta}=e^{4\pi i\rho^{i\mathbb{R}}}\stackreb{\rho^{i\mathbb{R}}=\rho_{\star}^{i\mathbb{R}}}{\sim}\ e^\mathfrak{r}\mathfrak{r}^{-(N+\frac12)},\quad
\frac{\sin2\pi(\eta-\sigma)}{\sin2\pi\sigma}=e^{\pi\nu^{i\mathbb{R}}}=(-1)^{N+1},
\end{equation}
then~\footnote{
We notice here that solving second equation one has to choose \(\eta=2\sigma+\frac{N+1}{2}\).}
\begin{equation}
2\cos2\pi\sigma = e^{-4\pi i\rho^{i\mathbb{R}}} + e^{4\pi i\rho^{i\mathbb{R}}}(1-e^{-2\pi\nu^{i\mathbb{R}}}) \stackreb{\eqref{eq:nuirho}}{\sim}\ e^{-\mathfrak{r}}\mathfrak{r}^{(N+\frac12)}.
\end{equation}
Hence, there are real solutions for \(\sigma\) only if \(\mathfrak{r}\to+\infty\), while for \(\mathfrak{r}\to-\infty\) it becomes necessarily complex. In other words, for \(\mathfrak{r}<0\) solution has complex quasi-period, or grows exponentially and cannot be normalized.
We therefore choose \(\mathfrak{r}>0\).
Together with quantization condition \eqref{eq:WKBquantization} it gives
\begin{equation}
\label{eq:nuQuantization}
\tilde{\nu}^{i\mathbb{R}}=-i \left( N+\frac12 \right),\quad N\in \mathbb{Z}_{\geq 0}.
\end{equation}
It has clear interpretation in the \(\mathfrak{r}\to+\infty\) limit turning into standard energy quantization for harmonic oscillator.
As we already found, \eqref{eq:nuQuantization} persists for generic asymptotically large~\(\mathfrak{r}\).
This condition describes positions of the asymptotically narrow bands in the spectrum of equation
\begin{equation}
\label{eq:Schrodinger}
\left(-\partial_x^2-\frac{\mathfrak{r}^2}{32}\cos x \right)\tilde{Y}=\mathcal{E}_{\cos}(N,\mathfrak{r})\tilde{Y}
\end{equation}
in terms of quasiclassical conformal blocks. Namely,
\begin{equation}
\begin{gathered}
\label{eq:bandEnergyExpansion}
\mathcal{E}_{\cos}(N,\mathfrak{r})=\frac{\mathfrak{r}}{4}\frac{\partial \mathcal{F}^{\infty}\left(-iN-\frac{i}{2},i\mathfrak{r}\right)  }{\partial \mathfrak{r}}=\\
=-\frac{\mathfrak{r}^2}{32}+\frac{\left( N+\frac12 \right)\mathfrak{r}}{4}-
\frac{4 \left(  N+\frac12\right)^2+1}{32}-
\frac{4 \left( N+\frac12 \right)^3+3 \left( N+\frac12 \right)}{64\mathfrak{r}}-\\-
\frac{80 \left( N+\frac12 \right)^4+136 \left( N+\frac12 \right)^2+9}{1024\mathfrak{r}^2}-
\frac{528 \left( N+\frac12 \right)^5+1640 \left( N+\frac12 \right)^3+405 \left( N+\frac12 \right)}{4096 \mathfrak{r}^3}-\\-
\frac{9 \left( 224 \left( N+\frac12 \right)^6+1120 \left( N+\frac12 \right)^4+654 \left( N+\frac12 \right)^2+27 \right)}{8192\mathfrak{r}^4}-\\-
\frac{33728 \left( N+\frac12 \right)^7+249872 \left( N+\frac12 \right)^5+276004 \left( N+\frac12 \right)^3+41607 \left( N+\frac12 \right)}{65536\mathfrak{r}^5}-\ldots 
\frac{}{} 
\end{gathered}
\end{equation}
is the energy of the \(N\)-th exponentially narrow band.
This formula matches well-known expressions for this energy, see \cite{DU,NISTMathieu} and references therein.
Differently, this expansion can be considered as a perturbation theory series for cosine potential, considered as oscillator with the infinite series of perturbative corrections.

Consider now solution of the spectral problem \eqref{eq:Schrodinger}:
\begin{equation}
\begin{gathered}
\label{eq:80}
\tilde{Y}(N,\mathfrak{r},x)=e^{-\frac{\mathfrak{r} x^2}{16}}\psi(N,\mathfrak{r},x),
\\
\psi(N,\mathfrak{r},x)=\psi^{(0)}(N,\mathfrak{r},x)+\sum_{n>0}\sum_{M\ge 0}C(n,M)\mathfrak{r}^{-n}\psi^{(0)}(M,\mathfrak{r},x)\,,
\end{gathered}
\end{equation}
where the functions \(\psi(N,\mathfrak{r},x)\) are given by perturbative series in \(\mathfrak{r}^{-1}\), with the coefficients given by inherited from oscillator Hermite polynomials \(\{\psi^{(0)}(N,\mathfrak{r},x)\}\).
Let us now replace simultaneously \(x\to iy\), \(\mathfrak{r}=-\mathfrak{r}^{\circ}\), it gives the new wave function
\begin{equation}
\label{eq:82}
\tilde{Y}(N,-\mathfrak{r}^{\circ},iy)=e^{-\frac{\mathfrak{r}^{\circ} y^2}{16}}\psi(N,-\mathfrak{r}^{\circ},iy)
\end{equation}
with the exponential factor \(e^{-\frac{\mathfrak{r}^{\circ} y^2}{16}}\) still decaying at both real infinities \(y\to\pm\infty\). The perturbative series \rf{eq:80} in oscillator wave functions turns therefore into another perturbative series for the solution 
\(\tilde{Y}(N,-\mathfrak{r}^{\circ},iy)\) of the problem for \(\cosh\) potential:
\begin{equation}
\label{eq:83}
\left( -\partial_y^2+\frac{\mathfrak{r}^2}{32}\cosh y \right)\tilde{Y}(N,-\mathfrak{r},iy)=
\mathcal{E}_{\cosh}(N,\mathfrak{r})\tilde{Y}(N,-\mathfrak{r},iy),
\end{equation}
with the energy
\begin{equation}
\label{eq:84}
\mathcal{E}_{\cosh}(N,\mathfrak{r})=-\mathcal{E}_{\cos}(N,-\mathfrak{r})
=-\frac{\mathfrak{r}}{4}\frac{\partial \mathcal{F}^{\infty}\left(iN+\frac{i}{2},i\mathfrak{r}\right)  }{\partial \mathfrak{r}}
\end{equation}
given almost by the same formula as in \rf{eq:bandEnergyExpansion}.
Notice, that this expression corresponds to the arguments \(\nu^{i\mathbb{R}}=iN\) and \(r=i\mathfrak{r}\) with $N\in \mathbb{Z}_{\geq 0}$ of the quasiclassical block at infinity~\footnote{Another option is to substitute instead \(\nu^{i\mathbb{R}}=-i-iN\) for \(r=-i\mathfrak{r}\), but for
	\(r=-i\mathfrak{r}\), one has to perform the Dehn twist twice, like in \eqref{eq:Dehn}, in order to get the correct monodromy mapping.
}, i.e. the position of the pole of corresponding Painlev\'e solution is determined by vanishing of the second tau function \(\tau_-\).

\subsection{Meaning of \(\tau_-(\check{\rho}^{i\mathbb{R}},0,i\mathfrak{r})\)
and spectral determinant}
\label{sec:Alba}

Let us now discuss the validity of above formulas.
It is easy to see that expression \eqref{eq:bandEnergyExpansion} is a divergent asymptotic series by design, since it describes the spectrum for cosine-potential, defined only up to exponentially small corrections.
The tau function of \citep{ILT} ``at infinity'' \(\tau_+^{\infty}(\rho,\nu,r)\) is just an asymptotic series at \(r\to +\infty\) on the real line, and this turns to be enough to
follow the same logic as for $t\to 0$ in order to define an analog of quasiclassical conformal block, the blow-up equations etc.
The
same could be true for \(\tau^{\infty}_-(\check{\rho}^{i\mathbb{R}},0,i\mathfrak{r})\), when we perform the
$r\to i\mathfrak{r}$ rotation, since the expansion over \(e^{-n\mathfrak{r}}\) with positive $n$'s at 
$\mathfrak{r}\to +\infty$ has better chances to define a reliable expression, than
an expansion over oscillating \(e^{inr}\) from \citep{ILT}.
Surprisingly at first glance, in order to find solution to the spectral problem \eqref{eq:Schrodinger} one needs to use \(\tau^{\infty}_+(\rho_\ast(\mathfrak{r}),0,i\mathfrak{r})\), even though we do not believe that this tau function \emph{with fixed \(\rho\)} defines any reasonable asymptotic series at \(\mathfrak{r}\to+\infty\).

This seeming contradiction can be nevertheless resolved in the following way.
We use, first, the zeroes of \(\tau^{\infty}(\rho,\nu,r)\) to find expansions \emph{around the pole} of solution, when this pole goes to \(\infty\), and not around \(r\to\infty\) itself.
These are actually different limits, since in contrast to \(\left.\tau^{\infty}(\rho,\nu,r)\right|_{r\to\infty}\) with fixed \(\rho\) and \(\nu\), we first substitute \(\rho=\rho_{\star}(\nu,r)\sim -\frac{r}{4\pi}+\ldots \), and only then send \(r\to\infty\).
This substitution cancels ``dangerous'' exponentials, and allows one to ``run off'' the real line, where \(\tau^{\infty}(\rho,\nu,r)\) has been originally defined.
It means that even though \(\tau_+\) does not define a solution to Painlev\'e III\(_3\) around \(\mathfrak{r}\to+\infty\), one can extract from it the spectral problem solution \eqref{eq:bandEnergyExpansion} in terms of the expansion of quasiclassical conformal block.
It is not therefore surprising that this expansion coincides with the well-known formula for the cos-Mathieu equation.


The situation with the second tau function \(\tau^{\infty}_-(\check{\rho}^{i\mathbb{R}},0,i\mathfrak{r})\) is indeed better.
According to \rf{eq:84} its vanishing determines the pole of solution, corresponding to the spectral problem for the cosh-Mathieu equation, and this is actually a well-known one-parametric family of solutions, discussed in the literature \cite{Novokshenov1986}. Moreover, this second tau function \(\tau_-\) can be identified with the spectral determinant from \cite{Zamolodchikov:1994uw,AlbaFermiGas}, giving rise to a Fermi-gas representation for particular PIII\(_3\) tau function and irregular blocks at infinity. 

This one-parametric family corresponds to
 \(s=e^{4\pi i\eta}=1\), and therefore one has to put in \eqref{eq:nuRhoIR} 
\(\eta=\frac{k+1}{2}\), getting for \(k=0\) just \(\nu^{i\mathbb{R}}=0\) and
\begin{equation}
\label{eq:49}
e^{4\pi i\check{\rho}^{i\mathbb{R}}}=\frac{i}{\pi}\cos 2\pi\sigma.
\end{equation}
Then, for the lower tau function from \eqref{eq:tauPM0}, using the formula for the structure constants \eqref{eq:Cpm} we get the following expression:
\begin{equation}\begin{gathered}
\label{eq:57}
\tau_-(\check{\rho}^{i\mathbb{R}},0,i\mathfrak{r}) =
\\
= const\cdot \mathfrak{r}^{\frac1{4}}e^{-\frac{\mathfrak{r}^2}{16}}\sum_{n=0}^{\infty}
\left( \frac{\cos2\pi\sigma}{2\pi} \right)^n G(1+n)2^{-n(n-1)}(2\pi)^{\frac{n}{2}}  \mathfrak{r}^{-\frac{n^2}{2}} e^{-n\mathfrak{r}} \mathcal{B}^{\infty}(in,i\mathfrak{r}),
\end{gathered}\end{equation}
where it is natural to put \(const=1\) in order to compare with the expression from \cite{Zamolodchikov:1994uw,AlbaFermiGas}, which reads
\begin{equation}
\label{eq:tauZam}
\tau_{Zam}=e^{-\frac{\mathfrak{r}^2}{16}}\mathfrak{r}^{\frac1{4}}\sum_{n=0}^{\infty}
\left( \frac{\cos2\pi\sigma}{2\pi} \right)^n \frac{1}{n!}\int_{-\infty}^{\infty}\prod_{i=1}^n dx_i e^{-\mathfrak{r}\sum_{i=1}^n \cosh x_i}\prod_{i<j}^n\tanh^2\left(\frac{x_i-x_j}{2}\right).
\end{equation}
Identification of these two expansions suggest, that the irregular block at infinity for imaginary  integer \(\nu\in i\mathbb{Z}_{>0}\) can be written as an eigenvalue integral~\footnote{
The l.h.s. in this relation is a divergent asymptotic series, and therefore is defined only up to a non-perturbative completion,
while the r.h.s. is a well-defined function. Hence, it gives a result of perhaps the only meaningful summation of the l.h.s., since we know after \cite{Zamolodchikov:1994uw}, that \eqref{eq:tauZam} corresponds to actual solution to PIII\(_3\), not just an asymptotic series in \(r^{-1}\).
}:
\begin{equation}
\label{eq:blockIntegral}
\mathcal{B}^{\infty}(in,i\mathfrak{r})=\frac{2^{n(n-1)}\mathfrak{r}^{\frac{n^2}{2}}e^{n\mathfrak{r}}}{(2\pi)^{\frac{n}{2}}G(n+2)}
\int_{-\infty}^{\infty}\prod_{i=1}^n dx_i e^{-\mathfrak{r}\sum_{i=1}^n \cosh x_i}\prod_{i<j}^n\tanh^2\left(\frac{x_i-x_j}{2}\right),
\end{equation}
and its expansion at \(\mathfrak{r}\to\infty\) corresponds to computation of this integral by saddle point approximation, e.g.
in the leading  asymptotics 
\begin{equation}
\label{eq:85}
\left.\mathcal{B}^{\infty}(in,i\mathfrak{r})\right|_{\mathfrak{r}\to\infty}=\frac{\mathfrak{r}^{\frac{n^2}{2}}}{(2\pi)^{\frac{n}{2}}G(n+2)}
\int_{-\infty}^{\infty}\prod_{i=1}^n dx_i e^{-\frac{\mathfrak{r}}{2}\sum_{i=1}^n x_i^2 }\prod_{i<j}^n(x_i-x_j)^2 = 1
\end{equation}
one gets unity from a standard computation of the Gaussian matrix integral.
Two first coefficients \rf{eq:blockIntegral} of the expansion \rf{eq:57}, \rf{eq:tauZam} are known special functions, for
\(n=1\) it is given by zeroth Macdonald function:
\begin{equation}
\label{eq:79}
\mathcal{B}^{\infty}(i,i\mathfrak{r})=\frac{\sqrt{\mathfrak{r}}e^{\mathfrak{r}}}{\sqrt{2\pi}G(3)}\int_0^{\infty}e^{-\mathfrak{r} \cosh x}dx=\sqrt{\frac{2\mathfrak{r}}{\pi}}e^{\mathfrak{r}}K_0(\mathfrak{r}),
\end{equation}
while the result for \(n=2\) was found in \cite{Bonelli:2017ptp} in terms of the Meijer G-function:
\begin{equation}
\label{eq:86}
\mathcal{B}^{\infty}(2i,i\mathfrak{r})=\pi^{-\frac12}\mathfrak{r}^2e^{2\mathfrak{r}}G^{30}_{13}\left(\left.\begin{array}{ccc}&\frac{3}{2}&\\0&0&0 \end{array}\right|\mathfrak{r}^2 \right).
\end{equation}
These formulas for \(c=1\) blocks at infinity can be even generalized to other values of central charges, and we present several explicit examples in sect.~\ref{sec:qpPainleve} below.

\section{Relation to conformal field theory}
\label{sec:CFT}

In this section we find the identification between regularized action functional and irregular conformal blocks, see also \cite{Awata:2009ur,Awata:2010bz,Piatek:2014lma}.
We work with conformal field theory with the central charge \(c=1+6Q^2\), where \(Q=b+b^{-1}\), and then take the limit \(b\to 0\).

\subsection{BPZ equations}
Consider two degenerate fields at level 2, \(\phi_{(1,2)}\) and \(\phi_{(2,1)}\), with dimensions
\begin{equation}
\label{eq:6}
\Delta_{(1,2)}=-\frac12-\frac{3}{4b^2},\qquad \Delta_{(2,1)}=-\frac12-\frac{3b^2}{4}.
\end{equation}
They satisfy the null-vector equations
\begin{equation}\begin{gathered}
\label{eq:nullVectors}
b^2 \partial^2 \phi_{(1,2)}(w)+\left(\mathcal{L}_{-2}\phi_{(1,2)}\right)(w)=0,\\
b^{-2}\partial^2\phi_{(2,1)}(z)+\left(\mathcal{L}_{-2}\phi_{(2,1)}\right)(z)=0,
\end{gathered}\end{equation}
where
\begin{equation}\begin{gathered}
\label{eq:9}
\mathcal{L}_{-2}\phi(z)=\oint_z \frac{dy}{2\pi i}\frac{T(y)}{y-z}\phi(z)=
\\
= \phi(z)\left( \frac{L_{-1}}{z}+\frac{L_0}{z^2}+\frac{L_1}{z^3}+\ldots  \right) - \left( L_{-2}+L_{-3}z+L_{-4}z^2+\ldots  \right)\phi(z).
\end{gathered}\end{equation}
Their fusion
\begin{equation}
\label{eq:7}
\phi_{(2,1)}(z) \phi_{(1,2)}(w)\sim (z-w)^{-\frac12}\phi_{(2,2)}(w)
\end{equation}
gives the field \(\phi_{(2,2)}\)  of conformal dimension \(\Delta_{(2,2)}=-\frac{3}{4}Q^2\), so that
consistency of the dimensions in \eqref{eq:7} requires that
monodromy of \(\phi_{(2,1)}\) around \(\phi_{(1,2)}\) is always \(-1\).

Consider now the following correlation functions of these degenerate fields:
\begin{equation}\begin{gathered}
\label{eq:8}
F_4(t,t')=\langle b^{-2}t',b^{-1}\sigma|b^{-2}t,b^{-1}\sigma\rangle,\\
F_{5,h}(w;t,t')=\langle b^{-2}t',b^{-1}(\sigma\pm 1/2)|\phi_{(1,2)}(w)|b^{-2}t,b^{-1}\sigma\rangle,\\
F_{5,l}(z;t,t')=\langle b^{-2}t',b^{-1}(\sigma\pm 1/2)|\phi_{(2,1)}(z)|b^{-2}t,b^{-1}\sigma\rangle,\\
F_6(z,w;t,t')=\langle b^{-2}t',b^{-1}(\sigma\pm1/2)\pm1/2|\phi_{(1,2)}(w)\phi_{(2,1)}(z)|b^{-2}t,b^{-1}\sigma\rangle,
\end{gathered}\end{equation}
where \(|b^{-2}t,b^{-1}\sigma\rangle\) are the Gaiotto--Whittaker vectors \cite{Gaiotto}, \cite{MMM} in the Verma module with highest weight \(\Delta(\sigma)=\frac1{4}\left( b+1/b \right)^2-b^{-2}\sigma^2\) (see Appendix~\ref{ss:cft} for notations and some details):
\begin{equation}\begin{gathered}\label{expGW}
L_2|b^{-2}t,b^{-1}\sigma\rangle=0,\quad
 L_1|b^{-2}t,b^{-1}\sigma\rangle=b^{-2}t\,|b^{-2}t,b^{-1}\sigma\rangle,\\
|b^{-2}t,b^{-1}\sigma\rangle=\sum_Y \left( b^{-2}t \right)^{\Delta(\sigma)+|Y|}Q_{\Delta}(Y,[1]^{|Y|})^{-1}L_{-Y}|b^{-1}
\sigma\rangle.
\end{gathered}\end{equation}
We indicate $b$-dependence explicitly, since in what follows it will be used, that in the \(b\to 0\) limit
\begin{equation}\begin{gathered}
\label{eq:classicalLimit}
F_4(b^{-2}t,b^{-2}t')\simeq e^{b^{-2}\left( \frac1{4}\log t+f_4(t,t') \right)},\\
F_6(z,w;t,t')\simeq\sqrt{z}e^{b^{-2}\left( \frac12 \log w+\frac1{4}\log t+f_5(w,t,t') \right)}\Psi(z,w,t,t'),\\
F_{5,l}(z;t,t')\simeq \sqrt{z}e^{b^{-2}\left( \frac1{4}\log t+f_4(t,t') \right)}\psi(z,t,t'),\\
F_{5,h}(w;t,t')\simeq e^{b^{-2}\left( \frac12 \log w+\frac1{4}\log t+f_5(w,t,t') \right)},
\end{gathered}\end{equation}
where it is taken into account that ``light'' \(\phi_{(2,1)}\) does not affect the ``classical action'' in contrast to the ``heavy field'' \(\phi_{(1,2)}(w)\).

It follows from (\ref{eq:nullVectors}), \eqref{eq:9} that the correlators \eqref{eq:8} satisfy
\begin{equation}\begin{gathered}
\label{eq:1}
\left( z^2\partial_z^2+\frac{t}{z}+b^2t\partial_t+z-b^2z\partial_z\right)F_{5,l}(z;t)=0,\\
\left( b^4w^2\partial_w^2-b^2w\partial_w+\frac{t}{w}+b^2t\partial_t+w \right)F_{5,h}(w;t)=0,\\
\left( z^2\partial_z^2+\frac{t}{z}+b^2t\partial_t+z -b^2z\partial_z\right)F_6(z,w;t)+
\\
+ b^2\left( \frac{w^2\Delta_{(1,2)}}{(z-w)^2}+\frac{w^2\partial_w+2\Delta_{(1,2)}w}{z-w}+(\Delta_{(1,2)}+w\partial_w) \right)F_6(z,w;t)=0,
\end{gathered}\end{equation}
where for simplicity we put \(t'=1\), and in the leading order at \(b^2\to 0\)
under \eqref{eq:classicalLimit} they turn into
the Mathieu equation
\begin{equation}
\label{eq:MathieuCFT}
\left( (z\partial_z)^2 + \frac{t}{z}+t\partial_tf_4(t)+z  \right)\psi(z,t,t')=0,
\end{equation}
the Hamilton--Jacobi equation
\begin{equation}
\label{eq:5pointsHeavy}
\frac{\partial f_5(w,t)}{\partial t}+\frac1{t}\left( w\frac{\partial f_5(w,t)}{\partial w} \right)^2+\frac{1}{w}+\frac{w}{t}=0,
\end{equation}
and
\begin{equation}\small
\label{eq:6points}
\left( (z\partial_z)^2+\frac{t}{z}+t\partial_tf_5(w,t)+z-\frac{3 w^2}{4(z-w)^2}+\frac{w^2\partial_wf_5(w,t)-w}{z-w}+w\partial_w f_5(w,t)-\frac1{4} \right)\Psi(z,w,t)=0.
\end{equation}
Notice that equation (\ref{eq:5pointsHeavy}) can also be obtained from the condition of (\(-1\))-monodromy around \(z=w\) for the equation (\ref{eq:6points}), following from \eqref{eq:7}, and it is exactly
the Hamilton--Jacobi equation with the Hamiltonian \eqref{eq:P3Hamiltonian1}
of PIII\(_3\) equation.

It is well-known that solution to the Hamilton--Jacobi equation  is given by the action functional:
\begin{equation}
\label{eq:f5Action}
f_5(w,t)=\int^tdt \mathcal{L}(w',w,t)
 =\int^tdt\left( \frac{t}{4}\left( \frac{w'}{w} \right)^2-\frac{w}{t}-\frac1{w}\right),
\end{equation}
where \(w=w(t)\), and one can express the momentum \(p=p(t)\) as
\begin{equation}
\label{eq:HJMomentum}
\partial_wf_5(w,t)=p=\frac{tw'(t)}{2w(t)^2}.
\end{equation}
Substituting (\ref{eq:HJMomentum}) into (\ref{eq:6points}) we get precisely \eqref{eq:scalarEquation}, if correlator with two degenerate fields is identified with \(\tilde{Y}(z)\) from (\ref{eq:tildeY}) as
\begin{equation}
\label{eq:14}
\Psi(z,w(t)|t)=\mathrm{const}(t)\cdot\frac{Y_1(z)}{\sqrt{A_{12}(z)}}.
\end{equation}
By explicit comparison between (\ref{eq:Mathieu1}) and (\ref{eq:MathieuCFT}) we conclude that
\begin{equation}
\label{eq:11}
\partial_tf_4(t_{\star})=-H(t_{\star}),
\end{equation}
where \(t=t_{\star}\) is pole of the solution: \(w(t)\stackreb{t\to t_{\star}}{\to}\infty\).

\subsection{Regularization of the action functional}

From the CFT point of view it is natural to identify \(f_4(t_{\star})\) with the regularized limit of \(f_5(w(t),t)\) when \(t\to t_{\star}\), 
exactly as it has been done in (\ref{eq:action}).
To do this we study first more complicated limit \(w\to\infty\), namely, we study the fusion of the degenerate field with the Gaiotto--Whittaker state.

\paragraph{Irregular limit.}

Consider expansion \eqref{expGW} of
\begin{equation}
\label{eq:12}
F_{5,h}(w,t)=\langle W_{b^{-2}}|\phi_{(1,2)}(w)|b^{-2}t,b^{-1}\sigma\rangle=
\sum_Y \left( b^{-2}t \right)^{\Delta(\sigma)+|Y|}\langle W_{b^{-2}}|\phi_{(1,2)}(w)L_{-Y}|b^{-1}\sigma\rangle,
\end{equation}
where \(\langle W_{b^{-2}}|\) is the dual vector to \(|b^{-2},b^{-1}(\sigma\pm 1/2)\rangle\), satisfying \(\langle W_{b^{-2}}|L_{-2}=0\) together with \(\langle W_{b^{-2}}|L_{-1}=\langle W_{b^{-2}}|b^{-2}\).
First, let us take the matrix element with the highest weight vector.
It satisfies the BPZ equation:
\begin{equation}
\label{eq:13}
\left( b^2\partial_w^2-\frac1{w}\partial_w+\frac{1}{b^2w}+\frac{\Delta(\sigma)}{w} \right)\langle W_{b^{-2}}|\phi_{(1,2)}(w)|b^{-1}\sigma\rangle=0.
\end{equation}
To study the behavior of this matrix element at \(w\to\infty\) we substitute
\begin{equation}
\label{eq:15}
\langle W_{b^{-2}}|\phi_{(1,2)}(w)|b^{-1}\sigma\rangle=w^{\frac{2+b^2}{4b^2}}e^{\mp \frac{2i}{b^2}\sqrt{w}}\Phi(x),
\end{equation}
with \(x= \pm \frac{ib^2}{4 \sqrt{w}} \), where \(\Phi\) satisfies now the following equation
\begin{equation}
\label{eq:16}
\left((x \partial_x)^2 +(\alpha+\beta)x\partial_x-\partial_x+\alpha\beta\right)\Phi(x)=0,
\end{equation}
with
\begin{equation}
\label{eq:18}
\alpha=\frac{1}{2}-\frac{2\sigma}{b^2},\qquad \beta=\frac12+\frac{2\sigma}{b^2}.
\end{equation}
Equation \eqref{eq:16} is solved in terms of Bessel functions of \(w\sim x^{-2}\), but for our purposes we rather need its asymptotic expansion in \(x\):
\begin{equation}
\label{eq:17}
\Phi(x)={}_2F_0(\alpha,\beta,x)=\sum_{n=0}^{\infty}\frac{(\alpha)_n(\beta)_n}{n!}x^n.
\end{equation}
Other matrix elements are expressed through \eqref{eq:15} by
\begin{equation}\small
\begin{gathered}
\label{eq:19}
\langle W_{b^{-2}}|\phi_{(1,2)}(w)L_{-l}^{k_l}\ldots L_{-1}^{k_1}|b^{-2}t,b^{-1}\sigma\rangle=
\\
=(w^{1-l}\partial_w+\Delta_{(1,2)}(1-l)w^{-l})^{k_l}\ldots (w^{-1}\partial_w-\Delta_{(1,2)}w^{-2})^{k_2}(\partial_w+b^{-2})^{k_1}\langle W_{b^{-2}}|\phi_{(1,2)}(w)|b^{-1}\sigma\rangle =
\\
=w^{\frac{2+b^2}{4b^2}}e^{\mp \frac{2i}{b^2}\sqrt{w}}\prod_{j=l}^2\left(w^{1-j}\partial_w+\Delta_{(1,2)}(1-j)w^{-j}+\frac{2+b^2}{4b^2}w^{-j}\mp \frac{2i}{b^2}w^{1/2-j}\right)^{k_j}
\times\\
\times\left(b^{-2}+\frac{2+b^2}{4b^2}w^{-1}\mp \frac{2i}{b^2}w^{-1/2}+\partial_w\right)^{k_1}{}_2F_0\left(\alpha,\beta,\pm \frac{ib^2}{4 \sqrt{w}}\right)=
\\
=
w^{\frac{2+b^2}{4b^2}}e^{\mp \frac{2i}{b^2}\sqrt{w}}\left(b^{-2 k_1} \prod_{j=2}^{\infty}\delta_{k_j,0}+O \left( \frac1{\sqrt{w}} \right) \right).
\end{gathered}\end{equation}
It means that the analog of OPE at \(w\to\infty\) in the irregular case is
\begin{equation}
\label{eq:20}
\langle W_{b^{-2}}|\phi_{(1,2)}(w)\mathop{=}_{w\to\infty}
w^{\frac{2+b^2}{4b^2}}e^{\mp \frac{2i}{b^2}\sqrt{w}}\left(\langle W_{b^{-2}}|+O \left( \frac1{\sqrt{w}} \right) \right).
\end{equation}
This leads to the following relation for the correlation functions:
\begin{equation}
\label{eq:22}
F_{5,h}(w,t)\mathop{=}_{w\to\infty}w^{\frac{2+b^2}{4b^2}}e^{\mp \frac{2i}{b^2}\sqrt{w}} \left(F_4(t)+const+O \left( \frac1{\sqrt{w}} \right)\right),
\end{equation}
and, by (\ref{eq:classicalLimit}) in the \(b\to 0\) limit we get desired
\begin{equation}
\label{eq:5}
f_5(w,t)\mathop{=}_{w\to\infty}\mp 2i\sqrt{w}+ f_4(t)+const+O \left( \frac1{\sqrt{w}} \right).
\end{equation}
Using expansion (\ref{eq:poleExpansion}) this is rewritten as
\begin{equation}
\label{eq:f5TStar}
f_5(w(t),t)\mathop{\sim}_{t\to t_{\star}} \frac{2t_{\star}}{t-t_{\star}}+f_4(t)+const.
\end{equation}
Here we have chosen the upper sign, since conformal block corresponding to the lower sign is exponentially small for real \(b\), \(t\), and \(t_{\star}\), see \eqref{eq:poleExpansion}.

\paragraph{Regular limit.}

Formula~(\ref{eq:f5Action}) can be used only to compute the difference of \(f_5(w(t),t)\) for two different times.
It is very convenient to choose \(t=0\) as initial time.
In this limit \(w\sim -\kappa t^{2\sigma}\), and correlation function can be rewritten as
\begin{equation}\begin{gathered}
\label{eq:24}
F_5(w(t),t)\sim t^{\Delta(\sigma)}\langle b^{-1}(\sigma+ 1/2)|\phi_{(1,2)}\left(-\kappa t^{2\sigma}\right)|b^{-1}\sigma\rangle\sim
\\
\sim
 t^{\Delta(\sigma)} \left( -\kappa t^{2\sigma} \right)^{\Delta(\sigma-1/2)-\Delta(\sigma)-\Delta_{(1,2)}}=const\cdot t^{Q^2/4+\sigma^2/b^2+\sigma/b^2+\sigma}.
\end{gathered}\end{equation}
Now again switch to the limit \(b\to 0\) using (\ref{eq:classicalLimit}):
\begin{equation}
\label{eq:25}
f_5(w(t),t)\mathop{\simeq}_{t\to 0}\sigma^2\log t+const.
\end{equation}
Taking into account this and (\ref{eq:f5TStar}) we conclude that \(f_5(w(t),t)-\frac{2t_{\star}}{t-t_{\star}}-\sigma^2\log t\) has both limits, at \(t=0\) and at \(t=t_{\star}\), so one can write it as integral:
\begin{equation}
\label{eq:21}
\int_0^{t_{\star}} \left(\mathcal{L}(w',w,t)dt- d\left(\frac{2 t_{\star}}{t-t_\star}+\sigma^2\log t\right)\right)=
-\sigma^2\log t_{\star} + f_4(t_{\star})+ const.
\end{equation}

\subsection{Connection problem for quasiclassical conformal blocks}

Let us finally explain the CFT meaning of the formula \eqref{eq:classicalFusion}.
As we know, conformal blocks in the limit \(b\to 0\) behave as in \eqref{eq:classicalLimit} after appropriate rescaling:
\begin{equation}
\begin{gathered}
\label{eq:3}
F(\sigma,t)=\exp\left(b^{-2}\mathcal{F}(\sigma,t)+O(1)\right),\\
F^{\infty}(\tilde{\nu},t)=\exp\left(-b^{-2}\mathcal{F}^{\infty}(\tilde{\nu},t)+O(1)\right).
\end{gathered}
\end{equation}
As usual, we assume that either \(F(\sigma,t)\), or \(F^{\infty}(\tilde{\nu},r)\), form bases in the space of conformal blocks, labelled by \(\sigma\) and \(\tilde{\nu}\), respectively.
Since each of these sets forms a basis, they should be related by a linear transformation:
\begin{equation}
\label{eq:fusion1}
F(\sigma,t)=\int d\tilde{\nu} K(\sigma,\tilde{\nu}) F^{\infty}(\tilde{\nu},t).
\end{equation}
We assume now that the kernel has the same \(b\to 0\) behavior as conformal blocks do, 
\begin{equation}
\label{eq:27}
K(\sigma,\tilde{\nu})=\exp\left(b^{-2}(\mathcal{C}-\mathcal{S}(\sigma,\tilde{\nu}))+O(1)\right),
\end{equation}
and check that this assumptions is self-consistent.
If so, in the $b\to 0$ limit the integral in \eqref{eq:fusion1} can be found by saddle point computation.
It means that first one should find the position of a saddle point in \(\tilde{\nu}\) by solving stationarity equation on \(\tilde{\nu}_s\):
\begin{equation}
\label{eq:saddle1}
\frac{\partial \mathcal{F}^{\infty}(\tilde{\nu}_s,r)}{\partial \tilde{\nu}_s}=-\frac{\partial \mathcal{S}(\sigma,\tilde{\nu}_s)}{\partial \tilde{\nu}_s}=-4\pi\rho(\sigma,\tilde{\nu}_s),
\end{equation}
which coincides with one of the formulas from \eqref{Finf}, being actually a defining relation for the Malgrange divisor, i.e.,
\begin{equation}
\label{eq:4}
\tilde{\nu}_s(\sigma,r)=\tilde{\nu}_{\star}(\sigma,r).
\end{equation}
The meaning of function \(\tilde{\nu}_{\star}(\sigma,r)\) is the following: the Malgrange divisor is a 2-dimensional submanifold in the 3-dimensional \(\mathcal{M}\times \mathbb{C}^{*}_t\).
\(\mathcal{M}\times \mathbb{C}^{*}_t\) is locally described by three coordinates, for example \(\sigma, \tilde{\nu}, t\), so that the divisor can be obtained just by expressing  one coordinate as a function of two others, and one of such expressions is given by \(\nu_{\star}(\sigma,r)\).

Completing the computation of the integral \eqref{eq:fusion1} we finally get
\begin{equation}
\label{eq:29}
\mathcal{F}(\sigma,t)+\mathcal{F}^{\infty}\left(\tilde{\nu}_{\star}(\sigma,r),r\right)+\mathcal{S}\left(\sigma,\tilde{\nu}_{\star}(\sigma,r)\right)=\mathcal{C},
\end{equation}
which coincides with \eqref{eq:classicalFusion}.

\section{Quantum Painlev\'e III\(_3\) at infinity and arbitrary central charge}

\label{sec:qpPainleve}

Up to now we have considered only the irregular conformal blocks at infinity with central charges \(c=1\) \eqref{eq:infinityBlock}, proposed in \cite{ILT}, and constructed  its quasiclassical analog \eqref{eq:infinityClassicalBock} with \(c\to\infty\).
These two expressions, \eqref{eq:infinityBlock} and \eqref{eq:infinityClassicalBock}, are naturally supposed to be just two avatars a generic irregular block at infinity with arbitrary \(c=1+6Q^2 = 1+6\frac{(\epsilon_1+\epsilon_2)^2}{\epsilon_2\epsilon_2}\), or two arbitrary \(\epsilon_{1,2}\)-parameters of \(\Omega\)-background.
We propose a definition of such generic irregular block below in this section.

In order to do this, let us remind, first, that \(c=1\) irregular blocks at infinity \eqref{eq:infinityBlock} were found in \cite{ILT} from the requirement that their Fourier transform \eqref{eq:tauILT} gives solution to Painlev\'e III\(_3\).
To generalize this idea for arbitrary central charges we use, after \cite{BGM1}, that generic \(\Omega\)-backgrounds correspond in the context of isomonodromy/CFT correspondence to the
\emph{quantization}~\footnote{
Not to be confused with the \(q\)-deformation.
Two \(\epsilon_{1,2}\)-parameters are expressed through the difference parameter \(q\) and multiplicative Planck constant \(p\).
The limits \(q\to 1\)  and  \(p\to 1\) are independent, so that one can get both quantum or classical differential equation, as well as quantum or classical \(q\)-difference equation.
Quantum equations of Painlev\'e type are already known for quite a long time, see \cite{hasegawa2007quantizing,kuroki2008quantum,nagoya2012symmetries}.
} of the original deautonomized integrable system. 

Hence, in order to construct general conformal blocks we are going to switch from sect.~\ref{sec:equationSystem} to quantum Painlev\'e III\(_3\) equation.
It is convenient to start from basic results of \cite{BGM1}, concerning quantum \(q\)-difference Painlev\'e III\(_3\), and then take the \(q\to 1\) limit of the minimal set of relations, which are sufficient to define generic irregular blocks at infinity.

\subsection{\(q\)-Painlev\'e III\(_3\) and \(q\to 1\) limit}

Quantum \(q\)-Painlev\'e  III\(_3\) equation \cite{BGM1} is actually a system of two algebraic relations on the operator-valued function \(\hat{G}(Z)\):
\begin{equation}
\begin{gathered}
\label{eq:qpPainleve}
\left\{\begin{aligned}
\hat{G}(Zq^{-1})^{\frac12}\hat{G}(Zq)^{\frac12}=\frac{\hat{G}(Z)+pZ}{\hat{G}(Z)+p},\\
\hat{G}(Z)\hat{G}(q^{-1}Z)=p^4\hat{G}(q^{-1}Z)\hat{G}(Z).
\end{aligned}\right.
\end{gathered}
\end{equation}
Its solution is given by ratio of the quantum tau functions
\begin{equation}
\begin{gathered}
\label{eq:qpSolution}
\hat{G}(Z)^{\frac12} = \pm p^{\frac12}Z^{\frac1{4}}\mathcal{T}_3^{-1}\mathcal{T}_1=
\\
=\pm ip^{\frac12}Z^{\frac1{4}}\left(\sum_{n\in \frac12+\mathbb{Z}} \hat{s}^n 
\mathsf{F}_{5d}\left(\hat{u}q_2^{4n},q_1q_2^{-1},q_2^2|Z\right) \right)^{-1}
\sum_{n\in \mathbb{Z}}\hat{s}^n\mathsf{F}_{5d}\left(\hat{u}q_2^{4n},q_1q_2^{-1},q_2^2|Z\right),
\end{gathered}
\end{equation}
where quantum tau functions look as
\begin{equation}\begin{gathered}
\label{eq:qpTauDef}
\mathcal{T}_1=\hat{a}\sum_{n\in \mathbb{Z}}\hat{s}^n\mathsf{F}_{5d}\left(\hat{u}q_2^{4n},q_1q_2^{-1},q_2^2|Z\right),\quad
\mathcal{T}_3=i\hat{a}\sum_{n\in \frac12+\mathbb{Z}}\hat{s}^n\mathsf{F}_{5d}\left(\hat{u}q_2^{4n},q_1q_2^{-1},q_2^2|Z\right),
\end{gathered}\end{equation}
with \(\mathsf{F}_{5d}\) being \(q\)-deformed irregular conformal blocks, or 5d partition functions of supersymmetric pure \(SU(2)\) gauge theory, including also classical and perturbative part, \(\hat{u}\) and \(\hat{s}\) are multiplicative quantum canonical variables, while \(q_1q_2^{-1}\) and \(q_2^2\) are parameters of the \(\Omega\)-background, so that
\begin{equation}
\label{eq:59}
\hat{u}\hat{s}=p^4\hat{s}\hat{u},\quad q=q_2^2,\quad p^2=q_1q_2, \quad 
q_2^2\hat{a}=\hat{a}q_1^{-1}q_2,\quad Z\hat{b}=q_1q_2\hat{b}Z,
\end{equation}
i.e. \(\hat{a}\) shifts \(\Omega\)-background parameters \(\{q_i\}\), and therefore the central charge, whereas \(\hat{b}\) shifts the \(q\)-isomonodromic time \(Z\),
and in the classical \(p=1\) limit one comes back to the self-dual \(\Omega\)-background.
To prove that \eqref{eq:qpSolution} is actually a solution of \eqref{eq:qpPainleve}, the quantum tau functions \eqref{eq:qpTauDef}
should satisfy some bilinear relations~\cite{BGM1}:
\begin{equation}\begin{gathered}
\label{eq:qpTauBlowUp}
\underline{\mathcal{T}_1}\overline{\mathcal{T}_1}=\mathcal{T}_1^2+p^2Z^{1/2}\mathcal{T}_3^2,\quad
\underline{\mathcal{T}_3}\overline{\mathcal{T}_3}=\mathcal{T}_3^2+p^2Z^{1/2}\mathcal{T}_1^2,
\quad \overline{\mathcal{T}_1}\mathcal{T}_1=\mathcal{T}_1\overline{\mathcal{T}_1},
\quad \overline{\mathcal{T}_3}\mathcal{T}_3=\mathcal{T}_3\overline{\mathcal{T}_3},
\end{gathered}\end{equation}
where the time shift operations are given by
\begin{equation}
\begin{gathered}
\label{eq:78}
\overline{(q_1,q_2,\hat{u},\hat{s},Z,\hat{a},\hat{b})}=
(q_1,q_2,\hat{u},\hat{s},q_2^2Z,\hat{a}\hat{b},\hat{b}),
\\
\underline{(q_1,q_2,\hat{u},\hat{s},Z,\hat{a},\hat{b})}=
(q_1,q_2,\hat{u},\hat{s},q_2^{-2}Z,\hat{a}\hat{b}^{-1},\hat{b}).
\end{gathered}
\end{equation}
Substituting \eqref{eq:qpTauDef} into \eqref{eq:qpTauBlowUp} and collecting coefficients at \(\hat{s}^k\) we get~\footnote{
There is a small distinction between odd and even \(k\),
but at the present case one can play with integer and half-integer powers and pack all equations into these two, see \cite{BShch3} for details.}
two bilinear equations for \(\mathsf{F}_{5d}\):
\begin{equation}\begin{gathered}
\label{eq:5dBlowUp1}
\sum_{2n\in \mathbb{Z}}\mathsf{F}_{5d}\left(\hat{u}q_1^{4n},q_1^2,q_1^{-1}q_2|q_1^2Z\right)
\mathsf{F}_{5d}\left(\hat{u}q_2^{4n},q_1q_2^{-1},q_2^2|q_2^2Z\right)=
\\=
\left(1-p^2Z^{1/2}\right)\sum_{2n\in \mathbb{Z}}\mathsf{F}_{5d}\left(\hat{u}q_1^{4n},q_1^2,q_1^{-1}q_2|Z\right)
\mathsf{F}_{5d}\left(\hat{u}q_2^{4n},q_1q_2^{-1},q_2^2|Z\right)
\end{gathered}\end{equation}
and
\begin{equation}\begin{gathered}
\label{eq:5dBlowUp2}
\sum_{2n\in \mathbb{Z}}\mathsf{F}_{5d}\left(\hat{u}q_1^{4n},q_1^2,q_1^{-1}q_2|q_1Z\right)
\mathsf{F}_{5d}\left(\hat{u}q_2^{4n},q_1q_2^{-1},q_2^2|q_2Z\right)=
\\=
\sum_{2n\in \mathbb{Z}}\mathsf{F}_{5d}\left(\hat{u}q_1^{4n},q_1^2,q_1^{-1}q_2|q_1^{-1}Z\right)
\mathsf{F}_{5d}\left(\hat{u}q_2^{4n},q_1q_2^{-1},q_2^2|q_2^{-1}Z\right).
\end{gathered}\end{equation}
Equation \eqref{eq:5dBlowUp1} was conjectured in \cite{BShch3}, and actually was the motivation for quantum deformation, other equations from \cite{BGM1} are now proven in \cite{Shch1}.
Equation \eqref{eq:5dBlowUp2} follows from commutativity of the tau functions,
and becomes trivial for \(q_2=q_1^{-1}\), but sill necessary in the general situation.
Relations \eqref{eq:5dBlowUp1}, \eqref{eq:5dBlowUp2} are called \(\mathbb{C}^2/\mathbb{Z}_2\) blow-up equations, not to be confused with the original \cite{NY1} Nakajima--Yoshioka \(\mathbb{C}^2\) blow-up equations~\footnote{
	It has been found however in \cite{BShch4,Shch1}, that sometimes equations of one type follow from equations of another type.
}.
To see the difference one can check, that in the commutative $p\to 1$ limit \(\mathbb{C}^2/\mathbb{Z}_2\) equations turn into bilinear relations for the \(c=1\) conformal blocks, being equivalent to Painlev\'e equations, while the Nakajima--Yoshioka equations turn into some relations including \(c=1\) and \(c\to\infty\) conformal blocks, as we discussed before.

In the 4d limit of \eqref{eq:qpPainleve}
\begin{equation}
\label{eq:51}
q_1=e^{\epsilon_1 l_5},\quad q_2=e^{\epsilon_2 l_5},\quad u=e^{2\sigma l_5},\quad Z=l_5^4t,\quad \hat{G}=l_5^2\hat{w},\quad l_5\to 0,\quad \epsilon_2<0,\quad\epsilon_1>0
\end{equation}
the blow-up equations \eqref{eq:5dBlowUp1} \eqref{eq:5dBlowUp2} acquire the form
\begin{equation}
\begin{gathered}
\label{eq:blowup4d1}
\sum_{2n\in \mathbb{Z}}
\mathrm{D}^2_{2\epsilon_1,2\epsilon_2}\big(\mathsf{F}(\sigma+2n\epsilon_1,2\epsilon_1,\epsilon_2-\epsilon_1|t),
\mathsf{F}(\sigma+2n\epsilon_2,\epsilon_1-\epsilon_2,2\epsilon_2|t)\big)
=\\=
-2t^{1/2}\sum_{2n\in \mathbb{Z}}
\mathsf{F}(\sigma+2n\epsilon_1,2\epsilon_1,\epsilon_2-\epsilon_1|t)
\mathsf{F}(\sigma+2n\epsilon_2,\epsilon_1-\epsilon_2,2\epsilon_2|t),
\end{gathered}
\end{equation}
and
\begin{equation}
\begin{gathered}
\label{eq:blowup4d2}
\sum_{2n\in \mathbb{Z}}
\mathrm{D}^1_{2\epsilon_{1,2},2\epsilon_2}\big(\mathsf{F}(\sigma+2n\epsilon_1,2\epsilon_1,\epsilon_2-\epsilon_1|t),
\mathsf{F}(\sigma+2n\epsilon_2,\epsilon_1-\epsilon_2,2\epsilon_2|t)\big)=0,
\end{gathered}
\end{equation}
respectively, with \(\mathrm{D}^k_{2\epsilon_1,2\epsilon_2}\) being the logarithmic non-symmetric Hirota derivatives in \(t\) defined via
\begin{equation}
\label{Ddef}
f\left(e^{2\epsilon_1\xi}t\right)g\left(e^{2\epsilon_2\xi}t\right)=\sum_{k=0}^{\infty}\frac{\xi^k}{k!}\mathrm{D}^k_{2\epsilon_1,2\epsilon_2}\big(f,g\big)(t).
\end{equation}
Equations~\eqref{eq:blowup4d1},~\eqref{eq:blowup4d2} were first derived yet in \cite{BShch1},
they are equivalent to Hirota bilinear relations on the (quantum) tau function and its B\"acklund--transformed.
It turns out that solving these two equations one can find iteratively coefficients of conformal blocks for \(\epsilon_2\neq -\epsilon_1\).

The quantum difference equation \eqref{eq:qpPainleve} turns in the 4d limit into the quantum differential equation:
\begin{equation}
\label{eq:pPainleve}
\left\{\begin{aligned}
4\epsilon_2^2t\frac{d}{d t}\left(t\frac{d \hat{w}}{d t}\cdot\hat{w}^{-1}\right)=\frac{2t}{\hat{w}}-2\hat{w},\\
\left[\hat{w}^{-1},t \frac{d \hat{w}}{dt}\right]=2(\epsilon_1+\epsilon_2).
\end{aligned}\right.
\end{equation}
Since classical version of this equation had expansion \eqref{eq:infinitySolution} at \(t\to\infty\), we expect a similar formula in the quantum case,
namely, that solution to \eqref{eq:pPainleve} is written as
\begin{equation}\begin{gathered}
\label{eq:pSolutionInf}
\hat{w}(r)^{\frac12}=\pm \frac{r}{8}\left( \mathcal{T}_3^{\infty} \right)^{-1}
\mathcal{T}_1^{\infty},
\end{gathered}\end{equation}
where
\begin{equation}\begin{gathered}
\label{eq:qTauInf}
\mathcal{T}_1^{\infty}=\hat{a}\sum_{n\in\mathbb{Z}}e^{4\pi i \hat{\rho}} 
\mathsf{F}^{\infty}\left(\hat{\nu}+2in\epsilon_2,\epsilon_1-\epsilon_2,2\epsilon_2|r\right),\\
\mathcal{T}_3^{\infty}=
\hat{a}\sum_{n\in\mathbb{Z}}(-1)^ne^{4\pi i \hat{\rho}} 
\mathsf{F}^{\infty}\left(\hat{\nu}+2in\epsilon_2,\epsilon_1-\epsilon_2,2\epsilon_2|r\right),
\end{gathered}\end{equation}
where the canonical co-ordinates on ``quantum'' $\mathcal{M}$ \eqref{syformA}, \(\{\hat{\rho},\hat{\nu}\}\) or \(\{\hat{\eta},\hat{\sigma}\}\), with \(\hat{s}=e^{4\pi i\hat{\eta}}\), corresponding to expansion at \(t\to 0\),
now satisfy the commutation relations:
\begin{equation}
\label{eq:54}
i \left[\hat{\sigma},\hat{\eta} \right]= \left[\hat{\nu},\hat{\rho} \right]=\frac{\epsilon_1+\epsilon_2}{2\pi}.
\end{equation}

\paragraph{Remark.} It would be interesting to compare the quantum cluster algebra from \cite{BGM1}, leading to \eqref{eq:54}, with quantization of monodromy data in \cite{ILTe}.
At the first glance they seem to be unrelated, since quantum torus in \cite{BGM1} has parameter \(p\), depending on the radius of the compact 5-th dimension \(l_5\), whereas quantum torus in \cite{ILTe} had quantum parameter like \(e^{i\pi\epsilon_1/\epsilon_2}\), depending on the central charge.
To see that these two constructions actually describe the same phenomenon, compute the monodromy of degenerate field \(\phi_{(1,2)}\) around general field with the charge \(\sigma\), which equals
by standard CFT arguments to \({\rm diag}(\hat{\mathsf{m}}_{\sigma},\hat{\mathsf{m}}_{\sigma}^{-1})\) with
\(\hat{\mathsf{m}}_{\sigma}=e^{i\pi\sigma/\epsilon_2}\).
Another constituent of all monodromy matrices
is the Fourier parameter \(\hat{s}\), and
one can check that they satisfy \(\hat{s}\hat{\mathsf{m}}_{\sigma}=e^{-2\pi i\,\epsilon_1/\epsilon_2} \hat{\mathsf{m}}_{\sigma}\hat{s}\).
It means that construction of \cite{BGM1} actually contains the quantum torus from \cite{ILTe}, i.e. they should be related to each other.

\subsection{\(\mathbb{C}^2/\mathbb{Z}_2\)--type blow-up relations and generic irregular blocks at infinity}

Formulas~\eqref{eq:qTauInf} together with the 4d limit of \eqref{eq:qpTauBlowUp} actually allow to write down the blow-up relations at \(t\to\infty\), they are quite similar to \eqref{eq:blowup4d1}, \eqref{eq:blowup4d2}, though with few important distinctions compare to \(t\to 0\) case.
First difference originates from a different relative sign between B\"acklund--transformed tau functions due to the different sign in \eqref{eq:infinitySolution} compare to \eqref{eq:tauRatio}, while the
second is that at \(t\to\infty\) the B\"acklund transformation is not a half-integer shift of the summation variable \(\mathbb{Z}\mapsto \frac12+\mathbb{Z}\), but insertion of an extra sign factor \((-1)^n\).
Moreover, now each independent relation decouples into a pair of equations:
\begin{equation}
\begin{gathered}
\label{eq:blowupInf1}
\sum_{n\in \mathbb{Z}}\mathrm{D}_{2\epsilon_1,2\epsilon_2}^1\big(
\mathsf{F}^{\infty}\left(\nu+2in\epsilon_1,2\epsilon_1,\epsilon_2-\epsilon_1|r\right),
\mathsf{F}^{\infty}\left(\nu+2in\epsilon_2,\epsilon_1-\epsilon_2,2\epsilon_2|r \right)\big)=0,
\\
\sum_{n\in \mathbb{Z}}\mathrm{D}_{2\epsilon_1,2\epsilon_2}^1\big(
\mathsf{F}^{\infty}\left(\nu+2in\epsilon_1+i(\epsilon_2-\epsilon_1),2\epsilon_1,\epsilon_2-\epsilon_1|r\right),
\mathsf{F}^{\infty}\left(\nu+2in\epsilon_2,\epsilon_1-\epsilon_2,2\epsilon_2|r \right)\big)=0,
\end{gathered}
\end{equation}
together with
\begin{equation}
\begin{gathered}
\label{eq:blowupInf2}
\sum_{n\in \mathbb{Z}}\mathrm{D}_{2\epsilon_1,2\epsilon_2}^2\big(
\mathsf{F}^{\infty}\left(\nu+2in\epsilon_1,2\epsilon_1,\epsilon_2-\epsilon_1|r\right),
\mathsf{F}^{\infty}\left(\nu+2in\epsilon_2,\epsilon_1-\epsilon_2,2\epsilon_2|r \right)\big)
=\\=
\frac{r^2}{32}\sum_{n\in \mathbb{Z}}
\mathsf{F}^{\infty}\left(\nu+2in\epsilon_1,2\epsilon_1,\epsilon_2-\epsilon_1|r\right)
\mathsf{F}^{\infty}\left(\nu+2in\epsilon_2,\epsilon_1-\epsilon_2,2\epsilon_2|r \right),
\\
\sum_{n\in \mathbb{Z}}\mathrm{D}_{2\epsilon_1,2\epsilon_2}^2\big(
\mathsf{F}^{\infty}\left(\nu+2in\epsilon_1+i(\epsilon_2-\epsilon_1),2\epsilon_1,\epsilon_2-\epsilon_1|r\right),
\mathsf{F}^{\infty}\left(\nu+2in\epsilon_2,\epsilon_1-\epsilon_2,2\epsilon_2|r \right)\big)
=\\=
-\frac{r^2}{32}\sum_{n\in \mathbb{Z}}
\mathsf{F}^{\infty}\left(\nu+2in\epsilon_1+i(\epsilon_2-\epsilon_1),2\epsilon_1,\epsilon_2-\epsilon_1|r\right)
\mathsf{F}^{\infty}\left(\nu+2in\epsilon_2,\epsilon_1-\epsilon_2,2\epsilon_2|r \right),
\end{gathered}
\end{equation}
with \(\mathrm{D}^k_{2\epsilon_1,2\epsilon_2}\) being the same logarithmic \(t\)-derivatives \eqref{Ddef}, rewritten as logarithmic \(r\)-derivatives using \eqref{eq:Cdef},
and the iterative procedure of finding their solution~\footnote{
	Namely, the \(t\to\infty\) equations mix terms at the same level, but with different shifts of \(\nu\).
	In order to solve it we use polynomial ansatz in \(\nu\), such that degree of a polynomial at level \(k\) is \(3k\), and then solve the linear system on coefficients at each level, but
	sometimes free term at level \(k\) can be obtained only from the equations for the level \(k+1\).
} is far more complicated.

To find the irregular block at infinity iteratively, we substitute into \eqref{eq:blowupInf1}, \eqref{eq:blowupInf2} the following ansatz:
\begin{equation}
\begin{gathered}
\label{eq:FinfFull}
\mathsf{F}^{\infty}(\nu,\varepsilon_1,\varepsilon_2|r)=
\mathsf{C}^{\infty}_{\rm cl}\left(\nu+i \bar{\varepsilon},\varepsilon_1,\varepsilon_2|r\right)
\mathsf{C}^{\infty}_{\rm pert}\left(\nu+i \bar{\varepsilon},\varepsilon_1,\varepsilon_2\right)
\mathsf{B}^{\infty}\left(\nu+i \bar{\varepsilon},\varepsilon_1,\varepsilon_2|r\right),
\end{gathered}
\end{equation}
where~\footnote{
	This shift by \(i\bar{\varepsilon}\) is directly related to using  \(\tilde{\nu}=\nu+\frac{i}{2}\) instead of \(\nu\) in many formulas of sect.~\ref{sec:infinity}.
}
\begin{equation}
\begin{gathered}
\label{eq:CclasspertInf}
\bar{\varepsilon}=\frac{\varepsilon_1+\varepsilon_2}{2},
\\
\mathsf{C}_{\rm cl}\left(\nu,\varepsilon_1,\varepsilon_2|r\right)=\exp \left( -\frac{r^2}{16\varepsilon_1\varepsilon_2}-\frac{\nu r}{\varepsilon_1\varepsilon_2}
+\frac{\varepsilon_1^2+4\varepsilon_1\varepsilon_2+\varepsilon_2^2-4\nu^2}{8\varepsilon_1\varepsilon_2}\log r \right),
\\
\mathsf{C}_{\rm pert} \left( \nu,\varepsilon_1,\varepsilon_2\right)=
e^{-\frac{\nu^2}{\varepsilon_1\varepsilon_2}\left(i\pi/4+\log 2\right)}
\mathsf{G}\left(i\nu+(\varepsilon_1-\varepsilon_2)/2,\varepsilon_1,-\varepsilon_2\right),
\end{gathered}
\end{equation}
and \(\mathsf{G}\) is a double Gamma function, defined by the following difference relations:
\begin{equation}\begin{gathered}
\label{eq:65}
\mathsf{G}(x+\omega_1,\omega_1,\omega_2)=\Gamma(x,\omega_2)\mathsf{G}(x,\omega_1,\omega_2),\\
\mathsf{G}(x+\omega_2,\omega_1,\omega_2)=\Gamma(x,\omega_1)\mathsf{G}(x,\omega_1,\omega_2),\\
\Gamma(x+\omega,\omega)=x\Gamma(x,\omega).
\end{gathered}\end{equation}
Solving \eqref{eq:blowupInf1} and \eqref{eq:blowupInf2}, we get the following expansion of \(\mathsf{B}^{\infty}\):
\begin{equation}
\label{eq:generalBlockInf}
\mathsf{B}^{\infty}(\nu,\varepsilon_1,\varepsilon_2|r)=1+
\sum_{k=1}^{\infty}\frac{\mathsf{N}_k(\nu,\varepsilon_1,\varepsilon_2)}{\mathsf{d}_k\cdot(\varepsilon_1\varepsilon_2r)^k},
\end{equation}
where denominators \(\mathsf{d}_k\) are some integers
\begin{equation}
\begin{gathered}
\label{eq:67}
\mathsf{d}_1=2^4,\quad \mathsf{d}_2=2^9,\quad \mathsf{d}_3=3\cdot 2^{13},\quad
\mathsf{d}_4=3\cdot 2^{19},\\ \mathsf{d}_5=15\cdot 2^{23},\quad \mathsf{d}_6=45\cdot 2^{28},\quad \mathsf{d}_7=35\cdot 9\cdot 2^{32},
\quad \ldots ,
\end{gathered}
\end{equation}
and \(\mathsf{N}(\nu,\varepsilon_1,\varepsilon_2)\) are homogeneous polynomials of \(\nu\), \(\varepsilon_1\), \(\varepsilon_2\) of total degree \(3k\) with two additional symmetries:
\begin{equation}
\label{eq:69}
\mathsf{N}_k(-\nu,\varepsilon_1,\varepsilon_2)=(-1)^k\mathsf{N}_k(\nu,\varepsilon_1,\varepsilon_2),\quad \mathsf{N}_k(\nu,\varepsilon_2,\varepsilon_1)=\mathsf{N}_k(\nu,\varepsilon_1,\varepsilon_2).
\end{equation}
The first numerator has the form
\begin{equation}
\label{eq:68}
\mathsf{N}_1(\varepsilon_1,\varepsilon_2,\nu)=3\varepsilon_1^2\nu+8\varepsilon_1\varepsilon_2\nu+3 \varepsilon_2^2\nu-4\nu^3,
\end{equation}
and other formulas can be found in the Appendix~\ref{app:confBlock}.
We find that in contrast to common irregular blocks at \(t\to 0\) (see Appendix~\ref{ss:cft}), here already the first non-trivial term of expansion depends on the central charge~\footnote{
The same phenomenon happens in the PIII\(_1\) and PIII\(_2\) cases, but does not happen for Painlev\'e IV and Painlev\'e V, we are grateful to H.~Nagoya for this comment.
}.

In both known limits formula \eqref{eq:generalBlockInf} reproduces the \(c=1\)
\begin{equation}
\label{eq:66}
\mathsf{B}^{\infty}\left(\nu,\varepsilon_1=1,\varepsilon_2=-1|r\right)=\mathcal{B}^{\infty}(\nu,r),
\end{equation}
expression, given by \eqref{eq:infinityBlock}, 
and \(c\to\infty\)
\begin{equation}
\label{eq:70}
\lim_{\varepsilon_2\to 0}\big(\varepsilon_2\log \mathsf{B}^{\infty}\left(\nu,\varepsilon_1=1,\varepsilon_2\right)\big)=-f^{\infty}(\nu,r)
\end{equation}
case of \eqref{eq:infinityClassicalBock}.
Another consistency check was performed in \cite{ShchUnpubl} for \(c=-2\) conformal blocks, using the formalism of \(c=-2\) tau functions from \cite{BShch4}.
It is also interesting to point out that classical and perturbative parts from \eqref{eq:CclasspertInf} almost coincide with those from \(c=1\) and \(c\to\infty\) expressions up to some trivial re-definitions.

Motivated by \eqref{eq:blockIntegral} from sect.~\ref{sec:Alba}
one can try to find similar integral formulas for other central charges.
In order to do this it is useful to combine the results of \cite{Grassi:2014uua}, where the spectral determinant was factorized into the product of two factors, corresponding to odd and even parts of spectra, with those from \cite{BShch4}, were these two factors were identified with \(c=-2\) tau functions.
This leads to explicit integral representations of generic irregular blocks \rf{eq:generalBlockInf} for \(c=-2\) at infinite series of special points $\nu\in \frac{i}{2}+ i\mathbb{Z}$. For \(n=1\) it gives a series of expressions:
\begin{equation}\begin{gathered}
\label{eq:87}
\mathsf{B}^{\infty}\left(3i/2,2,-1,i\mathfrak{r}\right)=
\sqrt{\frac{\mathfrak{r}}{8\pi}}e^{\mathfrak{r}}\int_{-\infty}^{\infty}dx e^{-\mathfrak{r}\cosh x}\left(1+\frac1{\cosh x}\right)=
\\=
\sqrt{\frac{\mathfrak{r}}{2\pi}}e^{\mathfrak{r}}\left((1-\mathfrak{r})K_0(\mathfrak{r})-\frac{\pi}{2}\left(\mathfrak{r}K_0(\mathfrak{r})\mathbf{L}_1(\mathfrak{r})+\mathfrak{r}K_1(\mathfrak{r})\mathbf{L}_0(\mathfrak{r})-1\right) \right),
\\
\mathsf{B}^{\infty}\left(5i/2,2,-1,i\mathfrak{r}\right)=
\\
=\sqrt{\frac{8\mathfrak{r}^3}{\pi}}e^{\mathfrak{r}}\left((1+\mathfrak{r})K_0(\mathfrak{r})+\frac{\pi}{2}\left(\mathfrak{r}K_0(\mathfrak{r})\mathbf{L}_1(\mathfrak{r})+\mathfrak{r}K_1(\mathfrak{r})\mathbf{L}_0(\mathfrak{r})-1\right) \right),
\\
\ldots
\end{gathered}
\end{equation}
where \(\mathbf{L}_{\alpha}(\mathfrak{r})\) is a modified Struve function, which appears after integration of \(K_0(\mathfrak{r})\), see \cite{besselInt}.
There are also some simple relations, like
\begin{equation}
\label{eq:89}
\mathsf{B}^{\infty}(0,1,-1,r)=\mathcal{B}^{\infty}(0,r)=1,\quad
\mathsf{B}^{\infty}(i/2,2,-1,r)=\mathsf{B}^{\infty}(-i/2,2,-1,r)=1.
\end{equation}

\subsection{Nakajima--Yoshioka--type blow-up relations at infinity}

Finally, let us check, that the generic irregular blocks at infinity \eqref{eq:FinfFull} also
satisfy, as their avatars from sect.~\ref{sec:infinity}, the analogs of Nakajima--Yoshioka blow-up equations. The corresponding non-homogeneous relation has the form
\begin{equation}
\label{eq:NYinf}
\sum_{n\in \mathbb{Z}}
\mathsf{F}^{\infty}(\nu+in\epsilon_1,\epsilon_1,\epsilon_2-\epsilon_1|r)
\mathsf{F}^{\infty}(\nu+in\epsilon_2,\epsilon_1-\epsilon_2,\epsilon_2|r)=
\mathrm{const}\cdot\mathsf{F}^{\infty}(\nu,\epsilon_1,\epsilon_2|r),
\end{equation}
(in \(\epsilon_2\to 0\) limit  \eqref{eq:NYinf} turns into \eqref{blowup_infty}),
where numeric constant in the r.h.s. depends on normalization of the double gamma functions. For \eqref{eq:FinfFull} it means in practice  the
following relation is satisfied by \eqref{eq:generalBlockInf}:
\begin{equation}
\begin{gathered}
\label{eq:61}
2 \mathsf{B}^{\infty}\left(\nu+i(\epsilon_1+\epsilon_2)/2,\epsilon_1,\epsilon_2|r\right)=
\sum_{n\in \mathbb{Z}}\ell_n(\nu,\epsilon_1,\epsilon_2)r^{-\frac12 n(n-1)}
\cdot\\\cdot
\mathsf{B}^{\infty}\left(\nu+i\epsilon_2/2  +in\epsilon_1,\epsilon_1,\epsilon_2-\epsilon_1|r\right)
\mathsf{B}^{\infty}\left(\nu+i \epsilon_1/2+in\epsilon_2,\epsilon_1-\epsilon_2,\epsilon_2|r\right),
\end{gathered}
\end{equation}
where
\begin{equation}
\label{eq:63}
\ell_n(\nu,\epsilon_1,\epsilon_2)=\frac{\mathsf{G}(i\nu+\epsilon_1-\epsilon_2-n\epsilon_1,\epsilon_1,\epsilon_1-\epsilon_2)\mathsf{G}(i\nu-\epsilon_2-n\epsilon_2,\epsilon_1-\epsilon_2,-\epsilon_2)}{\mathsf{G}(i\nu+\epsilon_1-\epsilon_2,\epsilon_1,\epsilon_1-\epsilon_2)\mathsf{G}(i\nu-\epsilon_2,\epsilon_1-\epsilon_2,-\epsilon_2)}2^{n-n^2}e^{i\pi\frac{n-n^2}{4}}
\end{equation}
are certain polynomials in \(\nu\), \(\epsilon_1\), \(\epsilon_2\).

There are three more Nakajima--Yoshioka type relations on generic blocks \eqref{eq:FinfFull}, namely:
\begin{equation}\begin{gathered}
\label{eq:NYinf1}
\sum_{n\in\mathbb{Z}}(-1)^n
\mathsf{F}^{\infty}(\nu+in\epsilon_1,\epsilon_1,\epsilon_2-\epsilon_1|r)
\mathsf{F}^{\infty}(\nu+in\epsilon_2,\epsilon_1-\epsilon_2,\epsilon_2|r)=0,
\end{gathered}\end{equation}
being the \(\epsilon\)-deformation of the tau function vanishing condition \eqref{tau1_zero}, together with
\begin{equation}\begin{gathered}
\label{eq:NYinf2}
\sum_{n\in\mathbb{Z}}\epsilon_1\partial_r
\mathsf{F}^{\infty}(\nu+in\epsilon_1,\epsilon_1,\epsilon_2-\epsilon_1|r)\cdot
\mathsf{F}^{\infty}(\nu+in\epsilon_2,\epsilon_1-\epsilon_2,\epsilon_2|r)
+\\+
\sum_{n\in\mathbb{Z}}
\mathsf{F}^{\infty}(\nu+in\epsilon_1,\epsilon_1,\epsilon_2-\epsilon_1|r)\cdot
\epsilon_2\partial_r
\mathsf{F}^{\infty}(\nu+in\epsilon_2,\epsilon_1-\epsilon_2,\epsilon_2|r)=0,
\end{gathered}
\end{equation}
which is the \(\epsilon\)-deformation of \eqref{d_r_logtau}, relating logarithmic derivative of \(c=1\) tau function to derivative of the classical conformal blocks~\footnote{
It should not be confused with \eqref{eq:blowupInf1} above, since the latter have different relation between the \(\epsilon\)-parameters of conformal blocks.
}, and
\begin{equation}\begin{gathered}
\label{eq:NYinf3}
2i(\epsilon_1-\epsilon_2)\sum_{n\in\mathbb{Z}}(-1)^n\partial_r
\mathsf{F}^{\infty}(\nu+in\epsilon_1,\epsilon_1,\epsilon_2-\epsilon_1|r)\cdot
\mathsf{F}^{\infty}(\nu+in\epsilon_2,\epsilon_1-\epsilon_2,\epsilon_2|r)
=\\=
\sum_{n\in\mathbb{Z}}
\mathsf{F}^{\infty}(\nu+in\epsilon_1,\epsilon_1,\epsilon_2-\epsilon_1|r)\cdot
\mathsf{F}^{\infty}(\nu+in\epsilon_2,\epsilon_1-\epsilon_2,\epsilon_2|r),
\end{gathered}\end{equation}
generalizing the formula \eqref{eq:bup1}, coming from the fact that the leading coefficient at the pole of solution depends on its position only.

However, unlike the formulas from sect.~\ref{ss:blowup} and sect.~\ref{sec:infinity}, this collection of Nakajima--Yoshioka--type equations does not define the generic irregular block, even if one substitutes the polynomial ansatz preserving all known symmetries.
For example, \eqref{eq:NYinf} for generic values of \(\Omega\)-background parameters is a relation on 3 different conformal blocks, with all different central charges, and the iterative procedure does not fix the coefficients.

Actually the exact form of the equations \eqref{eq:NYinf}, \eqref{eq:NYinf1}, \eqref{eq:NYinf2}, \eqref{eq:NYinf3} was found for already known functions  
\(\mathsf{F}^{\infty}\), so that they turn to be $\epsilon$-deformations of the relations from sect.~\ref{sec:infinity}. 
Nevertheless, is has been shown in \cite{Shch1} that \(\mathbb{C}^2/\mathbb{Z}_2\) blow-up equations follow from some extended collection on the Nakajima--Yoshioka relations.
However, in the \(t\to\infty\) limit we do not have at the moment any
basis for such relations, like quantum Painlev\'e equation, and the  \(t\to\infty\) analogs of this extended set remain to be among the open problems.

\section{Discussion}

There are actually many open questions.
We have used the setup from cluster varieties to understand the meaning of the parameters \(\rho\) and \(\nu\), instead of initial approach of \cite{Novokshenov1986}. The role and meaning of these cluster structures can go beyond just being a convenient technical tool.
For example, the Bohr-Sommerfeld quantization condition for the cluster variable \(\nu=-i(N+1)\) looks as a particular case of more general phenomenon.
Namely, it could describe more general spectral problems for potentials on the Stokes lines connecting turning points.
In particular, we expect something similar to happen in the Painlev\'e I and Painlev\'e II cases.

Let us also point out that expansions like \eqref{eq:tauILT} are known for the irregular limits of other Painlev\'e equations, see, e.g. \cite{Bonelli:2016qwg}.
In all these cases one can formally write down the tau function vanishing conditions like \eqref{eq:infinityRho}, and solve them up to certain order.
It will define (perhaps not completely) some new functions to be called ``quasiclassical conformal blocks'', and further study of these functions is an interesting open problem.

Generalization of our approach to the \(q\)-deformed case is yet unclear.
An illustration why it is problematic is already the fact, that the exact quantization conditions contain both quasiclassical conformal blocks, depending on \(\hbar\) and \(4\pi^2/\hbar\) \cite{Grassi:2017qee}, while the blow-up equations contain only one of them.

There are certainly tones of questions related to general conformal blocks at infinity \(\mathsf{B}^{\infty}(\nu,\varepsilon_1,\varepsilon_2|r)\) and to  corresponding tau functions.
The main question is what is the meaning of \(\mathsf{B}^{\infty}\) in terms of the supersymmetric gauge theory.
The fact that there is single Barnes function in the numerator of \(\mathsf{C}_{\rm pert}^{\infty}\) and only trivial poles at \(\varepsilon_{1,2}=0\) in the formula for \(\mathsf{B}^{\infty}\)  suggests that \(\mathsf{B}^{\infty}\) should be a result of some non-perturbative computation in the dual theory with single \(U(1)\) hypermultiplet (monopole or dyon).
However, we do not know what is this computation, and what is the meaning of \(\mathsf{N}_k(\nu,\varepsilon_1,\varepsilon_2)\) --- the strong-coupling analogs of Nekrasov functions. Our observations suggest, that there should be integral representations for all
\(\mathsf{B}^{\infty}\big(i(p+q)/2+inp+imq,p,-q,i\mathfrak{r}\big)\),
with \(p,q\in \mathbb{Z}\), being some analogs of the Dotsenko--Fateev integrals.
Existence of integral representations for the irregular conformal blocks at special points for different Painlev\'e equations is known, see for example~\cite{Grassi:2018spf,Itoyama:2019rgp}, the \(q\)-deformed versions of corresponding integrals can be found in~\cite{Bonelli:2017gdk}.
One can also try to use the approach of \cite{Bonelli:2017ptp} it order to go to the higher ranks.

Another related question is what is the representation-theoretical or geometric meaning of \(\mathsf{B}^{\infty}\) and the blow-up relations \eqref{eq:blowupInf1}, \eqref{eq:blowupInf2}, \eqref{eq:NYinf}, \eqref{eq:NYinf1}, \eqref{eq:NYinf2}, \eqref{eq:NYinf3}.
We also expect such relations to appear in all other Painlev\'e systems, that have domains with irregular behavior.
One may also ask what is the meaning of Nakajima--Yoshioka blow-up relations after quantization of the Painlev\'e equation, since before quantization they just describe the relation between matrix \(2\times 2\) system and scalar 2-nd order differential equation.

One more question is about the fusion matrix for irregular conformal blocks (and actually not only for them).
Namely, one can ask in general situation, what is the kernel \(\mathsf{K}\), relating conformal blocks at zero and infinity:
\begin{equation}
\label{eq:71}
\mathcal{B}(\sigma,\varepsilon_1,\varepsilon_2|t)=\int d\nu \, \mathsf{K}(\sigma,\nu,\varepsilon_1,\varepsilon_2) \mathsf{B}^{\infty}(\nu,\varepsilon_1,\varepsilon_2|r).
\end{equation}
In \(c=1\) limit it can be extracted from the connection constant from \cite{ILT}, exactly as it was done for Painlev\'e VI equation in \cite{ILTPVI}.
For \(c\to\infty\) this fusion matrix is \eqref{eq:fusion1}, while
for an arbitrary \(c\) we expect some variant of the Ponsot--Teschner formula \cite{Ponsot:1999uf}.
Keeping in mind the story about quantum tau functions, it would be interesting to derive such formula from a kind of quantization of the \(c=1\) connection constant (it looks natural to quantize classical dilogarithms appearing there). This can suggest a way to prove equivalence between the fusion kernels in \cite{ILTPVI} and in \cite{Ponsot:1999uf} at \(c=1\), by now being only checked numerically.

\section*{Acknowledgments}

We are grateful to M.~Bershtein, O.~Lisovyy, A.~Litvinov, N.~Iorgov, H.~Nagoya, A.~Naydiuk, N.~Nekrasov and A.~Shchechkin for useful discussions and correspondence.
We are especially grateful to A.~Grassi, whose comments on preliminary version of the paper actually initiated an extra sect.~\ref{sec:Alba}.
 A.M. is also indebted to V.~Bazhanov and the organizers of the ANZAMP-20 meeting,
where the preliminary results of this paper have been reported.
The work was partially carried out in Skolkovo Institute of Science and Technology under financial support of Russian Science Foundation within grant 19-11-00275.

\section*{Appendix}

\appendix	

\section{WKB parameterization of monodromies}

\label{sec:WKB}

We remind here some basics of the WKB approach to matrix linear systems.
For more detailed and rigorous explanation see \cite{iwaki2014exact1,iwaki2014exact2}.

\subsection{WKB gauge transformation}

Consider a linear system
\begin{equation}
\label{eq:31}
\hbar\frac{dY(z)}{dz}=A(z)Y(z).
\end{equation}
In the limit \(\hbar\to 0\) one can perform the gauge transformation \(U(z)=U_0(z)U_1(z)\) diagonalizing connection \(\hbar\frac{d}{dz}-A(z)\), where
\(U_0(z)\) diagonalizes \(A(z)\):
\begin{equation}
U_0(z)^{-1}A(z)U_0(z)=
\begin{pmatrix}
\lambda(z) & 0\\
0 & -\lambda(z)
\end{pmatrix}.
\end{equation}
One has
\begin{equation}
\label{eq:34}
U^{-1}\left(-\hbar\frac{d}{dz}+A(z)  \right)U=
U_1^{-1}
\begin{pmatrix}
\lambda  & 0\\
0 & -\lambda
\end{pmatrix}
U_1-\hbar U_1^{-1}U_0^{-1}U_0'U_1-\hbar U_1^{-1}U_1'=
\begin{pmatrix}
\lambda & 0\\
0 & -\lambda
\end{pmatrix},
\end{equation}
leading to equation on \(U_1(z)\):
\begin{equation}
\label{eq:masterWKB}
\left[
\begin{pmatrix}
\lambda & 0\\
0 & -\lambda
\end{pmatrix}, U_1 \right]=\hbar U_0^{-1}U_0'\cdot U_1+\hbar U_1'.
\end{equation}
To be able to solve this equation we first need to make sure that the matrix \(U_0^{-1}U_0'\) does not have diagonal components.
\(U_0\) is defined up to multiplications by diagonal matrices from the right.
Suppose that arbitrarily chosen \(U_0\) produces diagonal components in \(U_0^{-1}U_0'\).
Redefine it by
\begin{equation}
U_0\mapsto U_0
\begin{pmatrix}
\phi_1 & 0\\
0 & \phi_2
\end{pmatrix}
\end{equation}
and try to solve the matrix equation
\begin{equation}
\label{eq:36}
\begin{pmatrix}
\phi_1^{-1} & 0\\
0 & \phi_2^{-1}
\end{pmatrix}
U_0^{-1}U_0'
\begin{pmatrix}
\phi_1 & 0\\
0 & \phi_2
\end{pmatrix}
+
\begin{pmatrix}
\phi_1'\phi_1^{-1} & 0\\
0 & \phi_2'\phi_2^{-1}
\end{pmatrix}=
\begin{pmatrix}
0 & a(z)\\
b(z) & 0
\end{pmatrix},
\end{equation}
with \(a(z)\) and \(b(z)\) being some arbitrary functions, which become the off-diagonal elements of re-defined \(U_0^{-1}U_0'\).
This equation is equivalent to two ordinary differential equations:
\begin{equation}
\label{eq:37}
\phi_1'(z)=-\left[ U_0^{-1}U_0' \right]_{11} \phi_1,\quad \phi_2'(z)=-\left[ U_0^{-1}U_0' \right]_{22} \phi_2.
\end{equation}
They always have locally defined solutions.

Now we define the \(\hbar\)-expansion of \(U_1\):
\begin{equation}
\label{eq:38}
U_1=\mathbb{I}+\sum_{k=1}^{\infty}\hbar^k
\begin{pmatrix}
v_k^{(1)} & w_k^{(1)}\\
w_k^{(2)} & v_k^{(2)}
\end{pmatrix},
\end{equation}
substitute it into (\ref{eq:masterWKB}) and expand into the powers of \(\hbar\):
\begin{equation}
\label{eq:33}
2\lambda\begin{pmatrix}
0 & w_{k+1}^{(1)}\\
-w_{k+1}^{(2)} & 0
\end{pmatrix}=
\begin{pmatrix}
0 & a\\
b & 0
\end{pmatrix}
\begin{pmatrix}
v_k^{(1)} & w_k^{(1)}\\
w_k^{(2)} & v_k^{(2)}
\end{pmatrix}+
\frac{d}{dz}\begin{pmatrix}
v_k^{(1)} & w_k^{(1)}\\
w_k^{(2)} & v_k^{(2)}
\end{pmatrix}.
\end{equation}
Written in components it gives
\begin{equation}\begin{gathered}
\label{eq:39}
w_{k+1}^{(1)}=\frac1{2\lambda}\left( \frac{dw_k^{(1)}}{dz}+a v_k^{(2)} \right),\quad
w_{k+1}^{(2)}=\frac1{2\lambda}\left( \frac{dw_k^{(2)}}{dz}+a v_k^{(1)} \right),\\
\frac{dv_{k+1}^{(1)}}{dz}=-aw_{k+1}^{(1)},\quad \frac{dv_{k+1}^{(2)}}{dz}=-bw_{k+1}^{(2)}.
\end{gathered}\end{equation}
These equations, in principle, allow one to find the \(\hbar\)-expansion of \(U_1(z)\)
and after this get the solution of the initial system as
\begin{equation}
\label{eq:40}
Y(z)=U_0(z)U_1(z)
\begin{pmatrix}
\exp \left( \hbar^{-1}\int^z\lambda dz \right) & 0\\
0 & \exp \left(-\hbar^{-1}\int^z\lambda dz \right)
\end{pmatrix}.
\end{equation}

\subsection{Turning points}

All considerations of the previous section are applicable only in the region where \(\lambda(z)\neq 0\), but one  can consider separately the vicinity of the points where \(\det A(z)=0\), the analog of turning points in quantum mechanics.
We illustrate this in the model example~\footnote{
In this particular case the linear problem is solved in terms of Airy functions, but we do not use this exact solution.} with
 \(A(z)=
\begin{pmatrix}
0 & 1\\
z & 0
\end{pmatrix}\),
whose eigenvalues are \(\lambda=\pm\sqrt{z}\), and corresponding diagonalizing matrix is
\begin{equation}
\label{eq:41}
U_0(z)=
\begin{pmatrix}
z^{-\frac1{4}} & z^{-\frac1{4}}\\
z^{\frac1{4}} & -z^{\frac1{4}}
\end{pmatrix}.
\end{equation}
Since \(U_0^{-1}U_0'=\frac1{4z}
\begin{pmatrix}
0 & 1\\
1 & 0
\end{pmatrix}\)
has zeroes on diagonal, it satisfies our requirements for \eqref{eq:40}, and
the main asymptotic part (putting \(U_1(z)=\mathbb{I}\)) of the solution is
\begin{equation}
\label{eq:35}
Y(z)\simeq 
\begin{pmatrix}
z^{-\frac1{4}} & z^{-\frac1{4}}\\
z^{\frac1{4}} & -z^{\frac1{4}}
\end{pmatrix}
\begin{pmatrix}
\exp\left(\frac{2}{3\hbar}z^{\frac{3}{2}}\right) & 0\\
0 & \exp\left(-\frac{2}{3\hbar}z^{\frac{3}{2}}\right)
\end{pmatrix},
\end{equation}
or, if one decides to use the left action of monodromy matrices:
\begin{equation}
\label{eq:42}
\Psi(z)=Y(z)^T=
\begin{pmatrix}\psi_1(z)\\\psi_2(z)\end{pmatrix},
\end{equation}
with
\begin{equation}
\label{eq:AiryNormalization}
\psi_1(z)=
e^{\frac{2}{3\hbar}z^{\frac{3}{2}}}\begin{pmatrix} z^{-\frac1{4}} & z^{\frac1{4}}\end{pmatrix},\quad
\psi_2(z)=
e^{-\frac{2}{3\hbar}z^{\frac{3}{2}}}\begin{pmatrix} z^{-\frac1{4}} & -z^{\frac1{4}}\end{pmatrix}
\end{equation}
being two linearly independent solutions of the linear system.
Everywhere except three Stokes rays, \(z=r e^{\frac{2\pi i}{3}(k+\frac12)}\), \(r>0\) and \(k=0,1,2\), one of these solutions is asymptotically large (dominant), compared to another exponentially small one (sub-dominant),
and the dominant solution is actually defined only up to addition of the sub-dominant one.
On the Stokes rays both solutions can be defined uniquely, since both are oscillating, but when one goes from one Stokes ray to another one, some triangular transformation can emerge (this is called the \emph{Stokes phenomenon}).

We divide now complex plane by three anti-Stokes rays, \(z=r e^{\frac{2\pi i k}{3}}\), \(r>0\) and \(k=0,1,2\), so that solutions in each sector, bounded by the anti-Stokes rays, are given by analytic continuations of the solution on the corresponding Stokes ray.
We choose the easiest option to switch from one pair of solutions to another on the anti-Stokes rays, where both exponents in \eqref{eq:AiryNormalization} are real.

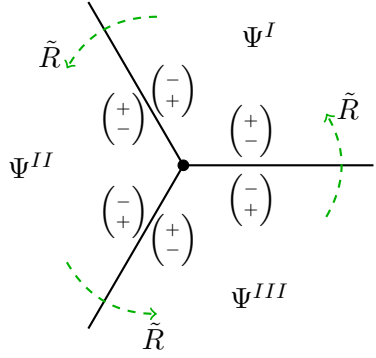
\begin{figure}[h]
\begin{center}
\begin{tikzpicture}
\draw[fill](0,0)circle[radius=0.7mm];
\draw[thick](0,0)--+(0:2.5) (0,0)--+(120:2.5) (0,0)--+(-120:2.5);
\draw[thick,green!70!black,dashed,->](-20:2) to[bend right] (20:2);
\draw[thick,green!70!black,dashed,->] (100:2) to[bend right] (140:2);
\draw[thick,green!70!black,dashed,->] (-140:2) to[bend right] (-100:2);
\node at(27:1) {\(\tiny\begin{pmatrix}+\\-\end{pmatrix}\)};
\node at(-27:1) {\(\tiny\begin{pmatrix}-\\+\end{pmatrix}\)};
\node at(142:1) {\(\tiny\begin{pmatrix}+\\-\end{pmatrix}\)};
\node at(98:1) {\(\tiny\begin{pmatrix}-\\+\end{pmatrix}\)};
\node at(-98:1) {\(\tiny\begin{pmatrix}+\\-\end{pmatrix}\)};
\node at(-142:1) {\(\tiny\begin{pmatrix}-\\+\end{pmatrix}\)};
\node at(20:2.3){\(\tilde{R}\)};
\node at(140:2.3){\(\tilde{R}\)};
\node at(260:2.3){\(\tilde{R}\)};
\node at(60:2){\(\Psi^{I}\)};
\node at(180:2){\(\Psi^{II}\)};
\node at(300:2){\(\Psi^{III}\)};
\end{tikzpicture}
\caption[]{Dominant and sub-dominant solutions.  \label{fig:dominantSubdominant}}
\end{center}
\end{figure}

The asymptotic solutions (\ref{eq:AiryNormalization}) contain also the factors \(z^{-\frac1{4}}\).
To make them single-valued one has to choose some branch cuts and fix jumps on these lines, so that their product is \((e^{2\pi i})^{-\frac1{4}}=-i\).
To simplify computations we choose 3 such cuts~\footnote{
Another possible option is to make a single branch cut with the jump \((e^{2\pi i})^{-\frac1{4}}=-i\) in some arbitrary way.}, coinciding with the anti-Stokes rays, all with jumps \(+i\),
and chose initial branch of this function so that \((r+i0)^{-\frac1{4}}>0\) for \(r>0\).
In this setup solutions \(\psi_1\) and \(\psi_2\) become dominant and sub-dominant on different sides of the anti-Stokes rays, see Fig.~\ref{fig:dominantSubdominant}, where ``\(+\)'' denotes dominant and ``\(-\)'' corresponds to subdominant.

When we cross the Stokes lines, solutions in the final sector are expressed as linear combinations of analytic continuations of solutions in the initial sector.
This linear transformation is described by the triangular matrix of general form
\begin{equation}
\label{eq:43}
\tilde{R}(\alpha)=
\begin{pmatrix}
\alpha & i\\
i & 0
\end{pmatrix},
\end{equation}
which takes into account the jumps of \(z^{-\frac1{4}}\) and the Stokes phenomenon, when dominant solution is defined up to adding subdominant in the basis as in Fig.~\ref{fig:dominantSubdominant}.
In principal, \(\alpha\)'s can be different for each Stokes transformation, but since solution \(\Psi\) is analytic at the turning point, corresponding total monodromy should satisfy \( \tilde{R}(\alpha_I)\tilde{R}(\alpha_{II})\tilde{R}(\alpha_{III})=\mathbb{I}\),
with the unique solution \(\alpha_I=\alpha_{II}=\alpha_{III}=-1\), so finally
\begin{equation}
\label{eq:tildeX}
\tilde{R}=
\begin{pmatrix}
-1 & i\\
i & 0
\end{pmatrix}.
\end{equation}

\subsection{WKB foliation and parameterization of monodromies}

\begin{figure}[h]
\begin{center}
\begin{tikzpicture}
\draw[fill] (0,1.5) node (v1) {} circle[radius=0.7mm];
\draw[fill](0,-1.25) node (v2) {} circle[radius=0.7mm];
\draw[thick] (v1.center) .. controls +(-150:1.5) and (-2.5,0.25) .. (-4.5,0.25);
\draw[thick] (v1.center) .. controls +(-30:1.5) and (2.5,0.25) .. (4.5,0.25);
\draw[thick] (v2.center) .. controls +(30:1.5) and (2.5,0) .. (4.5,0);
\draw[thick] (v2.center) .. controls +(150:1.5) and (-2.5,0) .. (-4.5,0);
\draw[thick] (v1.center) -- +(0,1.5);
\draw[thick] (v2.center) -- +(0,-1.5);
\draw[red,double] (-4.5,0.125)--(4.5,0.125);

\draw[red,->](-1,0)--(-1,0.25);

\draw[thick, green!70!black,dashed, ->](v2) to[bend right=15] (v1);
\node at (-0.35,0.6){\(\tilde{X}(x)\)};
\node at (0.3,1.7){\(P_1\)};
\node at (0.3,-1.45){\(P_0\)};
\end{tikzpicture}
\caption{Neighboring turning points. \label{fig:2points}}
\end{center}
\end{figure}
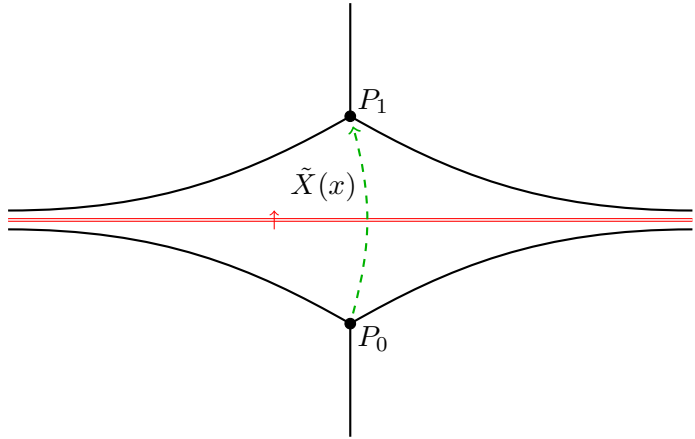

To extend this construction globally we have to start with global definition for the anti-Stokes rays, starting at the turning points.
This is done as follows: take the WKB differential \(\lambda dz\) and consider the anti-Stokes lines, where \(\lambda dz\in \mathbb{R}\), or \(\Im\ \lambda dz=0\).
Taken together, these lines define the \emph{WKB foliation}, but we are now interested only in the leaves of this foliation that start at the turning points.
These leaves divide the plain into domains, so that any solution, defined by its asymptotics at a turning point, can be continued to neighboring domain and compared with solutions at another turning point, see Fig. \ref{fig:2points}.

Since the WKB solutions \eqref{eq:40} behave like \(\exp(\pm\hbar^{-1}\int \lambda dz)\), the ``positive'' solution at one point $P_0$ maps to the negative one at the neighboring point $P_1$, and vice versa, so that the corresponding transition matrix is given uniquely by 
\begin{equation}
\label{eq:tildeX}
\tilde{X}(x)=
\begin{pmatrix}
0 & -ix\\
-i/x & 0
\end{pmatrix},
\end{equation}
where \(x\sim \exp(\hbar^{-1}\int_{P_0}^{P_1} \lambda dz)\) is one of the parameters parameterizing the monodromy data.
It is also known that \(x^2\sim \exp(\hbar^{-1}\oint\lambda dz)\) is a cluster variable.

To define the whole system of domains with chosen pair of solutions in each of them,
we add extra lines, separating neighboring turning points, and attach transition matrices 
\(\tilde{X}(x)\) to these lines, see Fig.~\ref{fig:2points}. The direction of transition through these lines, corresponding to the matrix \(X(x)\), is shown by extra tiny arrows, though due to \(\tilde{X}(x)^2=-1\) it affects only the signs.

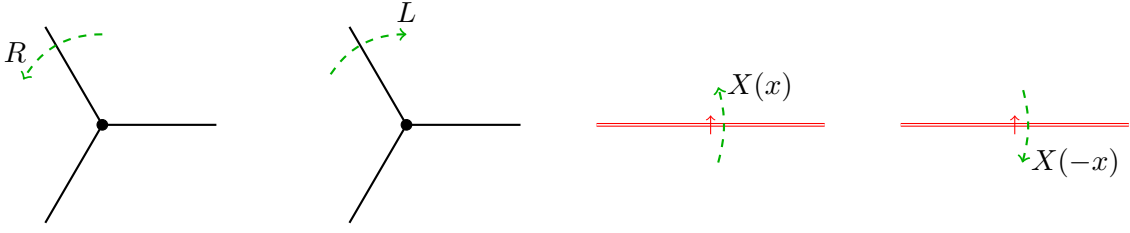
\begin{figure}[h!]
\begin{center}
\begin{tikzpicture}
\begin{scope}[xshift=-4cm]
\draw[fill](0,0)circle[radius=0.7mm];
\draw[thick](0,0)--+(0:1.5) (0,0)--+(120:1.5) (0,0)--+(-120:1.5);
\draw[thick,green!70!black,dashed,->] (90:1.2) to[bend right] (150:1.2);
\node at(140:1.5){\(R\)};
\end{scope}

\begin{scope}[xshift=0cm]
\draw[fill](0,0)circle[radius=0.7mm];
\draw[thick](0,0)--+(0:1.5) (0,0)--+(120:1.5) (0,0)--+(-120:1.5);
\draw[thick,green!70!black,dashed,<-] (90:1.2) to[bend right] (150:1.2);
\node at(90:1.5){\(L\)};
\end{scope}

\begin{scope}[xshift=4cm]
\draw[double,red](-1.5,0)--(1.5,0);
\draw[red,->](0,-0.125)--(0,0.125);
\node at(0.65,0.5){\(X(x)\)};
\draw[green!70!black,dashed,thick,->](0.1,-0.5) to[bend right=15](0.1,0.5);
\end{scope}

\begin{scope}[xshift=8cm]
\draw[double,red](-1.5,0)--(1.5,0);
\draw[red,->](0,-0.125)--(0,0.125);
\node at(0.8,-0.5){\(X(-x)\)};
\draw[green!70!black,dashed,thick,<-](0.1,-0.5) to[bend right=15](0.1,0.5);
\end{scope}

\end{tikzpicture}
\end{center}
\caption{Transition matrices. \label{fig:matrices}}
\end{figure}

Now we have all necessary ingredients, up to normalizations. By simultaneous conjugation one can e.g. remove the \(i\)-factors, so that
finally the transition matrices, corresponding to transitions, shown in 
Fig.~\ref{fig:matrices}, look as follows~\footnote{
Notice that our \eqref{eq:WKBmatrices} are different from similar matrices from \cite{CMR1},
probably since in Teichm\"uller case the group is \(PGL(2,\mathbb{R})\) and signs are inessential.}
\begin{equation}
\label{eq:WKBmatrices}
R=
\begin{pmatrix}
-1 & 1\\
-1 & 0
\end{pmatrix},\quad
L=R^{-1}=
\begin{pmatrix}
0 & -1\\
1 & -1
\end{pmatrix},\quad
X(x)=
\begin{pmatrix}
0 & -x\\
1/x & 0
\end{pmatrix}.
\end{equation}
For technical reason we also introduce the diagonal matrix 
\begin{equation}
\label{Da}
D(a)=
\begin{pmatrix}
a & 0\\
0 & 1/a
\end{pmatrix}
\end{equation}
to be used to adjust normalization in each sector.
There is also an obvious relation \(X(x)^{-1}=X(-x)=-X(x)\).

\section{Irregular conformal blocks and Barnes $G$-functions
\label{ss:cft}}

The Whittaker-Gaiotto vector $\ket{\Delta, \Lambda^2}\in \mathcal{H}_{\Delta,c}$ in the Virasoro module $\mathcal{H}_{\Delta,c}$ with the highest weight $\Delta$ and central charge $c$ is defined by
\begin{align}
	\begin{split}
		L_1 \ket{\Delta, \Lambda^2} &= \Lambda^2 \ket{\Delta, \Lambda^2} \\
		L_n \ket{\Delta, \Lambda^2} &= 0,\quad n \geq 2,
	\end{split}
\end{align}
(it is enough to require $L_2 \ket{\Delta, \Lambda^2} = 0$), which under initial condition
$\left.\ket{\Delta, \Lambda^2} \right|_{\Lambda=0}=|\Delta\rangle$ given at $\Lambda=0$ by the highest weight vector of $\mathcal{H}_{\Delta,c}$ can be presented
as an expansion \citep{MMM}
\begin{equation}
\label{GWexp}
\ket{\Delta, \Lambda^2} = \sum_Y \Lambda^{2|Y|}Q_{\Delta,c}(Y,[1]^{|Y|})^{-1}L_{-Y}|\Delta\rangle,
\end{equation}
where $Y=\{Y_1\geq Y_2\geq\ldots\geq Y_l>0\}$ is a partition, $|Y|=\sum_{k=1}^lY_k$, $L_{-Y}|\Delta\rangle=L_{-Y_1}\ldots L_{-Y_l}|\Delta\rangle\in \mathcal{H}_{\Delta,c}$ is a vector in the Virasoro module at the level $|Y|$, i.e. 
\begin{equation}
\label{level}
L_0L_{-Y}|\Delta\rangle=|Y|\cdot L_{-Y}|\Delta\rangle,
\end{equation} 
 and
\begin{equation} 
		 Q_{\Delta,c}(Y, Y') = \bra{\Delta} L_{Y} L_{-Y'} \ket{\Delta}
	\end{equation}
	is the Shapovalov form of $\mathcal{H}_{\Delta,c}$. In \eqref{GWexp} $|[1]^{|Y|}\rangle\in  \mathcal{H}_{\Delta,c}$ is a special vector $|[1]^{|Y|}\rangle=L_{-1}^{|Y|}|\Delta\rangle$ at level $|Y|$, corresponding to a column Young diagram of height $l$, and it follows from \eqref{GWexp} that Whittaker-Gaiotto vector satisfies
\begin{equation}
		L_0 \ket{\Delta, \Lambda^2} = \Big( \Delta + \frac{\Lambda}{2} \frac{\partial}{\partial \Lambda} \Big) \ket{\Delta, \Lambda^2}.
\end{equation}
The irregular 4-point conformal block is just a scalar product 
\begin{equation}
\label{scalar4}
	\braket{\Delta, 1| \Delta, \Lambda^2}
	= \langle b^{-2},b^{-1}\sigma|b^{-2}t,b^{-1}\sigma\rangle \equiv \langle b^{-1}\sigma|b^{-2}t,b^{-1}\sigma\rangle,
\end{equation}
where we have applied convenient parameterization
\begin{equation}
c=1+6(b+b^{-1})^2,\quad \Delta = \Delta(\sigma,b) = \frac{1-4\sigma^2}{4b^2}+\frac{1}{2}+\frac{b^2}{4}, \quad t=\Lambda^2/b^2.
\end{equation}
At $c=1$ or $b=i$ the irregular block from (\ref{eq:KyivFormula}) and below can be defined therefore by the following scalar product
\begin{equation}
	\mathcal{B}(\sigma, t) = \braket{i\sigma| -t,i\sigma} = 1 + \frac{t}{2\sigma^2} + \frac{t^2(1+8\sigma^2)}{4\sigma^2(1-4\sigma^2)^2} + O(t^3).
\end{equation}
Irregular conformal block \eqref{scalar4} can be obtained as a ``matter decoupling'' limit 
\begin{align} \label{4lim}
	\begin{split}
		\Delta_1 x &\to 0, \\
		\Delta_4 x &\to 0, \\
		(\Delta_2 - \Delta_1) \sqrt{x} &= \Lambda^2  = \text{const}, \\
		(\Delta_3 - \Delta_4) \sqrt{x} &= \Lambda^2 = \text{const} \\
	\end{split}
\end{align}
of the 4-point Virasoro conformal block
\begin{align}
	\begin{split} \label{4p}
		 	\mathcal{B}_\alpha(x) =  \sum_{|Y| = |Y'|} x^{|Y|} \gamma_{\Delta_4 \Delta_3 \Delta_\alpha}(Y) Q^{-1}_{\Delta_\alpha}(Y, Y') \gamma_{\Delta_1 \Delta_2 \Delta_\alpha}(Y'),
	\end{split}
\end{align}
where $\alpha$ parameterizes the intermediate dimension, the sum is over all pairs of Young diagrams, and
	\begin{equation} \label{gamma}
		\gamma_{\Delta_3 \Delta_2 \Delta_1}(Y) = \prod_{i=1}^{l(Y)} (Y^i \Delta_2 + \Delta_1 - \Delta_3 + \sum_{j < i} Y^j).
	\end{equation}
The five- and six-point blocks we discussed in sect.~\ref{sec:CFT} can be treated similarly, but explicit formulas are far more complicated, and therefore --- less useful. As an example we present here an explicit combinatorial expression for the 5-point case.

The analog of (\ref{4p}) for the five-point block is
	\begin{align}
		\begin{split} \label{5point_reg}
			\mathcal{B}_{\alpha, \beta}(z, x) = \sum_{\substack{|Y_1| = |Y_1'| \\ |Y_2| = |Y_2'|}} x^{|Y_1|} z^{|Y_2|} \gamma_{\Delta_1 \Delta_2 \Delta_\beta}(Y_1') Q_{\Delta_\beta}^{-1}(Y_1, Y_1') \gamma_{\Delta_\beta \Delta \Delta_\alpha} (Y_2) Q_{\Delta_\alpha}^{-1}(Y_2, Y_2') \times \\
			\sum_{Y \subset Y_1} z^{-|Y|} \gamma_{(\Delta_\alpha + |Y_2|) \Delta \Delta_\beta}(Y) \gamma_{\Delta_4 \Delta_3 \Delta_\alpha}(Y_2'+(Y_1\setminus Y)).
		\end{split}
	\end{align}
The last sum in (\ref{5point_reg}) is over the sub-collections $Y$ of rows of a diagram $Y_1$, and $Y_2'+(Y_1\setminus Y)$ denotes a tableau that is obtained by adding the remaining rows of $Y_1$ (that are not in $Y$) to the bottom of $Y_2'$. Note that although the resulting tableau is not necessarily a Young diagram, the definition of the corresponding gamma still makes sense. As compared to \cite{AM}, (\ref{5point_reg}) involves direct computation of the descendants three point functions.

Similarly to (\ref{4lim}) we now take the limit
\begin{equation} \label{5lim}
		\begin{gathered}
			z \to 0 , \quad \Delta_1 z^2 \to 0, \quad \Delta_4 z^2 \to 0,
			\\
			(\Delta_2 - \Delta_1)z = \Lambda_1^2 w = \text{const} , \quad
			(\Delta_3 - \Delta_4) z = \Lambda_2^2 w = \text{const}, \quad
			\frac{z^2}{x} = w^2 = \text{const},
		\end{gathered}
\end{equation}
when the gamma-factors (\ref{gamma}) behave as follows:
	\begin{align}
		\begin{split} \label{gammalim}
			\gamma_{\Delta_1 \Delta_2 \Delta_{\beta}}(Y_1') &= \prod_{i=1}^{l(Y_1')} \Big(\frac{\Lambda_1^2 w}{z} + (Y_{1}'^i-1) o(z^{-2})+O(1) \Big) = O \big(z^{-|Y_1'|} \big), \\
			\gamma_{\Delta_4 \Delta_3 \Delta_{\alpha}}(Y_2' + (Y_1 \setminus Y)) &= \prod_{i=1}^{l(Y_2' + (Y_1 \setminus Y))} \Big(\frac{\Lambda_2^2 w}{z} + ((Y_2' + (Y_1 \setminus Y))^i-1) o(z^{-2})+O(1) \Big) \\
			&= O \big(z^{-|Y_2'|-|Y_1|+|Y|} \big).
		\end{split}
	\end{align}
Note, that the estimates saturate when both diagrams under consideration ($Y_1'$ and $Y_2' + (Y_1 \setminus Y)$) are columns. The resulting power of $z$ in (\ref{5point_reg}) is $O(z^{2|Y_1|+Y_2-|Y|-|Y_1'|-|Y_2'|-|Y_1|+|Y|}) = O(1).$ It follows that for a given pair $(|Y_1|, |Y_2|)$ the only choice of $Y_1'$, $Y_2'$ and $Y_1 \setminus Y$ that contributes to the irregular limit is
	\begin{equation}
			Y_1' = [1^{|Y_1|}], \quad 
			Y_2' = [1^{|Y_2|}], \quad 
			Y_1 \setminus Y = [1^{|Y_1 \setminus Y|}],
	\end{equation}
where $[1^l]$ again denotes a Young diagram that is a column of height $l$.
		After these substitutions (\ref{5point_reg}) reduces to
	\begin{equation}  \label{irreg}
		\begin{gathered}
		w^{\Delta_\beta + \Delta - \Delta_\alpha} \bra{\Delta_\alpha, \Lambda_2^2} V_{\Delta}(w) \ket{\Delta_\beta, \Lambda_1^2} = \\
	=	\sum_{\substack{Y_1, Y_2, Y \subset Y_1 \\ Y_1\setminus Y \text{ is a column}}} w^{|Y_2| - |Y|} \Lambda_1^{2|Y_1|} \Lambda_2^{2|Y_2| + 2|Y_1 \setminus Y|} Q_{\Delta_\beta}^{-1}(Y_1, [1^{|Y_1|}]) Q_{\Delta_\alpha}^{-1}(Y_2, [1^{|Y_2|}])\cdot
	\\
	\cdot \gamma_{\Delta_\beta \Delta \Delta_\alpha} (Y_2) \gamma_{(\Delta_\alpha + |Y_2|) \Delta \Delta_\beta}(Y)
		\end{gathered}
	\end{equation}
When the dimension $\Delta$ is degenerate, the limit (\ref{5lim}) descends to the level of the corresponding BPZ equations, in particular for $\Delta = \Delta_{(1,2)}$ this limit corresponds to the Heun-Mathieu reduction into \eqref{eq:MathieuCFT}. For $\Delta = \Delta_{(2,1)}$ this limit turns the Painlev\'e VI Hamilton-Jacobi equation into the Painlev\'e III\(_3\) Hamilton-Jacobi equation \eqref{eq:5pointsHeavy} (together with reduction of Painlev\'e VI into Painlev\'e III\(_3\)).

\subsubsection*{Barnes functions}

The structure constants in (\ref{eq:KyivFormula}) are expressed in terms of the Barnes $G$-function. For completeness we collect here its most important properties we use in the main text. Namely,
\begin{equation}\begin{gathered}
G(x+1)=\Gamma(x)G(x), \quad G(1)=G(2)=1,
\\
G(1-k)=0,\quad k\in \mathbb{Z}_{>0}.
\end{gathered}\end{equation}
In sect.~\ref{sec:infinity} we have used an identity
\begin{equation}
\label{eq:BarnesDuality}
\frac{G(1+x-n)}{G(1+x)}=\frac{G(1-x+n)}{G(1-x)}\left( \frac{\sin\pi x}{\pi} \right)^n(-1)^{n(n-1)/2},
\end{equation}
which follows from the well-known formula
\begin{equation}
    \Gamma(z) \Gamma(1-z) = \frac{\pi}{\sin \pi z}
\end{equation}
for the Euler gamma-functions.

We also provide the integral which is used in the derivation of \eqref{eq:26} and (\ref{Finf})
\begin{equation}
\label{intlogamma}
	\int^z \log \Gamma(x) dx = \frac{z(1-z)}{2}+\frac{z}{2} \log 2\pi + (z-1) \log \Gamma(z) - \log G(z).
\end{equation}
The following identity for dilogarithm is used in the computation of the constant (\ref{eq:classicalFusion})
\begin{equation}
\label{eq:Li2zZinv}
    \mathrm{Li}_2(z) + \mathrm{Li}_2(1/z) = -\frac{\pi^2}{6} - \frac{1}{2} \log^2(-z).
\end{equation}
We also used another identity that relates dilogarithms to Barnes functions:
\begin{equation}
\label{eq:Li2Barnes}
\mathrm{Li}_2(e^{2\pi iz})=-2\pi i\log \frac{G(1+z)}{G(1-z)}-2\pi i z\log \frac{\sin\pi z}{\pi}-\pi^2z(1-z)+\frac{\pi^2}{6}.
\end{equation}

\section{The solution at infinity} \label{solution_infty}
When $t_{\star} \to \infty$, the solution of (\ref{eq:PIII}) can be expanded in the powers of $t_{\star}^{-1/4}$:
	\begin{align} \label{inflambda}
		\begin{split}
			w = &t_\star^{1/2} \mathfrak{t}^{1/2} \Big(\frac{1+\mathcal{X}}{1-\mathcal{X}} \Big)^2 - t_\star^{1/4} \mathfrak{t}^{1/4} \frac{(1+\mathcal{X})(2(\nu+i)\mathcal{X}^4 - (6i\nu^2 - 6\nu - i)\mathcal{X}^2 - 2\nu)}{16\mathcal{X}(1-\mathcal{X})^3} + \\
				& \frac{1}{2048 \mathcal{X}^2 (1-\mathcal{X})^4} \Big( 16(i+\nu)^2 \mathcal{X}^8 + 4(6i\nu^3-22\nu^2-27i\nu+11)\mathcal{X}^7 \\
				& - 16(6i\nu^3-12\nu^2-7i\nu+1)\mathcal{X}^6 - (36\nu^4+64i\nu^3-112\nu^2-128i\nu+53)\mathcal{X}^5 \\
				&- 4(36\nu^4+72i\nu^3-40\nu^2-4i\nu+1)\mathcal{X}^4 - (36\nu^4+80i\nu^3-136\nu^2-48i\nu+9)\mathcal{X}^3 \\
				&+ 16i\nu(6\nu^2+6i\nu-1)\mathcal{X}^2 - 4i\nu(6\nu^2-4i\nu-1)\mathcal{X} + 16\nu^2\Big) + O(t_\star^{-1/4})
		\end{split}
	\end{align}
	Here $\mathfrak{t} = t/t_\star$ is the re-scaled time (so the pole is at $\mathfrak{t} = 1$),
	\begin{equation}
	\begin{gathered}
		\mathcal{X} = \exp(8 i(\mathfrak{t}^{1/4}-1) t_\star^{1/4}) \mathfrak{t}^{i \tilde{\nu}/4} A(1, \tilde{\nu}),
	\\
		e^{4 \pi i  \rho_\star} = 2^{-5 i \tilde{\nu}} \exp \Big(\frac{\pi \tilde{\nu}}{2} \Big) \frac{\Gamma(i \tilde{\nu} + 1/2)}{\sqrt{2\pi}} t_\star^{-i \tilde{\nu}/4} \exp(-8 i t_\star^{1/4}) A(1, \tilde{\nu}),
	\\
		A(\mathfrak{t}, \nu) = 1 + \frac{3 i}{128} \Big(1 - 4 \nu^2 \Big) t_\star^{-1/4} \mathfrak{t}^{-1/4} - 
		\\
		-\frac{144 \nu^4 - 320 i \nu^3 - 72 \nu^2 + 272 i \nu + 9}{32768} t_\star^{-1/2} \mathfrak{t}^{-1/2} + O(t_\star^{-3/4}).
		\end{gathered}
	\end{equation}
Note that coefficients of $A(\mathfrak{t}, \nu) = 1 + A_1(\nu)  t_\star^{-1/4} \mathfrak{t}^{-1/4} + A_2(\nu) t_\star^{-1/2} \mathfrak{t}^{-1/2} + ...$ are not constrained by the ansatz (\ref{inflambda}). $A(\mathfrak{t}, \nu)$ has to be determined from the requirement that the pole is at $\mathfrak{t} = 1$, i.e. from vanishing of the tau function as in (\ref{eq:infinityRho}), or by re-summation of (\ref{inflambda}) near the pole:
	\begin{equation} \label{inflambda_pole}
		w = \frac{W_{-2}(\mathfrak{t})}{(\mathcal{X} - A(\mathfrak{t}, \tilde{\nu}))^2} + \frac{W_{-1}(\mathfrak{t})}{\mathcal{X} - A(\mathfrak{t}, \tilde{\nu})} + W_0(\mathfrak{t}) + ...
	\end{equation}
where $w(\mathfrak{t})$ indeed develops a pole at $\mathfrak{t} = 1$ since $\mathcal{X}(\mathfrak{t} = 1) = A(1, \tilde{\nu})$, and
	\begin{align}
		\begin{split}
			W_{-2}(\mathfrak{t}) &= -\Big( \Big(-\mathfrak{t} \frac{d}{d\mathfrak{t}} + 2i \mathfrak{t}^{1/4} t_\star^{1/4} + \frac{i \tilde{\nu}}{2} \Big) A(\mathfrak{t}, \tilde{\nu}) \Big)^2 ,\\
			W_{-1}(\mathfrak{t}) &= -\Big(-\mathfrak{t} \frac{d}{d\mathfrak{t}} + 2i \mathfrak{t}^{1/4} t_\star^{1/4} + \frac{i \tilde{\nu}}{2} \Big)^2 A(\mathfrak{t}, \tilde{\nu}).
		\end{split}
	\end{align}
	When expanded in powers of $(\mathcal{X}-1)$, (\ref{inflambda}) and (\ref{inflambda_pole}) should coincide, and it determines $A(\mathfrak{t}, \tilde{\nu})$ and coefficients $\{W_i(\mathfrak{t})\}$:
\begin{align}
		\begin{split}
			W_0(\mathfrak{t}) &= t_\star^{1/2} \mathfrak{t}^{1/2} + \frac{5i+6\nu}{8} t_\star^{1/4} \mathfrak{t}^{1/4} + \frac{5\nu^2+5i\nu-3}{128} + O(t_\star^{-1/4}), \\
			W_1(\mathfrak{t}) &= \frac{i}{8} t_\star^{1/4} \mathfrak{t}^{1/4} - \frac{6\nu^2-26i\nu+13}{512} + O(t_\star^{-1/4}), \\
			W_2(\mathfrak{t}) &= \frac{\nu}{8} t_\star^{1/4} \mathfrak{t}^{1/4} + O(1).
		\end{split}
	\end{align}
The value of $w_0$ follows from (\ref{eq:poleExpansion}):
	\begin{equation}
		w_0 = \frac{2 t_\star^{1/2}}{3} + \frac{2\nu+i}{3} t_\star^{1/4} + \frac{2\nu^2+2i\nu-5}{48} + O(t_\star^{-1/4}).
	\end{equation}
	To compute integral in (\ref{intinf}) for $t_\star \to \infty$ we substitute the solution (\ref{inflambda}). Up to the zeroth order in $t_\star$ the integral becomes
	\begin{equation}
	\label{eq:intfract}
		\int_1^\infty \Big(t_\star^{1/2} \mathfrak{t}^{-1/2} P_1(\hat{\mathcal{X}}) + t_\star^{1/4} \mathfrak{t}^{-3/4} P_2(\hat{\mathcal{X}}) +t_\star^{1/4} \mathfrak{t}^{-1/2} P_3(\hat{\mathcal{X}}) - \frac{2}{(\mathfrak{t}-1)^2} \Big) d\mathfrak{t},
	\end{equation}
	where
	\begin{equation}
	\begin{gathered}
	    \hat{\mathcal{X}} = \frac{\mathcal{X}}{A(1,\tilde{\nu})} = \exp(8 i(\mathfrak{t}^{1/4}-1) t_\star^{1/4}) \mathfrak{t}^{i \tilde{\nu}/4},
\quad 
		P_1(\mathcal{X}) = \frac{192\mathcal{X}^4}{(1-\mathcal{X}^2)^4},
\\
		P_2(\mathcal{X}) = -\frac{4\mathcal{X}^2((i+2\nu)\mathcal{X}^6 + 4i\mathcal{X}^5 - 18i(i+\nu)^2\mathcal{X}^4 - 8i\mathcal{X}^3 - 18i\nu^2\mathcal{X}^2 + 4i\mathcal{X} - i - 2\nu)}{(1-\mathcal{X}^2)^5},
\\
		P_3(\mathcal{X}) =- \frac{36i \mathcal{X}^4(1+\mathcal{X}^2)(2\nu^2+2i\nu-1)}{(1-\mathcal{X}^2)^5}.
	\end{gathered}	
	\end{equation}
The integration in \eqref{eq:intfract} is over $\mathfrak{t}$, while the exponent inside $\hat{\mathcal{X}}$ contains large parameter $t_\star$ which controls the order of subsequent integrations. We proceed via integration by parts, and the result reads
	\begin{equation}
		\begin{gathered}
		\label{eq:mainasympt}
		\int_1^\infty \Big(t_\star^{1/2} \mathfrak{t}^{-1/2} P_1(\hat{\mathcal{X}}) + t_\star^{1/4} \mathfrak{t}^{-3/4} P_2(\hat{\mathcal{X}}) +t_\star^{1/4} \mathfrak{t}^{-1/2} P_3(\hat{\mathcal{X}}) - \frac{2}{(\mathfrak{t}-1)^2} \Big) d\mathfrak{t} = \\
		= 4it_\star^{1/4} + \frac{9}{8} + \frac{i \nu}{2} - \log 2 + O(t_\star^{-1/4}).
		\end{gathered}
	\end{equation}

\section{Coefficients of the general conformal block at infinity}
\label{app:confBlock}

Here we present first seven terms of the expansion of generic irregular block \eqref{eq:generalBlockInf} at infinity:
\begin{equation}
\begin{gathered}
\label{eq:64}
\mathsf{B}^{\infty}(\nu,\varepsilon_1,\varepsilon_2)=1+
\frac{\mathsf{N}_1(\nu,\varepsilon_1,\varepsilon_2)}{2^4\cdot\varepsilon_1\varepsilon_2r}+
\frac{\mathsf{N}_2(\nu,\varepsilon_1,\varepsilon_2)}{2^9\cdot (\varepsilon_1\varepsilon_2r)^2}+
\frac{\mathsf{N}_3(\nu,\varepsilon_1,\varepsilon_2)}{3\cdot 2^{13}\cdot (\varepsilon_1\varepsilon_2r)^3}+
\frac{\mathsf{N}_4(\nu,\varepsilon_1,\varepsilon_2)}{3\cdot 2^{19}\cdot (\varepsilon_1\varepsilon_2r)^4}+
\\+
\frac{\mathsf{N}_5(\nu,\varepsilon_1,\varepsilon_2)}{15\cdot 2^{23}\cdot (\varepsilon_1\varepsilon_2r)^5}+
\frac{\mathsf{N}_6(\nu,\varepsilon_1,\varepsilon_2)}{45\cdot 2^{28}\cdot (\varepsilon_1\varepsilon_2r)^6}+
\frac{\mathsf{N}_7(\nu,\varepsilon_1,\varepsilon_2)}{35\cdot 9\cdot 2^{32}\cdot (\varepsilon_1\varepsilon_2r)^7}+\ldots 
\end{gathered}
\end{equation}
They are
\begin{equation*}
\label{eq:72}
\mathsf{N}_1(\nu,\varepsilon_1,\varepsilon_2)=3\varepsilon_1^{2}\nu + 8\varepsilon_1\varepsilon_2\nu + 3\varepsilon_2^{2}\nu - 4\nu^{3}
\end{equation*}

\begin{equation*}
\begin{gathered}
\label{eq:73}
\mathsf{N}_2(\nu,\varepsilon_1,\varepsilon_2)=9\varepsilon_1^{5}\varepsilon_2 + 48\varepsilon_1^{4}\varepsilon_2^{2} + 78\varepsilon_1^{3}\varepsilon_2^{3} + 48\varepsilon_1^{2}\varepsilon_2^{4} + 9\varepsilon_1\varepsilon_2^{5} +\\
9\varepsilon_1^{4}\nu^{2} - 88\varepsilon_1^{3}\varepsilon_2\nu^{2} - 238\varepsilon_1^{2}\varepsilon_2^{2}\nu^{2} - 88\varepsilon_1\varepsilon_2^{3}\nu^{2} + 
9\varepsilon_2^{4}\nu^{2} - 24\varepsilon_1^{2}\nu^{4} + 16\varepsilon_1\varepsilon_2\nu^{4} - 24\varepsilon_2^{2}\nu^{4} + 16\nu^{6}
\end{gathered}
\end{equation*}

\begin{equation*}\begin{gathered}
\label{eq:74}
\mathsf{N}_3(\nu,\varepsilon_1,\varepsilon_2)=81\varepsilon_1^{7}\varepsilon_2\nu - 2592\varepsilon_1^{6}\varepsilon_2^{2}\nu - 13425\varepsilon_1^{5}\varepsilon_2^{3}\nu - 21312\varepsilon_1^{4}\varepsilon_2^{4}\nu - \\
13425\varepsilon_1^{3}\varepsilon_2^{5}\nu - 2592\varepsilon_1^{2}\varepsilon_2^{6}\nu + 81\varepsilon_1\varepsilon_2^{7}\nu + 27\varepsilon_1^{6}\nu^{3} - 
1116\varepsilon_1^{5}\varepsilon_2\nu^{3} + 7057\varepsilon_1^{4}\varepsilon_2^{2}\nu^{3} + 18552\varepsilon_1^{3}\varepsilon_2^{3}\nu^{3} + \\
7057\varepsilon_1^{2}\varepsilon_2^{4}\nu^{3} - 1116\varepsilon_1\varepsilon_2^{5}\nu^{3} + 27\varepsilon_2^{6}\nu^{3} - 108\varepsilon_1^{4}\nu^{5} 
+ 1776\varepsilon_1^{3}\varepsilon_2\nu^{5} + 552\varepsilon_1^{2}\varepsilon_2^{2}\nu^{5} + 1776\varepsilon_1\varepsilon_2^{3}\nu^{5} - \\
108\varepsilon_2^{4}\nu^{5} + 144\varepsilon_1^{2}\nu^{7} - 576\varepsilon_1\varepsilon_2\nu^{7} + 144\varepsilon_2^{2}\nu^{7} - 64\nu^{9}
\end{gathered}\end{equation*}

\begin{equation*}\begin{gathered}
\label{eq:75}
\mathsf{N}_4(\nu,\varepsilon_1,\varepsilon_2)=243\varepsilon_1^{10}\varepsilon_2^{2} - 44064\varepsilon_1^{9}\varepsilon_2^{3} - 313740\varepsilon_1^{8}\varepsilon_2^{4} - \\
831456\varepsilon_1^{7}\varepsilon_2^{5} - 1124046\varepsilon_1^{6}\varepsilon_2^{6} - 831456\varepsilon_1^{5}\varepsilon_2^{7} - 
313740\varepsilon_1^{4}\varepsilon_2^{8} - 44064\varepsilon_1^{3}\varepsilon_2^{9} + 243\varepsilon_1^{2}\varepsilon_2^{10} + 486\varepsilon_1^{9}\varepsilon_2\nu^{2} \\
- 41040\varepsilon_1^{8}\varepsilon_2^{2}\nu^{2} + 808128\varepsilon_1^{7}\varepsilon_2^{3}\nu^{2} + 
3941328\varepsilon_1^{6}\varepsilon_2^{4}\nu^{2} + 6105012\varepsilon_1^{5}\varepsilon_2^{5}\nu^{2} + \\
3941328\varepsilon_1^{4}\varepsilon_2^{6}\nu^{2} + 808128\varepsilon_1^{3}\varepsilon_2^{7}\nu^{2} - 41040\varepsilon_1^{2}\varepsilon_2^{8}\nu^{2} 
+ 486\varepsilon_1\varepsilon_2^{9}\nu^{2} + 81\varepsilon_1^{8}\nu^{4} - 7776\varepsilon_1^{7}\varepsilon_2\nu^{4} + \\
206052\varepsilon_1^{6}\varepsilon_2^{2}\nu^{4} - 830176\varepsilon_1^{5}\varepsilon_2^{3}\nu^{4} - 
2213466\varepsilon_1^{4}\varepsilon_2^{4}\nu^{4} - 830176\varepsilon_1^{3}\varepsilon_2^{5}\nu^{4} + \\
206052\varepsilon_1^{2}\varepsilon_2^{6}\nu^{4} - 7776\varepsilon_1\varepsilon_2^{7}\nu^{4} + 81\varepsilon_2^{8}\nu^{4} - 
432\varepsilon_1^{6}\nu^{6} + 21312\varepsilon_1^{5}\varepsilon_2\nu^{6} - 210448\varepsilon_1^{4}\varepsilon_2^{2}\nu^{6} - \\
166656\varepsilon_1^{3}\varepsilon_2^{3}\nu^{6} - 210448\varepsilon_1^{2}\varepsilon_2^{4}\nu^{6} + 21312\varepsilon_1\varepsilon_2^{5}\nu^{6} - 
432\varepsilon_2^{6}\nu^{6} + 864\varepsilon_1^{4}\nu^{8} - 19968\varepsilon_1^{3}\varepsilon_2\nu^{8} + \\
33216\varepsilon_1^{2}\varepsilon_2^{2}\nu^{8} - 19968\varepsilon_1\varepsilon_2^{3}\nu^{8} + 864\varepsilon_2^{4}\nu^{8} - 
768\varepsilon_1^{2}\nu^{10} + 5632\varepsilon_1\varepsilon_2\nu^{10} - 768\varepsilon_2^{2}\nu^{10} + 256\nu^{12}
\end{gathered}\end{equation*}

\begin{equation*}
\begin{gathered}
\label{eq:76}
\mathsf{N}_5(\nu,\varepsilon_1,\varepsilon_2)=3645\varepsilon_1^{12}\varepsilon_2^{2}\nu - 942840\varepsilon_1^{11}\varepsilon_2^{3}\nu + 54505737\varepsilon_1^{10}\varepsilon_2^{4}\nu + \\
363517920\varepsilon_1^{9}\varepsilon_2^{5}\nu + 918147258\varepsilon_1^{8}\varepsilon_2^{6}\nu + 1221269040\varepsilon_1^{7}\varepsilon_2^{7}\nu + \\
918147258\varepsilon_1^{6}\varepsilon_2^{8}\nu + 363517920\varepsilon_1^{5}\varepsilon_2^{9}\nu + 54505737\varepsilon_1^{4}\varepsilon_2^{10}\nu - \\
942840\varepsilon_1^{3}\varepsilon_2^{11}\nu + 3645\varepsilon_1^{2}\varepsilon_2^{12}\nu + 2430\varepsilon_1^{11}\varepsilon_2\nu^{3} - 
374220\varepsilon_1^{10}\varepsilon_2^{2}\nu^{3} + 19525950\varepsilon_1^{9}\varepsilon_2^{3}\nu^{3} - \\
275671824\varepsilon_1^{8}\varepsilon_2^{4}\nu^{3} - 1287899580\varepsilon_1^{7}\varepsilon_2^{5}\nu^{3} - 
1960171080\varepsilon_1^{6}\varepsilon_2^{6}\nu^{3} - 1287899580\varepsilon_1^{5}\varepsilon_2^{7}\nu^{3} - \\
275671824\varepsilon_1^{4}\varepsilon_2^{8}\nu^{3} + 19525950\varepsilon_1^{3}\varepsilon_2^{9}\nu^{3} - 
374220\varepsilon_1^{2}\varepsilon_2^{10}\nu^{3} + 2430\varepsilon_1\varepsilon_2^{11}\nu^{3} + 243\varepsilon_1^{10}\nu^{5} - \\
43200\varepsilon_1^{9}\varepsilon_2\nu^{5} + 2537055\varepsilon_1^{8}\varepsilon_2^{2}\nu^{5} - 52037920\varepsilon_1^{7}\varepsilon_2^{3}\nu^{5} + 
113231582\varepsilon_1^{6}\varepsilon_2^{4}\nu^{5} + 333908800\varepsilon_1^{5}\varepsilon_2^{5}\nu^{5} + \\
113231582\varepsilon_1^{4}\varepsilon_2^{6}\nu^{5} - 52037920\varepsilon_1^{3}\varepsilon_2^{7}\nu^{5} + 
2537055\varepsilon_1^{2}\varepsilon_2^{8}\nu^{5} - 43200\varepsilon_1\varepsilon_2^{9}\nu^{5} + 243\varepsilon_2^{10}\nu^{5} - \\
1620\varepsilon_1^{8}\nu^{7} + 164160\varepsilon_1^{7}\varepsilon_2\nu^{7} - 4892880\varepsilon_1^{6}\varepsilon_2^{2}\nu^{7} +
34381120\varepsilon_1^{5}\varepsilon_2^{3}\nu^{7} + 36999432\varepsilon_1^{4}\varepsilon_2^{4}\nu^{7} + \\
34381120\varepsilon_1^{3}\varepsilon_2^{5}\nu^{7} - 4892880\varepsilon_1^{2}\varepsilon_2^{6}\nu^{7} + 164160\varepsilon_1\varepsilon_2^{7}\nu^{7} 
- 1620\varepsilon_2^{8}\nu^{7} + 4320\varepsilon_1^{6}\nu^{9} - 253440\varepsilon_1^{5}\varepsilon_2\nu^{9} + \\
3337120\varepsilon_1^{4}\varepsilon_2^{2}\nu^{9} - 2296320\varepsilon_1^{3}\varepsilon_2^{3}\nu^{9} + 
3337120\varepsilon_1^{2}\varepsilon_2^{4}\nu^{9} - 253440\varepsilon_1\varepsilon_2^{5}\nu^{9} + 4320\varepsilon_2^{6}\nu^{9} - \\
5760\varepsilon_1^{4}\nu^{11} + 171520\varepsilon_1^{3}\varepsilon_2\nu^{11} - 600320\varepsilon_1^{2}\varepsilon_2^{2}\nu^{11} + 
171520\varepsilon_1\varepsilon_2^{3}\nu^{11} - 5760\varepsilon_2^{4}\nu^{11} + 3840\varepsilon_1^{2}\nu^{13} - \\
40960\varepsilon_1\varepsilon_2\nu^{13} + 3840\varepsilon_2^{2}\nu^{13} - 1024\nu^{15}
\end{gathered}
\end{equation*}

\begin{equation*}
\begin{gathered}
\label{eq:77}
\mathsf{N}_6(\nu,\varepsilon_1,\varepsilon_2)=10935\varepsilon_1^{15}\varepsilon_2^{3} - 6123600\varepsilon_1^{14}\varepsilon_2^{4} + 1870567830\varepsilon_1^{13}\varepsilon_2^{5} + \\
15399920880\varepsilon_1^{12}\varepsilon_2^{6} + 51915211065\varepsilon_1^{11}\varepsilon_2^{7} + 
99666119520\varepsilon_1^{10}\varepsilon_2^{8} + 122548253940\varepsilon_1^{9}\varepsilon_2^{9} + \\
99666119520\varepsilon_1^{8}\varepsilon_2^{10} + 51915211065\varepsilon_1^{7}\varepsilon_2^{11} + 
15399920880\varepsilon_1^{6}\varepsilon_2^{12} + 1870567830\varepsilon_1^{5}\varepsilon_2^{13} - 6123600\varepsilon_1^{4}\varepsilon_2^{14} + \\
10935\varepsilon_1^{3}\varepsilon_2^{15} + 32805\varepsilon_1^{14}\varepsilon_2^{2}\nu^{2} - 11518200\varepsilon_1^{13}\varepsilon_2^{3}\nu^{2} + 
1355959926\varepsilon_1^{12}\varepsilon_2^{4}\nu^{2} - 48923552664\varepsilon_1^{11}\varepsilon_2^{5}\nu^{2} - \\
306706300965\varepsilon_1^{10}\varepsilon_2^{6}\nu^{2} - 743963673456\varepsilon_1^{9}\varepsilon_2^{7}\nu^{2} - 
976770129132\varepsilon_1^{8}\varepsilon_2^{8}\nu^{2} - 743963673456\varepsilon_1^{7}\varepsilon_2^{9}\nu^{2} - \\
306706300965\varepsilon_1^{6}\varepsilon_2^{10}\nu^{2} - 48923552664\varepsilon_1^{5}\varepsilon_2^{11}\nu^{2} + 
1355959926\varepsilon_1^{4}\varepsilon_2^{12}\nu^{2} - 11518200\varepsilon_1^{3}\varepsilon_2^{13}\nu^{2} + \\
32805\varepsilon_1^{2}\varepsilon_2^{14}\nu^{2} + 10935\varepsilon_1^{13}\varepsilon_2\nu^{4} - 2653560\varepsilon_1^{12}\varepsilon_2^{2}\nu^{4} + 
249573150\varepsilon_1^{11}\varepsilon_2^{3}\nu^{4} - 10060239960\varepsilon_1^{10}\varepsilon_2^{4}\nu^{4} + \\
101230015113\varepsilon_1^{9}\varepsilon_2^{5}\nu^{4} + 465473120976\varepsilon_1^{8}\varepsilon_2^{6}\nu^{4} + 
701077918020\varepsilon_1^{7}\varepsilon_2^{7}\nu^{4} + 465473120976\varepsilon_1^{6}\varepsilon_2^{8}\nu^{4} + \\
101230015113\varepsilon_1^{5}\varepsilon_2^{9}\nu^{4} - 10060239960\varepsilon_1^{4}\varepsilon_2^{10}\nu^{4} + 
249573150\varepsilon_1^{3}\varepsilon_2^{11}\nu^{4} - 2653560\varepsilon_1^{2}\varepsilon_2^{12}\nu^{4} + \\
10935\varepsilon_1\varepsilon_2^{13}\nu^{4} + 729\varepsilon_1^{12}\nu^{6} - 211896\varepsilon_1^{11}\varepsilon_2\nu^{6} + 
22012614\varepsilon_1^{10}\varepsilon_2^{2}\nu^{6} - 998646360\varepsilon_1^{9}\varepsilon_2^{3}\nu^{6} + \\
16573211767\varepsilon_1^{8}\varepsilon_2^{4}\nu^{6} - 12533430384\varepsilon_1^{7}\varepsilon_2^{5}\nu^{6} - 
54762121708\varepsilon_1^{6}\varepsilon_2^{6}\nu^{6} - 12533430384\varepsilon_1^{5}\varepsilon_2^{7}\nu^{6} + \\
16573211767\varepsilon_1^{4}\varepsilon_2^{8}\nu^{6} - 998646360\varepsilon_1^{3}\varepsilon_2^{9}\nu^{6} + 
22012614\varepsilon_1^{2}\varepsilon_2^{10}\nu^{6} - 211896\varepsilon_1\varepsilon_2^{11}\nu^{6} + 729\varepsilon_2^{12}\nu^{6} - \\
5832\varepsilon_1^{10}\nu^{8} + 1017360\varepsilon_1^{9}\varepsilon_2\nu^{8} - 61096680\varepsilon_1^{8}\varepsilon_2^{2}\nu^{8} + 
1387273920\varepsilon_1^{7}\varepsilon_2^{3}\nu^{8} - 7019554512\varepsilon_1^{6}\varepsilon_2^{4}\nu^{8} - \\
8853922464\varepsilon_1^{5}\varepsilon_2^{5}\nu^{8} - 7019554512\varepsilon_1^{4}\varepsilon_2^{6}\nu^{8} + 
1387273920\varepsilon_1^{3}\varepsilon_2^{7}\nu^{8} - 61096680\varepsilon_1^{2}\varepsilon_2^{8}\nu^{8} + \\
1017360\varepsilon_1\varepsilon_2^{9}\nu^{8} - 5832\varepsilon_2^{10}\nu^{8} + 19440\varepsilon_1^{8}\nu^{10} - 
2177280\varepsilon_1^{7}\varepsilon_2\nu^{10} + 73886400\varepsilon_1^{6}\varepsilon_2^{2}\nu^{10} - \\
694890240\varepsilon_1^{5}\varepsilon_2^{3}\nu^{10} + 82687392\varepsilon_1^{4}\varepsilon_2^{4}\nu^{10} - 
694890240\varepsilon_1^{3}\varepsilon_2^{5}\nu^{10} + 73886400\varepsilon_1^{2}\varepsilon_2^{6}\nu^{10} - \\
2177280\varepsilon_1\varepsilon_2^{7}\nu^{10} + 19440\varepsilon_2^{8}\nu^{10} - 34560\varepsilon_1^{6}\nu^{12} + 
2361600\varepsilon_1^{5}\varepsilon_2\nu^{12} - 38800640\varepsilon_1^{4}\varepsilon_2^{2}\nu^{12} + \\
78743040\varepsilon_1^{3}\varepsilon_2^{3}\nu^{12} - 38800640\varepsilon_1^{2}\varepsilon_2^{4}\nu^{12} + 
2361600\varepsilon_1\varepsilon_2^{5}\nu^{12} - 34560\varepsilon_2^{6}\nu^{12} + 34560\varepsilon_1^{4}\nu^{14} - \\
1259520\varepsilon_1^{3}\varepsilon_2\nu^{14} + 6643200\varepsilon_1^{2}\varepsilon_2^{2}\nu^{14} - 1259520\varepsilon_1\varepsilon_2^{3}\nu^{14} +
34560\varepsilon_2^{4}\nu^{14} - 18432\varepsilon_1^{2}\nu^{16} + \\
258048\varepsilon_1\varepsilon_2\nu^{16} - 18432\varepsilon_2^{2}\nu^{16} + 4096\nu^{18}
\end{gathered}
\end{equation*}

\begin{equation*}\small\begin{gathered}
\mathsf{N}_7(\nu,\varepsilon_1,\varepsilon_2)=229635\varepsilon_1^{17}\varepsilon_2^{3}\nu - 155539440\varepsilon_1^{16}\varepsilon_2^{4}\nu + 55884131793\varepsilon_1^{15}\varepsilon_2^{5}\nu \\
- 5391195581664\varepsilon_1^{14}\varepsilon_2^{6}\nu - 41384943697797\varepsilon_1^{13}\varepsilon_2^{7}\nu - 
132047280084240\varepsilon_1^{12}\varepsilon_2^{8}\nu - 245411586797391\varepsilon_1^{11}\varepsilon_2^{9}\nu - \\
298557581216832\varepsilon_1^{10}\varepsilon_2^{10}\nu - 245411586797391\varepsilon_1^{9}\varepsilon_2^{11}\nu - 
132047280084240\varepsilon_1^{8}\varepsilon_2^{12}\nu - 41384943697797\varepsilon_1^{7}\varepsilon_2^{13}\nu - \\
5391195581664\varepsilon_1^{6}\varepsilon_2^{14}\nu + 55884131793\varepsilon_1^{5}\varepsilon_2^{15}\nu - 
155539440\varepsilon_1^{4}\varepsilon_2^{16}\nu + 229635\varepsilon_1^{3}\varepsilon_2^{17}\nu + 229635\varepsilon_1^{16}\varepsilon_2^{2}\nu^{3} - \\
105632100\varepsilon_1^{15}\varepsilon_2^{3}\nu^{3} + 18965185533\varepsilon_1^{14}\varepsilon_2^{4}\nu^{3} - 
1530711243432\varepsilon_1^{13}\varepsilon_2^{5}\nu^{3} + 39264115062519\varepsilon_1^{12}\varepsilon_2^{6}\nu^{3} + \\
234925104022308\varepsilon_1^{11}\varepsilon_2^{7}\nu^{3} + 552197105280009\varepsilon_1^{10}\varepsilon_2^{8}\nu^{3} + 
717652495077840\varepsilon_1^{9}\varepsilon_2^{9}\nu^{3} + 552197105280009\varepsilon_1^{8}\varepsilon_2^{10}\nu^{3} + \\
234925104022308\varepsilon_1^{7}\varepsilon_2^{11}\nu^{3} + 39264115062519\varepsilon_1^{6}\varepsilon_2^{12}\nu^{3} - 
1530711243432\varepsilon_1^{5}\varepsilon_2^{13}\nu^{3} + 18965185533\varepsilon_1^{4}\varepsilon_2^{14}\nu^{3} - \\
105632100\varepsilon_1^{3}\varepsilon_2^{15}\nu^{3} + 229635\varepsilon_1^{2}\varepsilon_2^{16}\nu^{3} + 
45927\varepsilon_1^{15}\varepsilon_2\nu^{5} - 16186716\varepsilon_1^{14}\varepsilon_2^{2}\nu^{5} + \\
2366650629\varepsilon_1^{13}\varepsilon_2^{3}\nu^{5} - 174442592520\varepsilon_1^{12}\varepsilon_2^{4}\nu^{5} + 
5786660199159\varepsilon_1^{11}\varepsilon_2^{5}\nu^{5} - 38906707376292\varepsilon_1^{10}\varepsilon_2^{6}\nu^{5} - \\
182914338354531\varepsilon_1^{9}\varepsilon_2^{7}\nu^{5} - 274659838097904\varepsilon_1^{8}\varepsilon_2^{8}\nu^{5} - 
182914338354531\varepsilon_1^{7}\varepsilon_2^{9}\nu^{5} - 38906707376292\varepsilon_1^{6}\varepsilon_2^{10}\nu^{5} + \\
5786660199159\varepsilon_1^{5}\varepsilon_2^{11}\nu^{5} - 174442592520\varepsilon_1^{4}\varepsilon_2^{12}\nu^{5} + 
2366650629\varepsilon_1^{3}\varepsilon_2^{13}\nu^{5} - 16186716\varepsilon_1^{2}\varepsilon_2^{14}\nu^{5} + \\
45927\varepsilon_1\varepsilon_2^{15}\nu^{5} + 2187\varepsilon_1^{14}\nu^{7} - 959364\varepsilon_1^{13}\varepsilon_2\nu^{7} +
156244221\varepsilon_1^{12}\varepsilon_2^{2}\nu^{7} - 12286588104\varepsilon_1^{11}\varepsilon_2^{3}\nu^{7} +\\ 
465501666903\varepsilon_1^{10}\varepsilon_2^{4}\nu^{7} - 6261628992892\varepsilon_1^{9}\varepsilon_2^{5}\nu^{7} -
1711641220383\varepsilon_1^{8}\varepsilon_2^{6}\nu^{7} + 7163451093904\varepsilon_1^{7}\varepsilon_2^{7}\nu^{7} -\\ 
1711641220383\varepsilon_1^{6}\varepsilon_2^{8}\nu^{7} - 6261628992892\varepsilon_1^{5}\varepsilon_2^{9}\nu^{7} + 
465501666903\varepsilon_1^{4}\varepsilon_2^{10}\nu^{7} - 12286588104\varepsilon_1^{3}\varepsilon_2^{11}\nu^{7} + \\
156244221\varepsilon_1^{2}\varepsilon_2^{12}\nu^{7} - 959364\varepsilon_1\varepsilon_2^{13}\nu^{7} + 2187\varepsilon_2^{14}\nu^{7} - 
20412\varepsilon_1^{12}\nu^{9} + 5524848\varepsilon_1^{11}\varepsilon_2\nu^{9} - 559381032\varepsilon_1^{10}\varepsilon_2^{2}\nu^{9} + \\
25652744880\varepsilon_1^{9}\varepsilon_2^{3}\nu^{9} - 476390311684\varepsilon_1^{8}\varepsilon_2^{4}\nu^{9} + 
1668037353696\varepsilon_1^{7}\varepsilon_2^{5}\nu^{9} + 2332861063888\varepsilon_1^{6}\varepsilon_2^{6}\nu^{9} + \\
1668037353696\varepsilon_1^{5}\varepsilon_2^{7}\nu^{9} - 476390311684\varepsilon_1^{4}\varepsilon_2^{8}\nu^{9} + 
25652744880\varepsilon_1^{3}\varepsilon_2^{9}\nu^{9} - 559381032\varepsilon_1^{2}\varepsilon_2^{10}\nu^{9} + \\
5524848\varepsilon_1\varepsilon_2^{11}\nu^{9} - 20412\varepsilon_2^{12}\nu^{9} + 81648\varepsilon_1^{10}\nu^{11} -
15059520\varepsilon_1^{9}\varepsilon_2\nu^{11} + 968330160\varepsilon_1^{8}\varepsilon_2^{2}\nu^{11} - \\
24485180160\varepsilon_1^{7}\varepsilon_2^{3}\nu^{11} + 174548349024\varepsilon_1^{6}\varepsilon_2^{4}\nu^{11} + 
43770125952\varepsilon_1^{5}\varepsilon_2^{5}\nu^{11} + 174548349024\varepsilon_1^{4}\varepsilon_2^{6}\nu^{11} - \\
24485180160\varepsilon_1^{3}\varepsilon_2^{7}\nu^{11} + 968330160\varepsilon_1^{2}\varepsilon_2^{8}\nu^{11} - 
15059520\varepsilon_1\varepsilon_2^{9}\nu^{11} + 81648\varepsilon_2^{10}\nu^{11} - 181440\varepsilon_1^{8}\nu^{13} + \\
22498560\varepsilon_1^{7}\varepsilon_2\nu^{13} - 863143680\varepsilon_1^{6}\varepsilon_2^{2}\nu^{13} + 
10046032640\varepsilon_1^{5}\varepsilon_2^{3}\nu^{13} - 12006384768\varepsilon_1^{4}\varepsilon_2^{4}\nu^{13} + \\
10046032640\varepsilon_1^{3}\varepsilon_2^{5}\nu^{13} - 863143680\varepsilon_1^{2}\varepsilon_2^{6}\nu^{13} + 
22498560\varepsilon_1\varepsilon_2^{7}\nu^{13} - 181440\varepsilon_2^{8}\nu^{13} + 241920\varepsilon_1^{6}\nu^{15} - \\
18902016\varepsilon_1^{5}\varepsilon_2\nu^{15} + 372286208\varepsilon_1^{4}\varepsilon_2^{2}\nu^{15} - 
1227835392\varepsilon_1^{3}\varepsilon_2^{3}\nu^{15} + 372286208\varepsilon_1^{2}\varepsilon_2^{4}\nu^{15} - \\
18902016\varepsilon_1\varepsilon_2^{5}\nu^{15} + 241920\varepsilon_2^{6}\nu^{15} - 193536\varepsilon_1^{4}\nu^{17} + 
8343552\varepsilon_1^{3}\varepsilon_2\nu^{17} - 58533888\varepsilon_1^{2}\varepsilon_2^{2}\nu^{17} + \\
8343552\varepsilon_1\varepsilon_2^{3}\nu^{17} 
- 193536\varepsilon_2^{4}\nu^{17} + 86016\varepsilon_1^{2}\nu^{19} - 1490944\varepsilon_1\varepsilon_2\nu^{19} + 
86016\varepsilon_2^{2}\nu^{19} - 16384\nu^{21}
\end{gathered}
\end{equation*}

\bibliographystyle{JHEP}
\bibliography{P3block.bib}

\end{document}